\documentclass[A4paper;11pt]{article}

\usepackage[T1]{fontenc}
\usepackage[utf8]{inputenc}
\usepackage[a4paper]{geometry}
\usepackage{amsmath}
\usepackage{amsfonts}
\usepackage{amssymb} 
\usepackage{graphicx}
\usepackage{a4wide}
\usepackage{hepunits}
\usepackage{url}
\usepackage{supertabular}
\usepackage{xspace}
\usepackage{latexsym}
\usepackage{textcomp}
\usepackage{wrapfig}
\usepackage{subfig}
\usepackage{slashed}
\usepackage{hepunits}
\usepackage{setspace}
\usepackage[dvipsnames]{xcolor}
\usepackage{bm}
\usepackage[nottoc]{tocbibind}
\usepackage{subfig}
\usepackage[numbers,sort&compress]{natbib}

\newcommand{\eqfig}[2]{\vcenter{\hbox{\includegraphics[width=#1\textwidth]{#2}}}}

\newcommand{\refeq}[1]{(\ref{#1})}

\newcommand{\refcite}[1]{Ref.~\cite{#1}}
\newcommand{\refscite}[1]{Refs.~\cite{#1}}
\newcommand{\refsec}[1]{Sec.~\ref{#1}}

\newcommand{\ie}{\textit{i.e. }}
\newcommand{\eg}{\textit{e.g. }}
\newcommand{\etal}{\textit{et al.\xspace}}

\newcommand{\ds}{Dyson-Schwinger\xspace}
\newcommand{\dse}{Dyson-Schwinger equation\xspace}
\newcommand{\dses}{Dyson-Schwinger equations\xspace}

\newcommand{\bs}{Bethe-Salpeter\xspace}
\newcommand{\bse}{Bethe-Salpeter equation\xspace}

\newcommand{\ket}[1]{\ensuremath{\left|#1\right\rangle}\xspace}
\newcommand{\bra}[1]{\ensuremath{\left\langle #1\right|}\xspace}

\newcommand{\Dd}[0]{\ensuremath{\frac{\Delta}{2}}}

\newcommand{\Pn}[0]{\ensuremath{P \cdot n}}
\newcommand{\g}[0]{\ensuremath{\gamma}}
\newcommand{\G}[0]{\ensuremath{\Gamma}}

\newcommand{\cond}[1]{\ensuremath{\left\langle #1 \right\rangle}\xspace} 
\newcommand{\MM}[0]{\ensuremath{\mathcal{M}_m(\xi,t)}}
\newcommand{\vperp}[1]{\ensuremath{\textbf{#1}_\perp}}
\newcommand{\DDF}[0]{\ensuremath{F^q(\beta,\alpha,t)}}
\newcommand{\DDG}[0]{\ensuremath{G^q(\beta,\alpha,t)}}

\title{From Bethe-Salpeter Wave Functions\\
to Generalised Parton Distributions}

\author{C.~Mezrag$^1$\footnote{cmezrag@anl.gov}, H.~Moutarde$^2$\footnote{herve.moutarde@cea.fr},
 J.~Rodr\'iguez-Quintero$^3$\footnote{jose.rodriguez@dfaie.uhu.es}, \\
{\small \textit{$^1$ Physics Division, Argonne National Laboratory}} \\
{\small \textit{Argonne, Illinois 60439, USA}} \\
{\small \textit{$^2$ CEA, Centre de Saclay, IRFU/Service de Physique Nucl\'eaire}} \\
{\small \textit{F-91191 Gif-sur-Yvette, France}} \\
{\small \textit{$^3$ Departamento de F\'isica Aplicada, Facultad de Ciencias Experimentales}} \\
{\small \textit{Universidad de Huelva, Huelva 21071, Spain.}}
}

\begin{document}
\maketitle

\begin{abstract}
We review recent works on the modelling of Generalised Parton Distributions within the Dyson-Schwinger formalism. We highlight how covariant computations, using the impulse approximation, allows one to fulfil most of the theoretical constraints of the GPDs. Specific attention is brought to chiral properties and especially the so-called soft pion theorem, and its link with the Axial-Vector Ward-Takahashi identity. The limitation of the impulse approximation are also explained. Beyond impulse approximation computations are reviewed in the forward case. Finally, we stress the advantages of the overlap of lightcone wave functions, and possible ways to construct covariant GPD models within this framework, in a two-body approximation. 
\end{abstract}

\tableofcontents

\section{Introduction}
\label{sec:Intro}

Since they have been introduced in the 1990s, Generalised Parton Distributions \cite{Mueller:1998fv,Ji:1996nm,Radyushkin:1997ki} (GPDs) have been under a strong scientific investigation, both experimentally and theoretically, as testify the number of dedicated review papers \cite{Ji:1998pc,Goeke:2001tz,Diehl:2003ny,Belitsky:2005qn,Boffi:2007yc,Guidal:2013rya}. Recently, the interest for GPDs have even been increased by the publication of new experimental data \cite{Defurne:2015lna,Jo:2015ema}. In the near future, the upgrade of the Jefferson Laboratory facility from $6~\GeV$ to $12~\GeV$ should provide the GPD community with very accurate experimental data on a wide kinematic range in the so-called ``valence region''. In the kinematic region dominated by sea quarks and gluon contributions, COMPASS in a short time scale \cite{Sandacz:2015pka}, and EIC in a longer time scale \cite{Accardi:2012qut} should also provide brand new experimental data.

On the theoretical side, the GPD framework is now well established. Evolution equations are known up to next-to-leading order, and some of the higher-twist corrections have been implemented successfully up to twist-four \cite{Braun:2012bg,Braun:2012hq}. From this situation, several phenomenological models have flourished \cite{Guidal:2004nd,Goloskokov:2005sd,Polyakov:2008aa,Kumericki:2009uq,Goldstein:2010gu}, allowing the description of worldwide available experimental data, with a reasonable accuracy. Using those phenomenological models, extraction of GPDs from experiments has been performed, providing a three-dimensional sketch of the nucleon, but still in the limit of the considered approximations, like for instance, computations at leading order in $\alpha_s$, the strong coupling constant (see \eg \refcite{Guidal:2013rya}). If few attempts have been made to compute GPDs beyond phenomenological models \cite{Polyakov:1999gs,Anikin:1999pf,Broniowski:2003rp, Broniowski:2007si,Dorokhov:2011ew}, they have not encountered the successes of the latter ones, when compared to experimental data. Nonetheless, it should be emphasised that such a task plays a key role in the ambitious purpose of validating our basic understanding of the strong interactions by the use of GPDs. 

As GPDs encode non-perturbative information on hadrons, any model built within a bottom-up approach must be intrinsically non-perturbative. Among the different possible ways, we focus here on the \dses \cite{Dyson:1949ha, Schwinger:1951ex, Schwinger:1951hq} (DSEs) and on the \bse \cite{Salpeter:1951sz, GellMann:1951rw, Schwinger:1951ex, Schwinger:1951hq, Schwinger:1953tb} (BSE). DSEs consist in coupled, self-consistent equations, relating the Green functions of QCD among themselves. The BSE is also a self-consistent equation, coupled to the Green functions of QCD, allowing one to compute the so-called \bs wave function for a given meson. In the recent years, significant progresses have been made in order to solve consistently these equations in a symmetry-preserving approximation scheme \cite{Chang:2009zb}. It is now possible to compute within the DSEs framework, a significant number of observables, like mesons masses and decay constants (see \eg \refcite{Chang:2011ei}) or form factors \cite{Roberts:2011wy} with a reasonable accuracy.  

The idea of computing hadron structure within a DSEs-BSE framework is not new, since one can tally different approaches to compute GPDs from DSEs-BSE developed during the last fifteen years \cite{Tiburzi:2002tq, Theussl:2002xp, Bissey:2003yr, VanDyck:2007jt, Frederico:2009fk}. Nevertheless, it was not possible at that time to rely on proper symmetry-preserving truncation schemes. This has recently changed with the computation of the pion Distribution Amplitude (DA) \cite{Chang:2013pq}, paving the way for similar developments on more complex objects, especially on PDFs \cite{Chang:2014lva} and GPDs \cite{Mezrag:2014tva,Mezrag:2014jka}.

Giving technical details on GPDs and DSEs would require a large number of pages. \emph{A contrario}, the present work focuses on a specific aim: highlighting progresses made in modelling GPDs in a DSEs framework. Therefore, the reader looking for extended proofs is invited to refer to the original work or to the following reviews \refscite{Diehl:2003ny,Belitsky:2005qn} for GPDs and \refscite{Maris:2003vk,Cloet:2013jya} for DSEs. 

The present review is organised as it follows. In section \ref{sec:HadronStructure} the main properties of GPDs are outlined. In addition, a related object, called Double Distribution (DD) is introduced. Then, in section \ref{sec:DSEs}, basic facts on DSEs-BSE are given, including insight of techniques and truncation schemes used to solved the coupled equations. Section \ref{sec:CovariantComputations} highlights the advantages and drawbacks of computing pion GPDs in a covariant framework.
Light is also shed on chiral properties of the chiral-even pion GPD in section \ref{sec:SPT}, showing that the rainbow ladder truncation scheme is consistent with the soft pion theorem. Section \ref{sec:BeyondImpulse} shows the limitations of the so-called impulse approximation, mainly used in covariant approaches, and the way to go beyond it in the forward case. In section \ref{sec:LightconeFormalism}, the possibility to compute GPDs using the lightcone formalism is emphasised, and the consequences in terms of DSEs are explored. Finally, the conclusion is given in section \ref{sec:Conclusion}.


\section{Hadron structure}
\label{sec:HadronStructure}
This section is devoted to introduce the definition, basic properties and main features related to GPDs. Then, we pay special attention to the modelling approach based on the so-called double distributions that have the merit to guarantee, by construction, the fulfilling of the main properties resulting from the observation of fundamental symmetries for the GPDs. 
In further sections, this fulfilling  will be, as should be, a cornerstone for the building of a GPD model based on the computational framework provided by DSE and BSE.

\subsection{GPD}

Starting by the definition of GPDs\footnote{We stick on chiral-even GPDs only. Transversity GPDs are described in \refcite{Diehl:2001pm}.}, we discuss then their main properties and features, as their evolution with the factorisation scale or extraction from exclusive processes.  

\subsubsection{Definition}

Formally, GPDs are defined as the Fourier transform of a non-local matrix element. In the case of a spin-$1/2$ hadron, two GPDs are required to fully parameterise the following matrix element:
\begin{eqnarray}
  \label{eq:GPDDef}
 F^q & = & \frac{1}{2} \int \frac{\textrm{d}z^-}{2\pi}e^{ixP^+ z^-} \left. \bra{p_2}\bar{\psi}^q\left(-\frac{z}{2}\right) \g^+ \psi^q\left( \frac{z}{2} \right) \ket{p_1} \right|_{z^+=z_\perp = 0} \nonumber \\
 & = & \frac{1}{2P^+}\left[ H^q(x,\xi,t)\bar{u}\left(p_2\right)\g^+ u\left(p_1\right) + E^q(x, \xi, t)\bar{u}\left(p_2\right)\frac{i \sigma^{+ \mu }\Delta_\mu }{2M}u\left(p_1\right)  \right],
\end{eqnarray}
with $\psi^q$ being the quark field of flavour $q$, $p_1$ and $p_2$ are respectively the momenta of the incoming and outgoing hadron, $P = \frac{p_1+p_2}{2}$ and $\Delta = p_2-p_1$. The lightcone variables are defined in appendix \ref{sec:AppendixA}.  $\xi = -\frac{\Delta^+}{2 P^+}$ is called the skewness, $t$ is the Mandelstam variable such that $t=\Delta^2$, and $M$ is the mass of the considered hadron. Working in the lightcone gauge, the Wilson line $\left[-\frac{z}{2};\frac{z}{2}\right]$ reduces to $1$ and is thus omitted here. Additional quark GPDs can be defined through:
\begin{eqnarray}
  \label{eq:tGPDDef}
   \tilde{F}^q & = & \frac{1}{2} \int \frac{\textrm{d}z^-}{2\pi}e^{ixP^+ z^-} \left. \bra{p_2}\bar{\psi}^q\left(-\frac{z}{2}\right) \g^+ \g_5 \psi^q\left( \frac{z}{2} \right) \ket{p_1} \right|_{z^+=z_\perp = 0} \nonumber \\
 & = & \frac{1}{2P^+}\left[ \tilde{H}^q(x,\xi,t)\bar{u}\left(p_2\right)\g^+ \g_5  u\left(p_1\right) + \tilde{E}^q(x, \xi, t)\bar{u}\left(p_2\right)\frac{\g_5\Delta^+ }{2M}u\left(p_1\right)  \right],
\end{eqnarray}
as well as gluon GPDs:
\begin{eqnarray}
  \label{eq:GluonGPDDef}
   F^g & = & \frac{1}{P^+} \int \frac{\textrm{d}z^-}{2\pi}e^{ixP^+ z^-} \left. \bra{p_2}G^{+\mu}\left(-\frac{z}{2}\right) G_\mu^{~+}\left( \frac{z}{2} \right) \ket{p_1} \right|_{z^+=z_\perp = 0} \nonumber \\
 & = & \frac{1}{2P^+}\left[ H^g(x,\xi,t)\bar{u}\left(p_2\right)\g^+ u\left(p_1\right) + E^g(x, \xi, t)\bar{u}\left(p_2\right)\frac{i \sigma^{+ \mu }\Delta_\mu }{2M}u\left(p_1\right)  \right], \\
 \label{eq:tGluonGPDDef}
 \tilde{F}^g & = & \frac{1}{2} \int \frac{\textrm{d}z^-}{2\pi}e^{ixP^+ z^-} \left. \bra{p_2}G^{+\mu}\left(-\frac{z}{2}\right) \tilde{G}_\mu^{~+}\left( \frac{z}{2} \right) \ket{p_1} \right|_{z^+=z_\perp = 0} \nonumber \\
 & = & \frac{1}{2P^+}\left[ \tilde{H}^g(x,\xi,t)\bar{u}\left(p_2\right)\g^+ \g_5  u\left(p_1\right) + \tilde{E}^g(x, \xi, t)\bar{u}\left(p_2\right)\frac{\g_5\Delta^+ }{2M}u\left(p_1\right)  \right],
\end{eqnarray}
where $G^{\mu\nu}$ is the gluon field strength and $\tilde{G}^{\mu\nu} = \frac{1}{2}\epsilon_{\mu\nu\rho\sigma}G^{\rho\sigma}$. In the case of the pion, which will be our main topic later on, spinlessness and discrete symmetries restrict the parameterisation of these matrix elements to two GPDs: 
\begin{eqnarray}
  \label{eq:PionGPDDef}
  H_\pi^q(x,\xi,t) & = & \frac{1}{2} \int \frac{\textrm{d}z^-}{2\pi}e^{ixP^+ z^-} \left. \bra{p_2}\bar{\psi}^q\left(-\frac{z}{2}\right) \g^+ \psi^q\left( \frac{z}{2} \right) \ket{p_1} \right|_{z^+=z_\perp = 0} ,\\
  \label{eq:PionGluonGPDDef}
  H_\pi^g(x,\xi,t) & = & \frac{1}{P^+} \int \frac{\textrm{d}z^-}{2\pi}e^{ixP^+ z^-} \left. \bra{p_2}G^{+\mu}\left(-\frac{z}{2}\right) G_\mu^{~+}\left( \frac{z}{2} \right) \ket{p_1} \right|_{z^+=z_\perp = 0}.
\end{eqnarray}
In the following, unless explicitly said, we will focus on the pion quark GPD $H_\pi^q$.

\subsubsection{Support and symmetry properties}

\begin{figure}[t]
  \centering
  \includegraphics[width=0.25\textwidth]{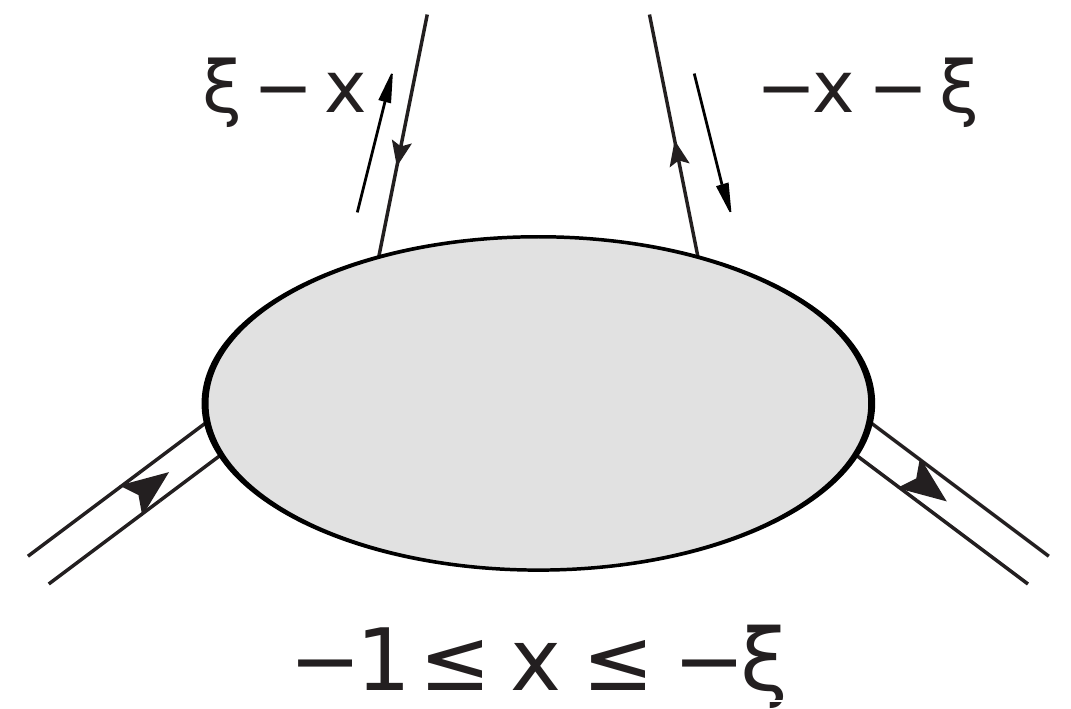}
  \quad
  \includegraphics[width=0.25\textwidth]{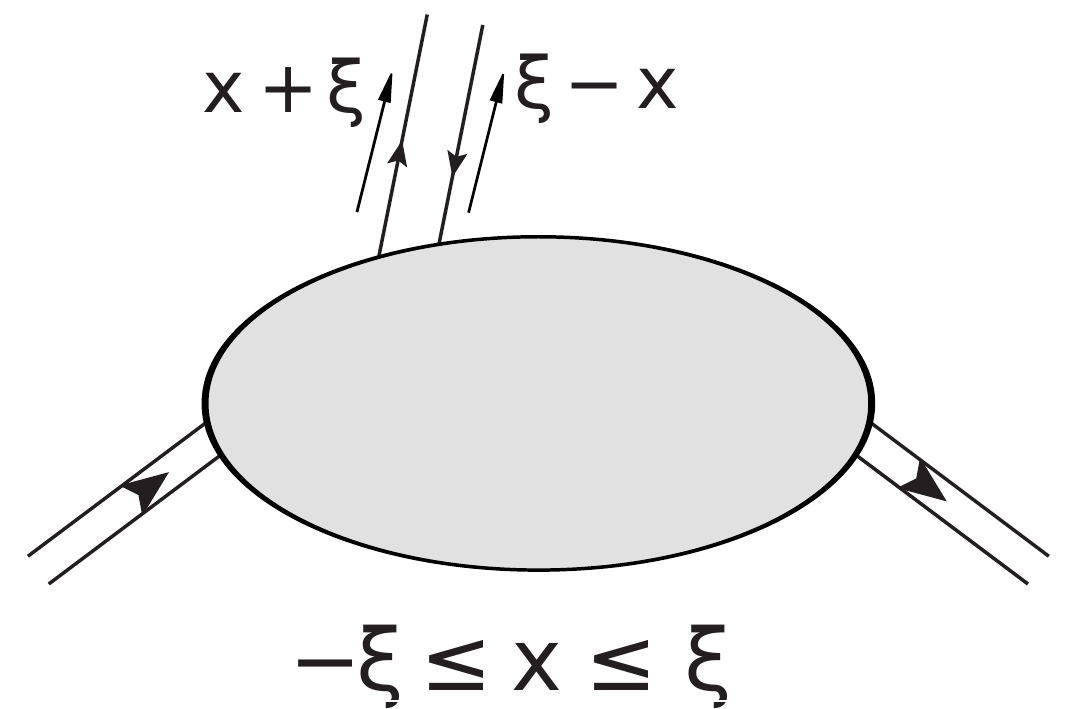}
  \quad
  \includegraphics[width=0.25\textwidth]{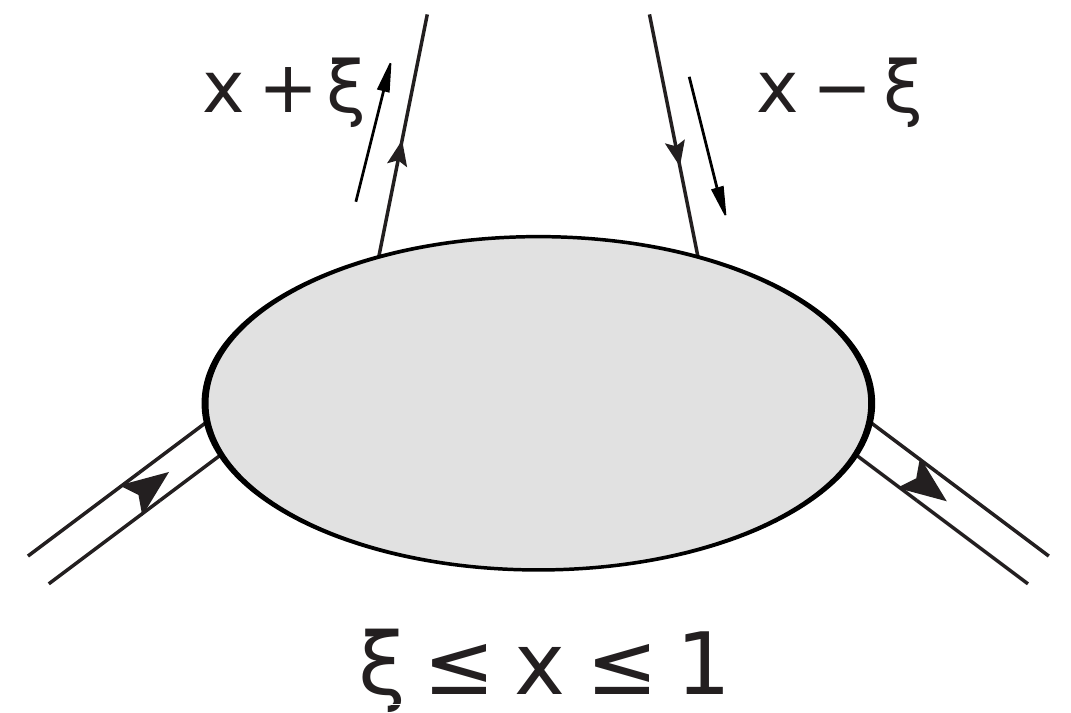}
  \caption{Different interpretations of the GPD. Left-hand side: kinematic corresponding to an antiquark; centre: kinematic corresponding to a quark antiquark pair; right-hand side: kinematic corresponding to a quark.}
  \label{fig:Interpretation}
\end{figure}

From the definition of GPDs in terms of matrix elements, it is possible to derive theoretical properties coming from field theory. First of all, one has to deal with the so-called support properties of the GPDs, stating that:
\begin{equation}
  \label{eq:SupportProperties}
  (x,\xi) \in \left[-1;1\right]^2,
\end{equation}
which can be obtained from amplitude computations \cite{Diehl:1998sm}. Depending on respective values of $x$ and $\xi$, one has different possible interpretations of the GPDs as shown on figure \ref{fig:Interpretation}.
The two lines $x =\pm \xi$ play a crucial role. They define the central region, often called ERBL, to the outer one called the DGLAP region. In order to ensure the factorisation of processes between a hard part computed using pQCD and soft parts including GPDs, the latter must be continuous on the lines $x = \pm \xi$ (see section \ref{sec:Exclusive}). In addition to $x$, $\xi$ and $t$, GPDs also depend on a factorisation scale $\mu_F$, and a renormalisation scale $\mu_R$, usually taken equal. These dependences are omitted for brevity but will be discussed below in this section.

GPDs are also constrained by discrete symmetries. Gluons being their own antiparticles, gluon GPDs $H^g$ and $E^g$ are even in $x$, whereas $\tilde{H}^g$ and $\tilde{E}^g$ are odd. Quark GPDs generally do not have any symmetry properties with respect to $x$, but still, combinations can be defined as:
\begin{eqnarray}
  \label{eq:SingletGPD}
  H^{q(+)}(x,\xi,t) & = & H^q(x,\xi,t)-H^q(-x,\xi,t), \nonumber \\
  \tilde{H}^{q(+)}(x,\xi,t) & = & \tilde{H}^q(x,\xi,t)+\tilde{H}^q(-x,\xi,t) ,
\end{eqnarray}
and correspond to an exchange of a charge $C=1$ in the $t$-channel. These combinations are often called ``singlet'' and can be extended to $E^q$ and $\tilde{E}^q$. The complementary combinations, labelled $H^{(-)}$, are called non-singlet and correspond to the exchange of a charge $C=-1$ in the $t$-channel:
\begin{eqnarray}
  \label{eq:NonSingletGPD}
  H^{q(-)}(x,\xi,t) & = & H^q(x,\xi,t)+H^q(-x,\xi,t), \nonumber \\
  \tilde{H}^{q(-)}(x,\xi,t) & = & \tilde{H}^q(x,\xi,t)-\tilde{H}^q(-x,\xi,t).
\end{eqnarray}

Contrary to charge conjugation, time reversal invariance has direct consequences on each GPD. For most of them, it implies that the GPD is even in $\xi$. This is true for both $H^q$ and $E^q$:
\begin{equation}
  \label{eq:TimeReversal}
  H^q(x,\xi,t) = H^q(x,-\xi,t) 
  \quad , \quad
  E^q(x,\xi,t) = E^q(x,-\xi,t).
\end{equation}
The different possible cases are detailed in \refcite{Belitsky:2005qn} with computation guidelines. Hermiticity of the theory is also responsible for an important constraint on GPDs. Indeed, taking the hermitian conjugate of equation \refeq{eq:GPDDef} leads to:
\begin{equation}
  \label{eq:GPDHermiticity}
  \left[ H(x,\xi,t) \right]^* = H(x,-\xi,t) 
  \quad , \quad 
  \left[ E(x,\xi,t) \right]^* = E(x,-\xi,t).
\end{equation}
Injecting the constraint coming from time reversal invariance of equation \refeq{eq:TimeReversal}, one can conclude that GPDs are real. This latter statement is true for any GPD.

Due to isospin symmetry, the pion GPDs have additional symmetry properties. In fact, isovector ($I=1$) and isoscalar ($I=0$) GPDs can be defined and related to ``quark'' GPDs as:
\begin{eqnarray}
  \label{eq:RelationIsoscalarChargedPions}
  H^{I=0}(x,\xi,t) & = & H^u_{\pi^\pm}(x,\xi,t) + H^d_{\pi^\pm}(x,\xi,t)    \\
  \label{eq:RelationIsoscalarNeutralPions}
  & = & H^u_{\pi^0}(x,\xi,t) + H^d_{\pi^0}(x,\xi,t),  \\  
  \label{eq:RelationIsovectorPiplus} 
  H^{I=1}(x,\xi,t)  & = & H^u_{\pi^+}(x,\xi,t) - H^d_{\pi^+}(x,\xi,t)  \\
  \label{eq:RelationIsovectorPiminus}
  & = & - \big( H^u_{\pi^-}(x,\xi,t) - H^d_{\pi^-}(x,\xi,t)  \big),  \\
  \label{eq:RelationIsovectorNeutralPions}
  0 & = & H^u_{\pi^0}(x,\xi,t) - H^d_{\pi^0}(x,\xi,t).
\end{eqnarray}
Using the previous symmetry properties, $H^u$ and $H^d$ can be related so that for the $\pi^+$, one gets:
\begin{eqnarray}
H^{I=0}
& = & H^u(x,\xi,t) - H^u(-x,\xi,t), \label{eq:H_isoscalar_odd} \\
H^{I=1}
& = & H^u(x,\xi,t) + H^u(-x,\xi,t). \label{eq:H_isoscalar_even}
\end{eqnarray}  
And thus isovector and isoscalar GPDs correspond to non-singlet and singlet GPDs respectively.

It is also possible to highlight the chiral properties of the pion GPD, especially the so-called the soft pion theorem. Defining the pion distribution amplitude (DA) $\varphi_\pi$ as:
\begin{equation}
  \label{eq:PionDADef}
  f_\pi \varphi_\pi(u) = \left. \int \frac{\textrm{d}z^-}{2\pi}e^{i(2u-1)P^+z^-/2}\bra{\pi, P}\bar{\psi}(-\frac{z}{2})\g\cdot n\g_5\psi(\frac{z}{2})\ket{0}\right|_{z^+ = z_\perp =0}, 
\end{equation}
where $f_\pi$ stands for the pion decay constant, PCAC and crossing symmetry allow one to relate the pion GPDs to the pion DA in the kinematic limit $t \rightarrow 0$ and $\xi \rightarrow 1$ through \cite{Polyakov:1998ze}:
\begin{equation}
  \label{eq:SoftPionTheorem}
  H^{I=0}(x,1,0) = 0  \quad\textrm{ and } \quad H^{I=1}(x,1,0) = \varphi_\pi \left( \frac{1+x}{2} \right).
\end{equation}

\subsubsection{Forward limit, positivity and polynomiality}

In the so-called forward limit, \ie when both $\xi$ and $t$ vanish, the GPD $H^q$ reduces itself to the usual parton distribution function. More precisely:
\begin{eqnarray}
  \label{eq:ForwardLimit}
  H^q(x,0,0) & = & q(x) \quad \textrm{for }x \ge 0, \\
  H^q(x,0,0) & = & -\bar{q}(-x) \quad \textrm{for }x \le 0,
\end{eqnarray}
where $q(x)$ and $\bar{q}(x)$ are respectively the PDF of a quark and antiquark of flavour $q$.

As it is detailed in section \ref{sec:LightconeFormalism}, the GPDs can be seen on the lightcone as an overlap of lightcone wave-functions \cite{Diehl:2000xz}. It leads to an additional theoretical property of the GPDs called the positivity property. Indeed, one can consider the GPD $H$ of a scalar hadron at vanishing $t$ as:
\begin{equation}
  \label{eq:PositivityWFD}
  H(x,\xi) = \sum_{S} \bra{\Psi_{\textrm{out}}(x,\xi,S)}\Psi_{\textrm{in}}(x,\xi,S) \rangle ,
\end{equation}  
with $\Psi_{\textrm{in}}(x,\xi,S)$ being the probability amplitude for the hadron to split in a quark carrying a momentum $(x+\xi)P^+$ along the lightcone, and a spectator denoted $S$. In the same way, $\Psi_{\textrm{out}}(x,\xi,S)$ corresponds to the probability amplitude to generate the considered hadron from a spectator $S$ and a quark carrying a momentum $(x-\xi)P^+$ along the lightcone. Then the Cauchy-Schwartz inequality yields \cite{Pire:1998nw}:
\begin{align}
  \label{eq:CauchySchwartz}
  \left| \sum_{S} \bra{\Psi_{\textrm{out}}(x,\xi,S)}\Psi_{\textrm{in}}(x,\xi,S) \rangle \right|^2 \le &\sum_{S} \bra{\Psi_{\textrm{out}}(x,\xi,S)}\Psi_{\textrm{out}}(x,\xi,S) \rangle  \nonumber \\ & \sum_{S'} \bra{\Psi_{\textrm{in}}(x,\xi,S')}\Psi_{\textrm{in}}(x,\xi,S') \rangle .
\end{align}
In terms of GPDs and PDFs, this result can be seen as:
\begin{equation}
  \label{eq:Positivity}
  H^q(x,\xi) \le \sqrt{q(x_1)q(x_2)},
\end{equation}
where
\begin{equation}
  \label{eq:X1X2Def}
  x_1 = \frac{x-\xi}{1-\xi} 
  \quad, \quad
  x_2 = \frac{x+\xi}{1+\xi}.
\end{equation}
The same kind of inequality can be derived for gluon distributions:
\begin{equation}
  \label{eq:PositivityGluons}
  H^g(x,\xi) \le \sqrt{(1-\xi^2)x_1x_2g(x_1)g(x_2)},
\end{equation}
as well as for other GPDs (including for non-scalar hadrons) \cite{Diehl:2000xz,Pobylitsa:2001nt,Pobylitsa:2002gw} or for the so-called impact parameter space GPDs \cite{Diehl:2002he}.

Last but not least, the Mellin moments $\MM$ of the GPDs, defined as:
\begin{equation}
  \label{eq:DefMellinMoments}
  \mathcal{M}_m^{H}(\xi,t) = \int_{-1}^{1}\textrm{d}x~ x^m H(x,\xi,t)\quad , \quad \mathcal{M}_m^{E}(\xi,t) = \int_{-1}^{1}\textrm{d}x~ x^m E(x,\xi,t)
\end{equation}
shows interesting properties. First, the Mellin moments computed for $m=0$ give precisely the contributions of the flavour $q$ to the hadron form factors $F_1$ and $F_2$, \ie:
\begin{equation}
  \label{eq:SumRulesFF}
  \int_{-1}^1 \textrm{d}x~H^q(x,\xi,t) = F_1^q(t) \quad, \quad \int_{-1}^1 \textrm{d}x~E^q(x,\xi,t) = F_2^q(t) .
\end{equation}
Then, the Mellin moments can be related to local matrix elements through:
\begin{equation}
  \label{eq:MMandLocalMatrixElement}
  \mathcal{M}_m^{H}(\xi,t) = \frac{1}{2\left(P^+ \right)^{m+1}} \bra{p_2}\bar{\psi}^q(0)\g^+ (i\overleftrightarrow{\partial}^+)^m\psi^q(0) \ket{p_1},
\end{equation}
for a spinless hadron and:
\begin{equation}
  \label{eq:MMandLocalMatrixElementSpin}
  \bar{u}(p_2)\left[\g^+\mathcal{M}_m^{H}(\xi,t)+\frac{\sigma^{+\nu}\Delta_\nu}{2M}\mathcal{M}_m^{E}(\xi,t)\right]u(p_1) = \frac{1}{2\left(P^+ \right)^{m+1}} \bra{p_2}\bar{\psi}^q(0)\g^+ (i\overleftrightarrow{\partial}^+)^m\psi^q(0) \ket{p_1},
\end{equation}
for spin one-half hadrons, where $\overleftrightarrow{\partial}^+ = \frac{\overrightarrow{\partial}^+-\overleftarrow{\partial}^+}{2}$. Equations \refeq{eq:MMandLocalMatrixElement} and \refeq{eq:MMandLocalMatrixElementSpin} are valid in the lightcone gauge, but the generalisation to any gauge is straightforward using the covariant derivative $\overleftrightarrow{D}$ instead of the partial one. They can be also seen in terms of covariant local twist-two operators defined as:
\begin{equation}
  \label{eq:LocalOperator}
  O^{\{ \mu \mu_1... \mu_m\}} = \bar{\psi}\g^{\{\mu} i\overleftrightarrow{\partial}^{\mu_1}...i\overleftrightarrow{\partial}^{\mu_m \}}\psi
\end{equation}
where $\{\dots\}$ indicates that the operator is totally symmetric and traceless. Projected on the lightcone, the twist-two operators of equation \refeq{eq:LocalOperator} give the ones of equation \refeq{eq:MMandLocalMatrixElementSpin}, as expected from the OPE formalism \cite{Wilson:1969zs,Anikin:1978tj,Mueller:1998fv}.
 Covariance allows one to parameterise the matrix element of the local twist-two operators for a spin-$1/2$ hadron as:
\begin{eqnarray}
  \label{eq:ParameterisingLocalOperators}
  \bra{p_2}O^{\{ \mu \mu_1... \mu_m\}}\ket{p_1} & = & \bar{u}(p_2)\g^{\{\mu}u(p_1) \sum_{i=0}^{\left[\frac{m}{2}\right]}A^q_{m+1,2i}(t)\left(-\Dd^{\mu_1}\right)...\left(-\Dd^{\mu_{2i}}\right)P^{\mu_{2i+1}}...P^{\mu_m\}} \nonumber \\
  & & +\bar{u}(p_2)\frac{\sigma^{\{\mu \alpha}\Delta_\alpha}{2M}u(p_1) \sum_{i=0}^{\left[\frac{m}{2}\right]}B^q_{m+1,2i}(t)\Delta^{\mu_1}...\Delta^{\mu_{2i}}P^{\mu_{2i+1}}...P^{\mu_m\}} \nonumber \\
  & & - \textrm{mod}(2,m)\bar{u}(p_2)\frac{\Delta^{\{\mu}}{2M}u(p_1) C^q_{m+1}(t)\left(-\Dd^{\mu_1}\right)...\left(-\Dd^{\mu_m\}}\right). \nonumber \\
\end{eqnarray}
The even power of $\Delta$ is a consequence of time reversal invariance as seen in equation \refeq{eq:TimeReversal}. In addition, $[\cdot]$ denotes the floor function and mod$(2,m)$ vanishes when $m$ is even, and is equal to 1 when $m$ is odd. Projecting equation \refeq{eq:ParameterisingLocalOperators} on the lightcone and using the Gordon identity (see appendix \ref{sec:AppendixA}) to reduce from three Dirac structure to two, one can show that:
\begin{eqnarray}
  \label{eq:PolynomialityH}
  \int \textrm{d}x~x^m H(x,\xi,t) & = & \sum_{j=0}^{\left[\frac{m}{2}\right]} \xi^{2j} A^q_{m+1,2j}(t) + \textrm{mod}(m,2)\xi^{m+1}C^q_{m+1}(t), \\
  \label{eq:PolynomialityE}
  \int \textrm{d}x~x^m E(x,\xi,t) & = & \sum_{j=0}^{\left[\frac{m}{2}\right]} \xi^{2j}B^q_{m+1,2j}(t) - \textrm{mod}(m,2)\xi^{m+1}C^q_{m+1}(t).
\end{eqnarray}
Therefore, the Mellin moments of GPDs are polynomials in $\xi$. $A^q$, $B^q$ and $C^q$ are sometimes called generalised form factors in relation with equation \refeq{eq:SumRulesFF}.

\subsubsection{Exclusive processes and GPD extraction}
\label{sec:Exclusive}

GPDs play a key role in the computation of exclusive processes amplitudes. Therefore, processes like Deeply Virtual Compton Scattering (DVCS), Time-like Compton Scattering (TCS) and Deeply Virtual Meson Production (DVMP) are experimental channels allowing to access GPDs. The amplitudes can be split in a ``hard'' part, corresponding to the interaction of the active parton with the probe, and a ``soft'' part, describing non-perturbative phenomena, \ie GPDs and also DAs in the case of DVMP. This is known as the \emph{factorisation theorem} \cite{Collins:1996fb,Ji:1998xh,Collins:1998be,Mueller:1998fv,Radyushkin:1996nd,Radyushkin:1997ki}.
Focusing on DVCS, the total amplitude can be split into four\footnote{Twelve when taking into account twist-three and chiral-odd GPDs.} leading-twist Compton Form Factors (CFF) denoted $\mathcal{F}$ which are related to the chiral-even GPDs through:
\begin{equation}
  \label{eq:CFFDef}
  \mathcal{F}(\xi,t) = \int_{-1}^{1} \textrm{d}x~ C(x,\xi) F(x,\xi,t),
\end{equation}
where $F$ is the GPD associated with the CFF $\mathcal{F}$. $C$ corresponds to the coefficient coming from the computation of the perturbative part of the process associated with the GPD $\mathcal{F}$. The dependencies in terms of the photon virtuality $Q^2$, renormalisation and factorisation scales are omitted for brevity. Details will be given in section \ref{sec:Evolution}. In the specific case of the contribution related to the GPD $H$, with a perturbative coefficient computed at leading order (LO), the associated CFF $\mathcal{H}$ can be written as:
\begin{equation}
  \label{eq:CFFH}
  \mathcal{H}(\xi,t) =  \int_{-1}^{1} \textrm{d}x~ \left( \frac{1}{\xi-x-i\epsilon} - \frac{1}{\xi+x-i\epsilon}\right)H(x,\xi,t).
\end{equation}
At this point, the continuity of the GPD on the lines $x=\pm\xi$ is required to ensure that the CFF is finite. Still, due to the singularities, the CFF is a complex number, its real and imaginary parts can be computed at LO through:\begin{eqnarray}
  \label{eq:ReCFFH}
  \left. \Re\left( \mathcal{H}(\xi,t) \right)\right|_{\textrm{LO}} & = & \textrm{p.v.}\int_{-1}^1 \textrm{d}x \left( \frac{1}{\xi-x} - \frac{1}{\xi+x}\right)H(x,\xi,t), \\
  \label{eq:ImCFFH}
  \left. \Im \left( \mathcal{H}(\xi,t) \right)\right|_{\textrm{LO}} & = & \pi\left( H(\xi,\xi,t) -H(-\xi,\xi,t) \right),
\end{eqnarray} 
where p.v. is the Cauchy principal value prescription. The imaginary part of this CFF is, for instance, the main contribution to observables depending on beam spin helicities.

\begin{figure}[t]
  \centering
  \includegraphics[width=0.3\textwidth]{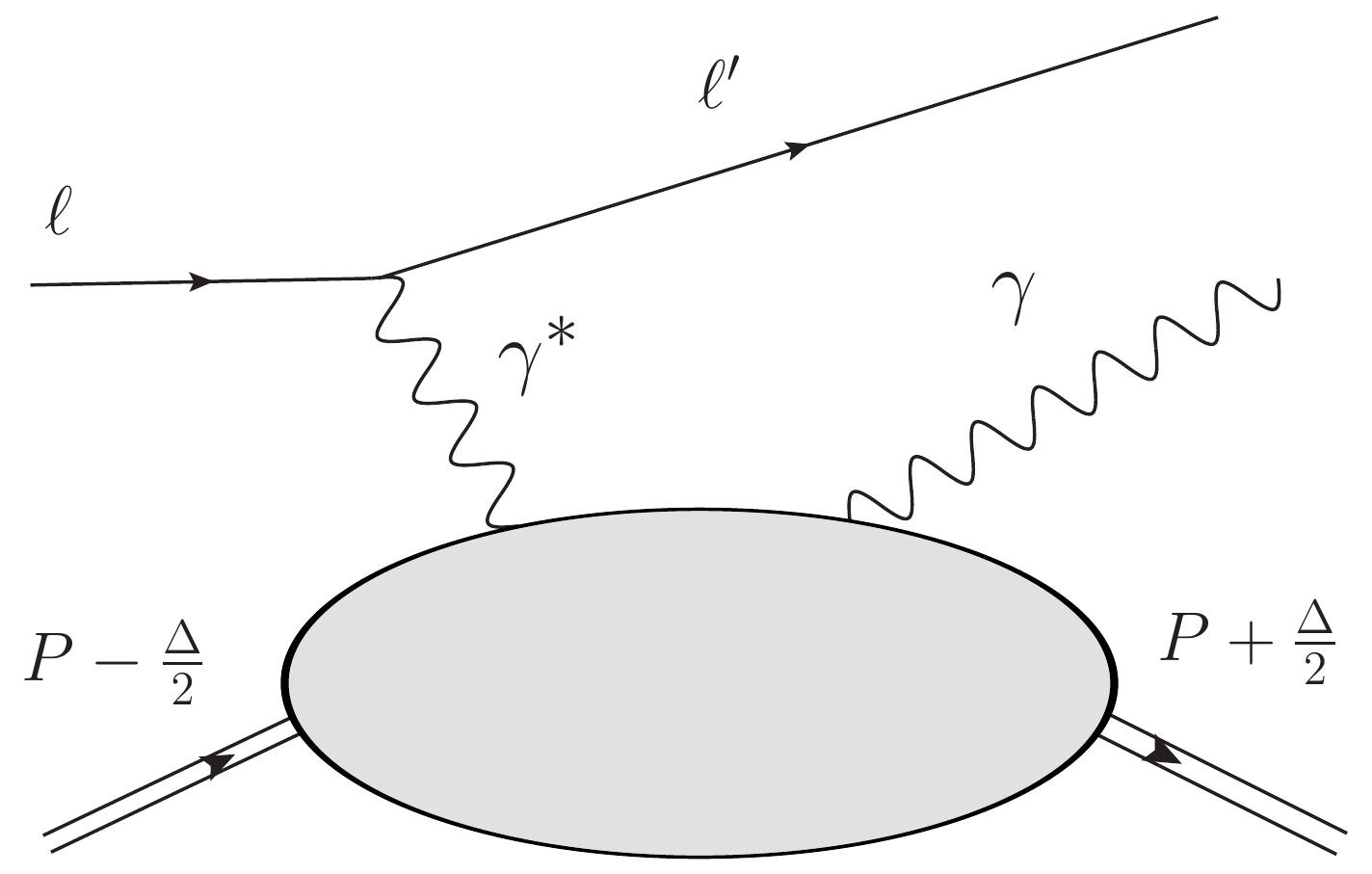}
  \quad
  \includegraphics[width=0.3\textwidth]{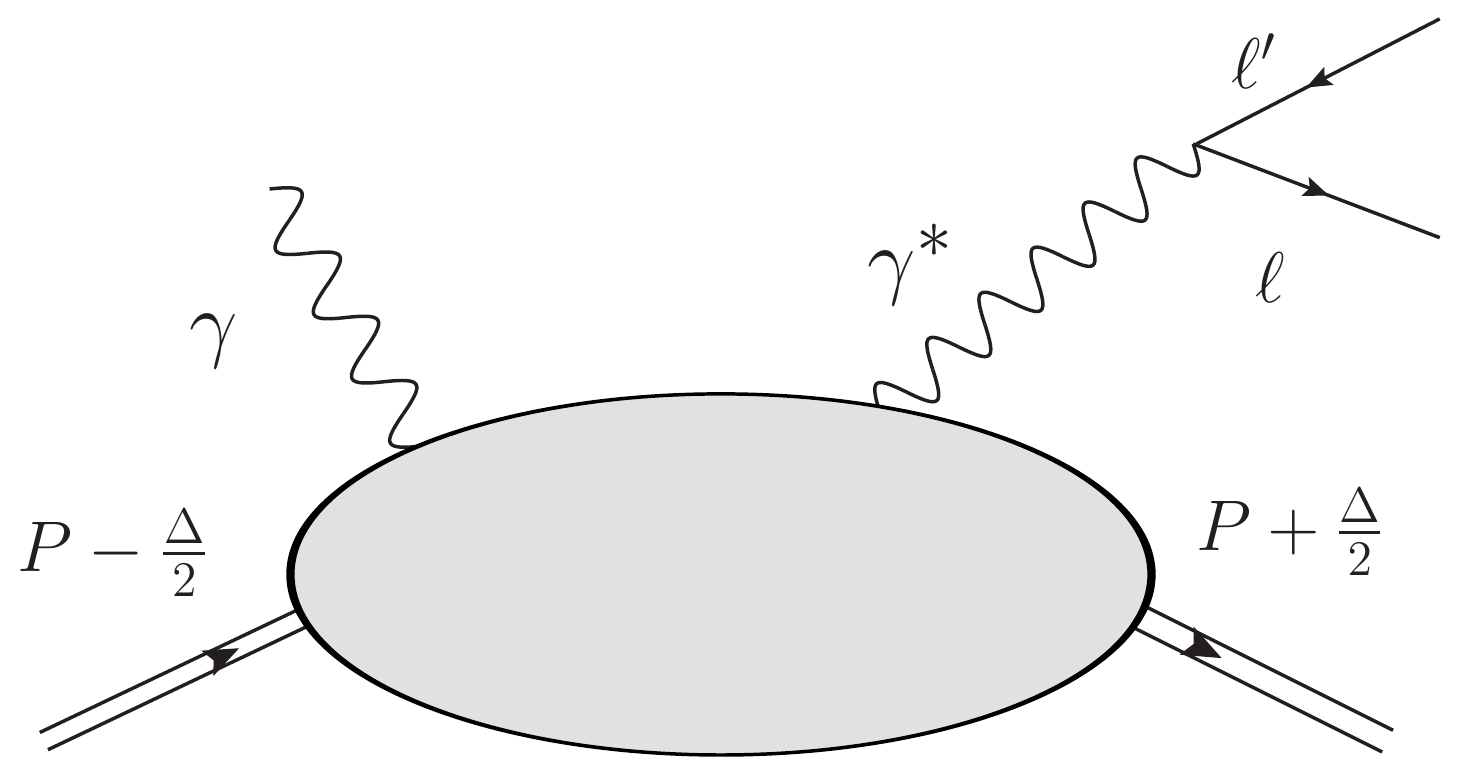}
  \quad
  \includegraphics[width=0.3\textwidth]{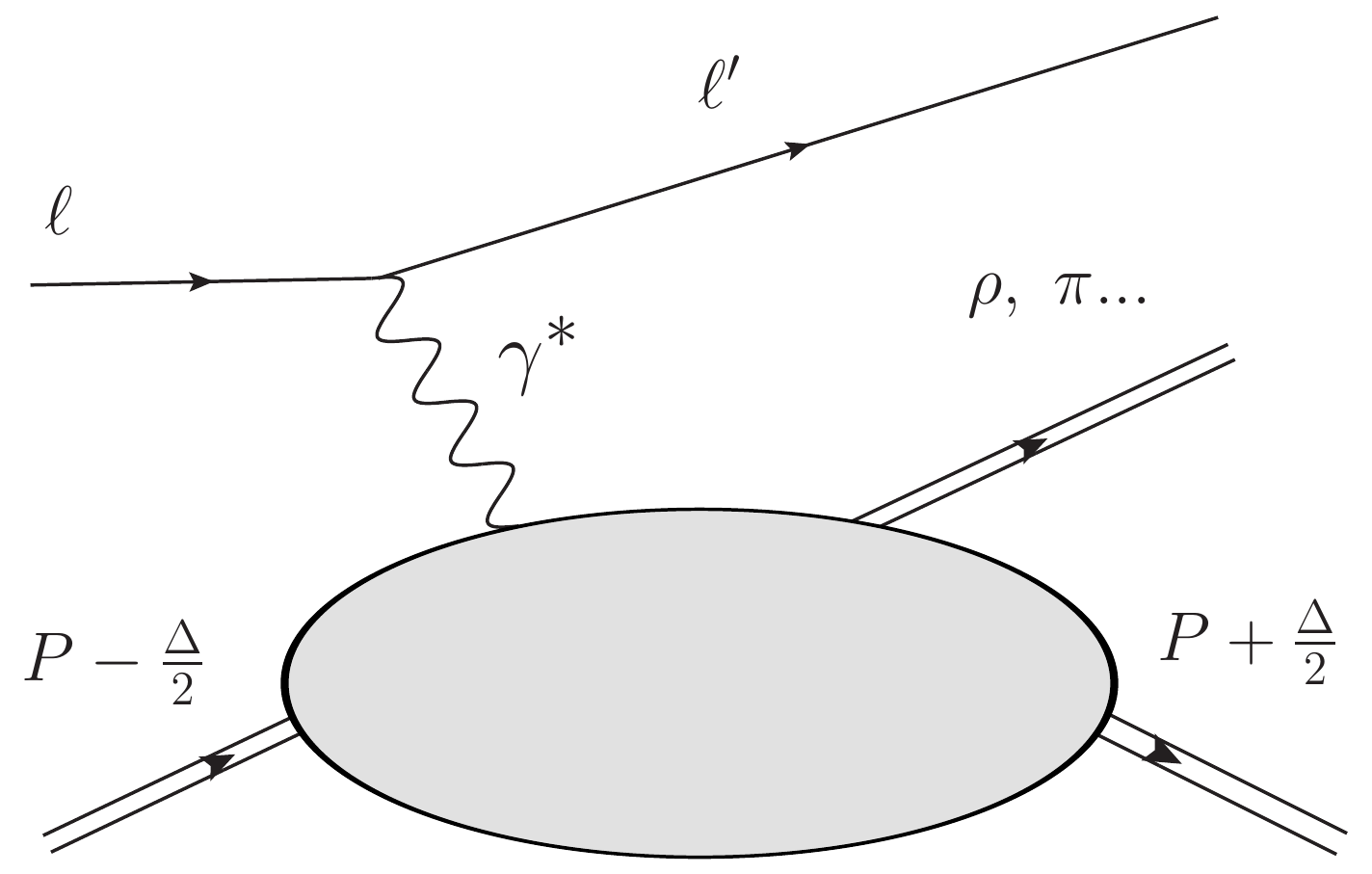}
  \caption{Exclusive processes. Left-hand side: DVCS; center: TCS; right-hand side: DVMP.}
  \label{fig:ExclusiveProcesses}
\end{figure}

Extracting the different CFFs from experimental data reveals itself to be a hard task, partially because few data are available today. In valence region, \ie for large $\xi$, both Jefferson Laboratory (JLab) Hall A and CLAS collaborations have released DVCS cross-sections \cite{Defurne:2015lna,Jo:2015ema}, as well as asymmetries \cite{Girod:2007aa}. In the medium $\xi$ region, the HERMES collaboration has released data for the DVCS asymmetries but not for the cross-sections (see \eg \cite{Murray:2013gg}). At low $\xi$, collider data have been released both by H1 and ZEUS collaborations \cite{Chekanov:2003ya,Aktas:2005ty}. Despite this significant experimental effort to measure DVCS observables, a large part of the kinematic domain remains uncovered. Thus several techniques have been developed to extract CFFs, including local fits \cite{Guidal:2008ie}, global fits \cite{Moutarde:2009fg,Kumericki:2009uq,Kumericki:2013lia} and extension to neural network are seriously considered today \cite{Kumericki:2011rz}.

Interpretations of experimental data are even harder due to the higher order corrections in perturbative QCD, and higher twist corrections. Indeed, if next-to-leading order corrections are well-known for exclusive processes \cite{Ji:1997nk,Mankiewicz:1997bk,Belitsky:1999sg,Freund:2001hd,Freund:2001hm,Freund:2001rk,Pire:2011st}, it has been shown recently that they have a significant effect on CFFs \cite{Moutarde:2013qs}. In the same way, target mass (\ie higher twist) and finite $t$ corrections \cite{Braun:2012bg,Braun:2012hq} are presumably required to fully understand experimental results at intermediate $Q^2$ \cite{Defurne:2015lna}.

\subsubsection{Evolution}
\label{sec:Evolution}

Factorisation scale dependence has mainly be omitted until now in this review. Yet, the reader should keep in mind that the splitting of a experimental process between a perturbative part and a non-perturbative object is arbitrary, and that no observable may depend on such a scale. Therefore, from this idea, in the same way than for PDFs, the scale dependence of GPDs can be computed perturbatively. Singlet and non-singlet quark GPDs defined in equations \refeq{eq:SingletGPD} and \refeq{eq:NonSingletGPD} have different dynamics with respect to evolution. If the former mixes with gluon GPDs, the latter does not. Thus, non-singlet quark GPDs fulfil the following evolution equation:
\begin{equation}
  \label{eq:EvolutionNS}
  \mu_F^2 \frac{\partial}{\partial \mu_F^2}H^{-}(x,\xi,t) = \int \textrm{d} y \frac{1}{|\xi|} K_{NS}\left(\frac{x}{\xi},\frac{y}{\xi} \right)H^{-}(y,\xi,t),
\end{equation}
where $K_{NS}$ is the so-called non-singlet evolution kernel. In the singlet case, it is convenient to define the singlet vector $\bm{H}$ as:
\begin{equation}
  \label{eq:DefVectorSinglet}
  \bm{H} = 
  \begin{pmatrix}
    (2n_f)^{-1} \sum_q H^q(x,\xi,t) -H^q(-x,\xi,t) \\
    H^g(x,\xi,t)
  \end{pmatrix}
  ,
\end{equation}
where $n_f$ stands for the number of active flavours. $\bm{H}$ fulfills a differential equation equivalent to equation \refeq{eq:EvolutionNS} using the matrix singlet kernel $\bm{K}$:
\begin{equation}
  \label{eq:KernelSinglet}
  \bm{K} =
  \begin{pmatrix}
    K^{qq}\left(\frac{x}{\xi},\frac{y}{\xi} \right) & K^{qg}\left(\frac{x}{\xi},\frac{y}{\xi} \right) \\
    K^{gq}\left(\frac{x}{\xi},\frac{y}{\xi} \right) & K^{gg}\left(\frac{x}{\xi},\frac{y}{\xi} \right)
  \end{pmatrix}
  .
\end{equation}
The evolution kernels $K_{NS}$ and $\bm{K}$ can be computed perturbatively and are known at LO \cite{Geyer:1985vw,Braunschweig:1985nr,Dittes:1988xz,Mueller:1998fv,Ji:1996nm,Radyushkin:1997ki,Balitsky:1997mj,Radyushkin:1998es,Blumlein:1997pi,Blumlein:1999sc} and at NLO \cite{Belitsky:1998vj,Belitsky:1998gc,Belitsky:1999gu,Belitsky:1999fu,Belitsky:1999hf}. As GPDs reduce to PDFs in the forward limit (\ie $\xi \rightarrow 0$), the off-forward evolution kernels $K_{NS}$ and $\bm{K}$ also reduce to the non-singlet and singlet DGLAP kernels respectively. While when $\xi \rightarrow 1$, the off-forward evolution kernels reduce to the famous ERBL kernels \cite{Efremov:1978rn,Chernyak:1977fk,Farrar:1979aw,Lepage:1979zb,Lepage:1980fj} which describe the evolution of DAs. Several techniques have been employed to solve the evolution equations, from numerical approaches \cite{Vinnikov:2006xw} or analytic techniques based on the fact that, at LO, the conformal moments diagonalises the evolution equations (see \eg \refscite{Diehl:2003ny,Belitsky:2005qn} and references therein for details). 

\subsubsection{3D picture of hadrons}

\begin{figure}[h]
  \centering
  \includegraphics[width=0.45\textwidth]{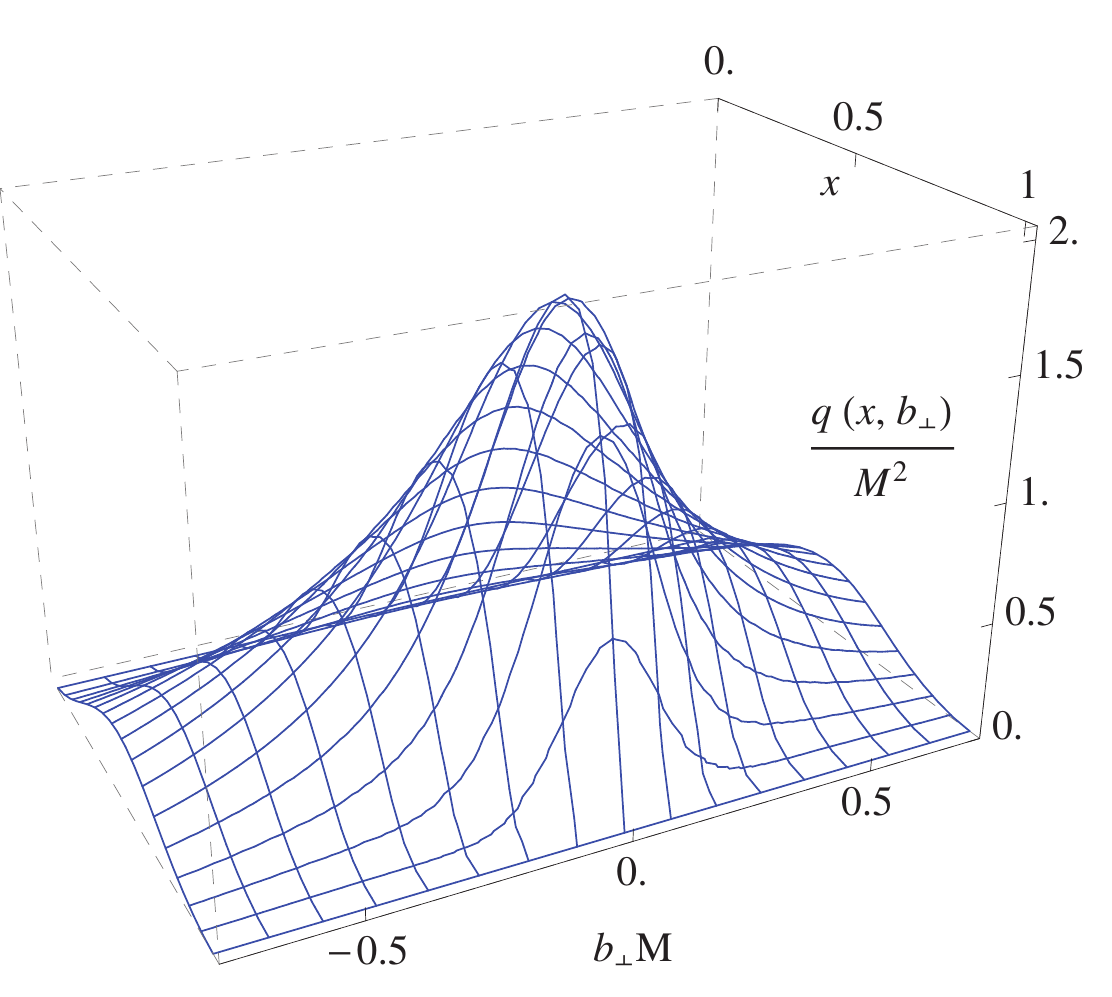}
  \caption{Transverse plane density of the pion GPD $H^q$. The 3D-plot of $\rho^q(x,\vperp{b})$ comes from \refcite{Mezrag:2014jka}. Acknowledging its rotational invariance, the probability density is only plotted in one of the transverse-plane axes for all $x$.}
  \label{fig:3DGPDPicture}
\end{figure}

GPDs present an interesting physical interpretation in the kinematic region $\xi=0$. For instance, in the case of the pion, the quantity $\rho$ defined as:
\begin{equation}
  \label{eq:3DDensity}
  \rho^q(x,\vperp{b}) = \int \frac{\textrm{d}^2\Delta_\perp}{(2\pi)^2} e^{-i\vperp{b}\Delta_\perp}H^q(x,0,-\Delta_\perp^2),
\end{equation} 
is the probability density to find a quark carrying a momentum fraction $x$ at a given position $\vperp{b}$ in the transverse plane \cite{Burkardt:2000za}. Thus, one gets a {\it picture} of the pion within a transverse {\it slice}, perpendicular to the light-cone direction, as can be seen in the example shown in figure \ref{fig:3DGPDPicture}.

\subsection{Double Distributions}

\subsubsection{Definition and relation to GPDs}

\begin{figure}[t]
  \centering
  \includegraphics[width=0.35\textwidth]{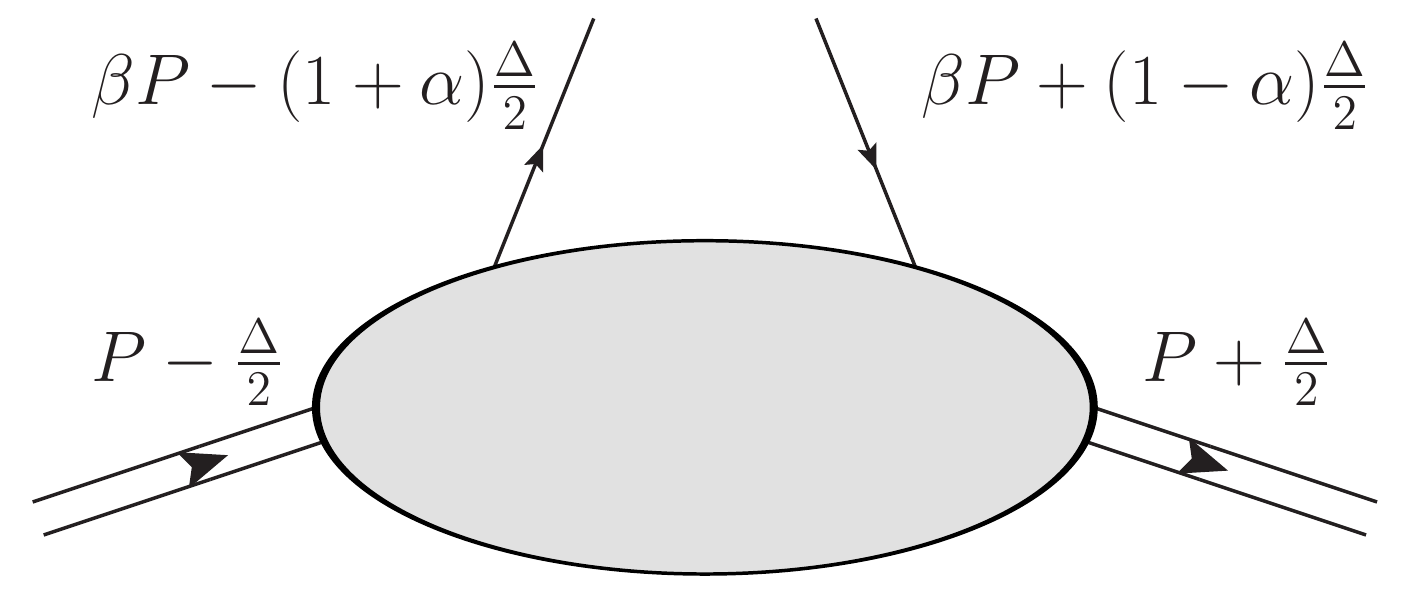}
  \quad \quad
  \includegraphics[width=0.20\textwidth]{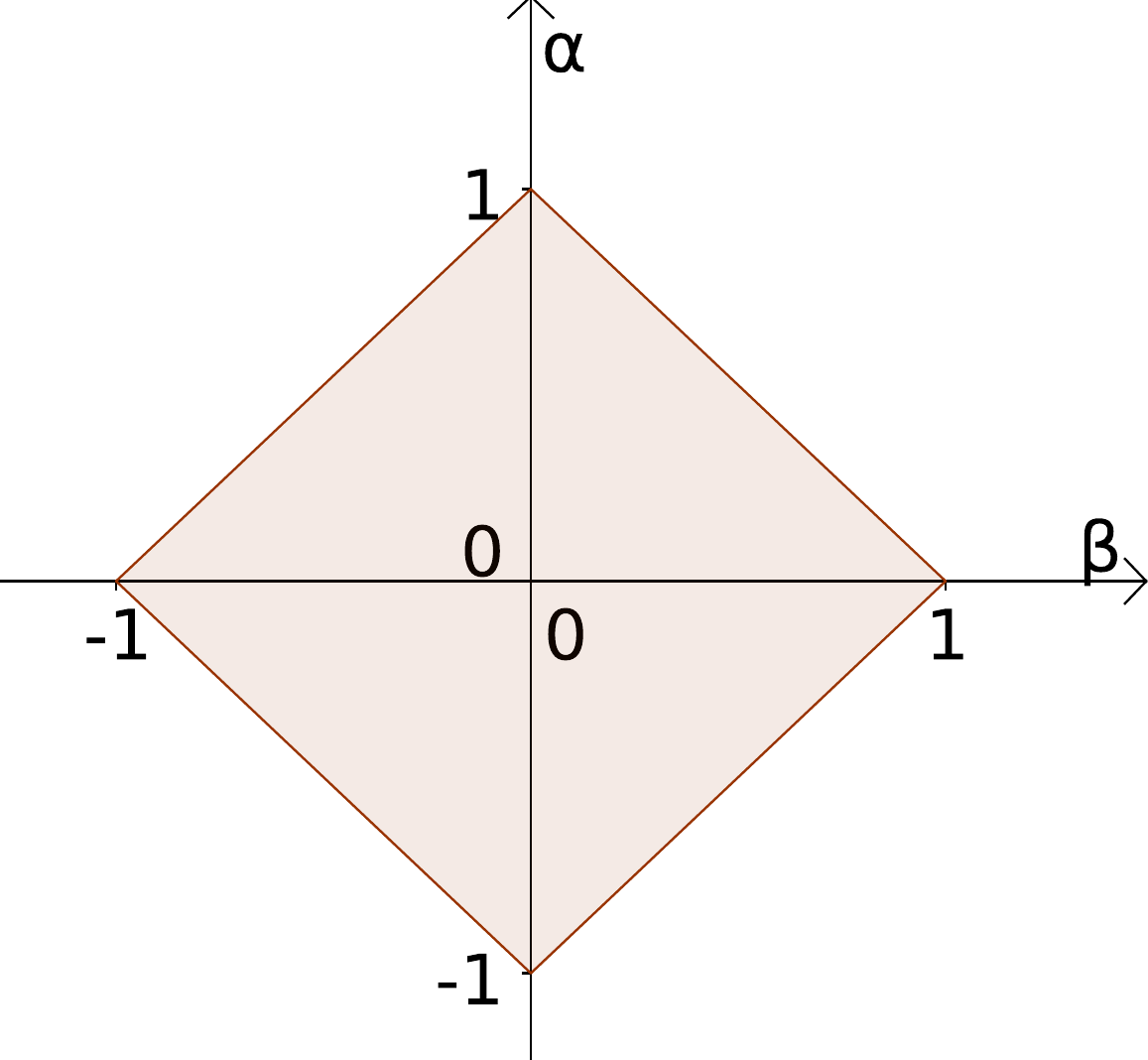}
  \caption{Left-hand side: Momenta associated with hadrons and partons within a DD framework. Right-hand side: DD support.}
  \label{fig:DDKinematics}
\end{figure}

Originally introduced in \cite{Mueller:1998fv,Radyushkin:1996nd,Radyushkin:1996ru,Radyushkin:1997ki}, Double Distributions (DDs) are an alternative way to encode information contained in non-local matrix elements. In the case of a scalar hadron, two DDs $\DDF$ and $\DDG$ can be defined as:
\begin{eqnarray}
\left .\bra{P+\frac{\Delta}{2}} \bar{q}\left( -\frac{z}{2} \right) \gamma_\mu q \left( \frac{z}{2} \right) \ket{P-\frac{\Delta}{2}}\right|_{z^2=0} 
& = & 2P_{\mu}\int_{\Omega} \mathrm{d}\beta\mathrm{d}\alpha \, e^{- i \beta (P \cdot z) + i \alpha \frac{(\Delta \cdot  z)}{2}} F^q( \beta, \alpha, t ) \nonumber \\
& & \, - \Delta_{\mu}\int_{\Omega} \mathrm{d}\beta\mathrm{d}\alpha \, e^{- i \beta (P \cdot z) + i \alpha \frac{(\Delta \cdot z)}{2}} G^q( \beta, \alpha, t ) \nonumber \\
& & +\text{ higher twist terms}.
\label{eq:ScalarDD}
\end{eqnarray}
The name of the variables follows the conventions of \refcite{Goeke:2001tz}. $\beta$ and $\alpha$ have a physical interpretation in terms of parton momenta as shown on figure \ref{fig:DDKinematics}.
Discrete symmetries also constrain DDs. For instance, time reversal invariance determines their parity in $\alpha$: $F$ is even in $\alpha$ whereas $G$ is odd. The DD support, usually denoted $\Omega$, presents a specific rhombus shape:
\begin{equation}
  \label{eq:DDSupport}
  \Omega = \left\{(\beta,\alpha),\ |\beta|+|\alpha|\le 1  \right\},
\end{equation}
as shown on figure \ref{fig:DDKinematics}. Being non-perturbative objects, DDs depend on a factorisation scale $\mu_F$ and obey to evolution equations \cite{Mueller:1998fv,Radyushkin:1996nd,Radyushkin:1996ru,Radyushkin:1997ki}. 
The matrix elements defining GPDs \refeq{eq:PionGPDDef} and DDs \refeq{eq:ScalarDD} are very similar, and thus one
can show that DDs and GPDs are related through the so-called Radon transform: GPDs are actually the Radon transform of DDs. Practically, one can compute GPDs from DDs through:
\begin{equation}
  \label{eq:RelationDDGPDH}
  H^q(x,\xi,t)  =  \int_{\Omega} \mathrm{d}\beta\mathrm{d}\alpha \, \left ( F^q( \beta, \alpha, t )+\xi G^q(\beta,\alpha,t) \right) \delta(x-\beta -\alpha \xi).
\end{equation}
Inverting the Radon transform is not an easy task, and thus DD are scarcely computed from GPDs.
The polynomiality property can be seen as a direct consequence of equation \refeq{eq:RelationDDGPDH}. Computing the Mellin Moment $\MM$ in terms of DD leads to:
\begin{equation}
  \label{eq:GPDMellinMomentViaDD}
  \MM = \int_{\Omega} \mathrm{d}\beta\mathrm{d}\alpha \, (\beta+\xi \alpha)^m\left ( F^q( \beta, \alpha, t )+\xi G^q(\beta,\alpha,t) \right),
\end{equation}
which is a polynomial in $\xi$ of degree at most $m+1$.

\subsubsection{Ambiguity and schemes}

Historically, the DD $\DDG$ has been overlooked, and introduced only latter in \refcite{Polyakov:1999gs} as the so-called $D$-term $\DDG = \delta(\beta)D(\alpha,t)$. In fact, as shown in \refcite{Teryaev:2001qm} for DDs vanishing on the edges of the rhombus $\Omega$, this $D$-term corresponds to a \emph{specific scheme} of DD. Indeed, projecting the matrix element of equation \refeq{eq:ScalarDD} on a light-like vector $z^\mu$, it yields:
\begin{equation}
  \label{eq:EffectiveDD}
  \left .\bra{P+\frac{\Delta}{2}} \bar{q}\left( -\frac{z}{2} \right)z \cdot  \gamma q \left( \frac{z}{2} \right) \ket{P-\frac{\Delta}{2}}\right|_{z^2=0} 
 =  -2i\int_{\Omega} \mathrm{d}\beta\mathrm{d}\alpha \, e^{- i \beta (P z) + i \alpha \frac{(\Delta z)}{2}} N^q( \beta, \alpha, t ) 
\end{equation}
with
\begin{equation}
  \label{eq:DefEffectiveDDN}
   N^q( \beta, \alpha, t ) = \frac{\partial F^q}{\partial \beta}(\beta,\alpha,t) +\frac{\partial G^q}{\partial \alpha}(\beta,\alpha,t).
\end{equation}
Therefore, any function $\sigma$ such that:
\begin{equation}
  \label{eq:ParitySigma}
  \sigma^q (\beta,-\alpha,t) = - \sigma^q(\beta,\alpha,t).
\end{equation}
can be used to modify the DDs $\DDF$ and $\DDG$:
\begin{eqnarray}
  \label{eq:DDTransformationF}
  F^q(\beta,\alpha,t) & \rightarrow & F^q(\beta,\alpha,t) + \frac{\partial \sigma^q}{\partial \alpha }(\beta,\alpha,t), \\
  \label{eq:DDTransformationG}
  G^q(\beta, \alpha,t) & \rightarrow & G^q(\beta, \alpha,t) - \frac{\partial \sigma^q}{\partial \beta }(\beta,\alpha,t).
\end{eqnarray}
These transformations leave the effective DD $N^q(\beta,\alpha,t)$ unchanged\footnote{It is sometimes called the DD ``gauge transformation'' due to the possible analogy with electromagnetism \cite{Teryaev:2001qm}.}. Therefore, both $\DDF$ and $\DDG$ are not uniquely defined. This has been generalised in \refcite{Tiburzi:2004qr} for DD not vanishing on the rhombus edges. Different DD schemes have been used, among them the so-called DD+D, which is the original scheme of \refcite{Polyakov:1999gs}, and the one-component DD scheme 1CDD. In the DD+D scheme, $\DDF$ and $\DDG$ are given by:
\begin{eqnarray}
  \label{eq:FinDD+D}
  \DDF & \rightarrow & F^q_{\textrm{DD+D}}(\beta,\alpha,t),\\
  \label{eq:GasDterm}
  \DDG & \rightarrow & D(\alpha,t)\delta(\beta) .
\end{eqnarray}
Therefore, all the information carried by the variable $\beta$ is contained in the DD $F$. The 1CDD scheme introduced in \refcite{Belitsky:2000vk} proceeds from an opposite philosophy as:
\begin{eqnarray}
F^q_{1\textrm{CDD}}( \beta, \alpha ) & = & \beta f^q( \beta, \alpha ), \label{eq:F-1CDD} \\
G^q_{1\textrm{CDD}}( \beta, \alpha ) & = & \alpha f^q( \beta, \alpha ), \label{eq:G-1CDD}
\end{eqnarray}
and thus the information content is somehow balanced between the two DDs.

\subsection{Phenomenological models in a nutshell}

Many different phenomenological models of GPDs have been elaborated so far. To close this section, 
since phenomenology is not the core of the present work, only a brief --surely not exhaustive-- enumeration of available phenomenological models will be given in the following. More details can be found for instance in \refcite{Guidal:2013rya}.

Double Distributions have been widely used to build phenomenological models, for they are a simple way to ensure the polynomiality property of GPDs as shown in equation \refeq{eq:GPDMellinMomentViaDD}. The main available models are the so-called ``Goloskokov-Kroll'' model \cite{Goloskokov:2005sd, Goloskokov:2007nt, Goloskokov:2009ia}, modifications of it \cite{Mezrag:2013mya}, and the Guidal-Guichon-Vanderhaeghen model \cite{Goeke:2001tz, Vanderhaeghen:1998uc, Guichon:1998xv, Vanderhaeghen:1999xj, Guidal:2004nd}. Agreement with available experimental data is satisfying \cite{Kroll:2012sm} but may be challenged by the forthcoming experiments. The so-called dual model \cite{Polyakov:1998ze,Polyakov:2002wz,Polyakov:2008aa} consists in expanding the GPDs on Gegenbauer polynomials in the $t$-channel, and then modeling the coefficients. It has been shown recently \cite{Muller:2014wxa} that this approach is equivalent to another one, called the Mellin-Barnes model \cite{Mueller:2005ed,Kumericki:2009uq}. The latter consists in modeling the conformal moments of a GPD, and then use the Mellin-Barnes inverse transform to compute the GPD itself. Models relying on the Mellin-Barnes parameterisation are in good agreement with available data. Another approach relying on a spectator reggeized diquark model \cite{Goldstein:2010gu} has also been compared to experimental data.

Other approaches have also been developed to model the pion GPD. Polyakov and Weiss \cite{Polyakov:1999gs}, and Anikin \etal~ \cite{Anikin:1999pf}, discussed the effect of an instanton vaccum by means of a effective nonlocal quark-hadron lagrangian. Chiral symmetry is also central in the developments of Broniowski \etal~ \cite{Broniowski:2003rp, Broniowski:2007si} in the framework of the Nambu-Jona-Lasinio model (see the reviews \refcite{RuizArriola:2002wr, Christov:1995vm} and references therein).
 Choi \etal~ \cite{Choi:2001fc, Choi:2002ic}, then Mukherjee and Radyushkin \cite{Mukherjee:2002gb}, proposed light-front calculations with gaussian or power-law wavefunctions in a triangle diagram approximation.
Furthermore, GPD modeling in the \bs framework has enjoyed several studies \cite{Tiburzi:2002tq, Theussl:2002xp, Bissey:2003yr, VanDyck:2007jt, Frederico:2009fk}, usually with simple \bs vertices and with computations of triangle diagrams. 
GPD modeling for large, or moderately large, $t$, was investigated by Bakulev \etal~ \cite{Bakulev:2000eb}, Vogt \cite{Vogt:2001if} and Hoodboy \etal~ \cite{Hoodbhoy:2003uu}. Amrath \etal~ \cite{Amrath:2008vx} modeled the GPD $H$ in the framework of the popular Radyushkin Double Distribution Ansatz \cite{Musatov:1999xp} and discussed the experimental access to pion GPDs through DVCS on a virtual pion target. At last, let us mention the computation of the generalized form factor in chiral perturbation theory at one-loop order by Diehl \etal~ \cite{Diehl:2005rn}, although only focusses on applications to lattice QCD and does not proceed further to a complete model of the pion GPD.


\section{Dyson-Schwinger equations}
\label{sec:DSEs}

As mentioned above, our main goal is to pave the way for the building of a GPD model within a framework which relies as much as possible on QCD. This way, we are willing to understand how the dynamical features of the strong interaction generate the structure of hadrons. The framework we will focus on is based on the DSEs and BSEs. The computational scheme relies on the extraction from QCD equations of motion, namely the DSEs, of the basic ingredients required for obtaining the GPDs within different approaches. Moreover, as one deals with the structure of bound-states formed by quarks, their amplitudes derived from BSEs are also required. Both, DSEs and BSEs are a tower of self-consistent non-perturbative integral equations which should be properly and consistently solved by invoking a particular truncation scheme.   

One must also keep in mind that \dses are solved in Euclidean space. Therefore the question of going from Euclidean space to Minkowskian space must be raised. In the following, unless stated otherwise, Euclidean time will be used for computations, \ie Schwinger functions are considered instead of Green functions. We also assume that the measure fulfils the good properties, so that it is possible to get Wightman and Green functions by analytic continuation \cite{Streater:1989vi, Seiler:1982,Glimm:1987ng}. Therefore, comparison with experimental data will be done in Minkowskian space after analytic continuation of the Schwinger functions.

\subsection{Basic equations and truncation schemes}

The significant progresses achieved in the last two decades in describing observable properties of hadrons from continuum-QCD~\cite{Bashir:2012fs,Eichmann:2012zz,Cloet:2013jya} are partly due to the application of symmetry-preserving truncation schemes to QCD's DSE~\cite{Munczek:1994zz,Bender:1996bb,Qin:2011dd,Qin:2011xq}. The basics of a consistent and symmetry-preserving scheme for DSE and BSE will be briefly discussed below. The first ingredient needed to compute hadron properties from QCD's DSEs is the quark propagator that is to be obtained from the so-called Gap equation.

\subsubsection{The Gap equation}

\begin{figure}[h]
  \centering
  \includegraphics[width=0.5\textwidth]{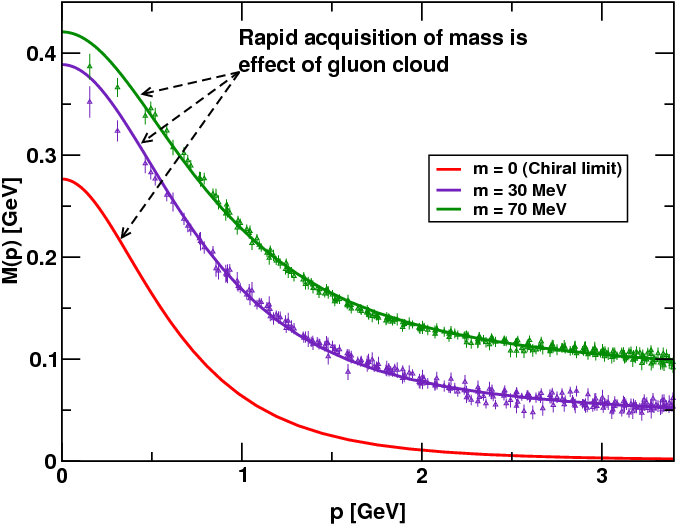}
  \caption{Computations of the dressed quark mass $M_q$ with respect to the quark momentum $p$ within the \dse framework compared to lattice-QCD results. The colours correspond to different bare quark masses. At low $p$, QCD dynamics generates a significant mass term, even in the chiral limit. Dyson-Schwinger results come from \refscite{Bhagwat:2003vw,Bhagwat:2006tu} whereas lattice data are taken from \refcite{Bowman:2005vx}. This figure comes from \refcite{Bashir:2012fs}.}
  \label{fig:RunningQuarkMass}
\end{figure}

The Gap equation relates the quark propagator $S_q(p)$ with the gluon propagator $D^{\mu\nu}(p)$ and the quark-gluon vertex $\G^{\mu}_q(k,p)$ through:
\begin{equation}
  \label{eq:GapEquation}
  S^{-1}_q(p) = Z_2 (i \gamma \cdot p +m_q^{0}) + Z_1 \int_\Lambda \frac{\textrm{d}^4k}{(2 \pi )^4} g^2 D_{\mu\nu}(p-k)\gamma^\mu \frac{\lambda^a}{2}S_q(k)\frac{\lambda_a}{2} \Gamma^\nu_q(k,p), 
\end{equation}
where $m_q^0$ is the considered bare current quark mass and $\Lambda$ signals that the present integral is well-regularised (this is practically achieved through Pauly-Villars techniques see \eg \refcite{Holl:2005vu}). $\lambda^a$ are the Gell-Mann matrices. In spite of the omission of the renormalisation scale $\mu_R$ in equation \refeq{eq:GapEquation}, the reader should keep in mind that the Green functions depend on $\mu_R$, and so do $Z_1(\mu_R,\Lambda)$ and $Z_2(\mu_R,\Lambda)$ which are respectively the quark-gluon vertex and the quark wave function renormalisation constants. Thus the Gap equation is well-defined only when a renormalisation condition is given. This condition is usually taken as:
\begin{equation}
  \label{eq:RConditions}
  S^{-1}(p)\Big|_{p^2=\mu_R^2} = i\g \cdot p + m_q(\mu_R^2),
\end{equation}
with $p$ deeply spacelike. The renormalised current quark mass $m_q(\mu_R^2)$ is related to $m^0_q$ through:
\begin{equation}
  \label{eq:RenormalisedCQM}
  m_q(\mu_R^2) = Z_m^{-1}Z_2 m_q^{0},
\end{equation}
$Z_m$ being the renormalisation constant associated with the Lagrangian mass term. At very large values of $\mu_R$, one expects to recover the bare quark mass.
The solution of the Gap equation can be written as:
\begin{equation}
  \label{eq:Propagator}
  S_q(p) = -i \gamma \cdot p \sigma_{Vq}(p^2,\mu_R^2) + \sigma_{Sq}(p^2,\mu_R^2) = (i \gamma \cdot p A_q(p^2,\mu_R^2) + B_q(p^2,\mu_R^2))^{-1}.
\end{equation}
Within those notations, when assuming multiplicative renormalisability, it is possible to define the dressed running quark mass $M_q(p^2)$ as:
\begin{equation}
  \label{eq:MassFunction}
  M_q(p^2) = \frac{B_q(p^2,\mu_R^2)}{A_q(p^2,\mu_R^2)},
\end{equation}
which does not depend on the renormalisation scale. The dressed quark mass and the renormalised current quark mass are related one to each other through the renormalisation condition \refeq{eq:RConditions}:
\begin{equation}
  \label{eq:MassRelation}
  m_q(\mu_R^2) = M_q(\mu_R^2).
\end{equation}
Computation of the dressed quark mass have been performed in the DSE formalism and are in very good agreement with lattice QCD predictions. As shown on figure \ref{fig:RunningQuarkMass}, most of the mass comes from dynamical processes at low energy. This phenomenon is emphasised in the chiral limit, where the entire mass is dynamically generated, and this suggests that DCSB is the origin for most of the visible mass of the Universe.

\subsubsection{The Bethe-Salpeter equation}

Beyond the quark propagators, as said above, the dynamics of bound states is also studied through non-perturbative objects. Focusing on the pion case, the so-called \bs wave function $\chi(x_1,x_2)$ \cite{Salpeter:1951sz, GellMann:1951rw, Schwinger:1951ex, Schwinger:1951hq, Schwinger:1953tb} provides a relativistic description of the bound state in terms of effective quarks and antiquarks, 
\begin{equation}
\label{eq:BSWFCoordinates}
\chi(x_1,x_2) = \langle 0 | T \left[ \psi(x_1) \bar{\psi}(x_2) \right] | \pi, K \rangle 
\end{equation}
where, for the sake of simplicity, flavour and Dirac indices have been omitted. Owing to translational invariance, it only depends on $x_2-x_1$ and, in momentum space, reads:
\begin{eqnarray}
\label{eq:BSWF}
\chi(k,K) = \delta(K-k_1-k_2) \int d^4x e^{ik\cdot x} \langle 0 | T \left[ \psi(\eta x) \bar{\psi}(-(1-\eta)x) \right] | \pi, K \rangle
\end{eqnarray}
with $x=x_2-x_1$ and
\begin{equation}
  \begin{aligned}
    \label{eq:DefXeta}
    K & =& &k_1+k_2& , \\
    k &=& &(1-\eta)k_1-\eta k_2& \ .
  \end{aligned}
\end{equation}
The pion \bs amplitude $\G_\pi(p,P)$, related to the \bs wave function through:
\begin{equation}
  \label{eq:BSADef}
  \G_{\pi}(k,K) = S^{-1}(-k_2)\ \chi(k,K) \ S^{-1}(k_1),
\end{equation}
is the solution of the homogeneous \bs equation:
\begin{equation}
  \label{eq:BSE}
  \G_{\pi; ij}(p,P) = \int \frac{\textrm{d}^4k}{(2\pi)^4} \left[ S(k_{\bar{\eta}})\G_\pi (k,P)S( k_\eta) \right]_{ab} K^{ab}_{ij}(k,p,P),
\end{equation}
where:
\begin{equation}
  \begin{aligned}
    \label{eq:Defketa}
    &k_\eta =& k_1 &=k+\eta P,& \\
    &k_{\bar{\eta}}  =& -k_2 &=k-(1-\eta)P ,& 
  \end{aligned}
\end{equation}
and $K^{ab}_{ij}(k,p,P)$ is the quark-antiquark scattering kernel. As the choice of the value of $\eta\in[0,1]$ is arbitrary, no observable should depend on it. In the case of the pion, the solution of equation \refeq{eq:BSE} is usually projected on a Dirac basis through \cite{LlewellynSmith:1969az}:
\begin{equation}
  \label{eq:PionDiracStructure}
   \G_{\pi}(p,P) = \g_5[iE(p,P) + \g \cdot P~F(p,P) + p\cdot P~p\cdot \g G(p,P)+\sigma_{\mu\nu}p^{\mu}P^{\nu}H(p,P)].
\end{equation}

The normalisation of the \bs amplitude can be derived in a covariant way as \cite{Lurie:1965,Maris:1999nt}:
\begin{eqnarray}
  \label{eq:BSNormalisation}
2 P^\mu \ & = &  {\rm Tr}_{\rm CDF} \left[\int \frac{d^4k}{2\pi^4} 
\bar{\Gamma}_\pi\left(k,P\right) \frac{\partial S(k_\eta)}{\partial P_\mu}  \Gamma_\pi\left(k,P\right) S(k_{\bar{\eta}}) \right]\nonumber \\
 & & + {\rm Tr}_{\rm CDF} \left[\int \frac{d^4k}{2\pi^4} 
\bar{\Gamma}_\pi\left(k,P\right)  S(k_\eta) \Gamma_\pi\left(k,P\right) \frac{\partial S(k_{\bar{\eta}})}{\partial P_\mu} \right]\nonumber \\
& &  + {\rm Tr}_{\rm CDF} \left[\int \frac{d^4k}{2\pi^4} \frac{d^4q}{2\pi^4} \bar{\chi}_\pi(q,P) \frac{\partial K(q,k,P)}{\partial P^\mu} \chi_\pi(k,P) \right]
\end{eqnarray} 
${\rm Tr}_{\rm CDF}$ denoting the trace on colour, Dirac, and flavour indices.

In principle, injecting equation \refeq{eq:PionDiracStructure} in equation \refeq{eq:BSE} provides a systems of coupled equations. Yet, solving it requires the knowledge of both the fully dressed quark propagator $S_q(p)$, which can be computed from the gap equation \refeq{eq:GapEquation}, and the kernel $K^{ab}_{ij}(k,p,P)$, the exact expression of which remains unknown. Therefore, approximations have to be introduced.

\subsubsection{Truncation schemes and symmetries}

The main idea is to solve consistently the \bs and Gap equations. To do so, truncation schemes have been developed. Any truncation scheme needs to be consistent with the fact that the pion is both a QCD bound state and a Goldstone mode of chiral symmetry breaking. So any of them should therefore respects the underlying symmetries and the way some of them are broken. For instance, chiral symmetry is explicitly broken in the so-called Axial-Vector Ward-Takahashi Identity (AVWTI), relating the axial vector vertex $\G_{5\mu}^j(p,P)$ with the pseudoscalar vertex $\G_5^j(p,P)$ through:
\begin{equation}
  \label{eq:AVWTI}
  P^\mu\G_{5\mu}^j(p,P) = \frac{\tau^j}{2}\left(S^{-1}(k_\eta)i\g_5 + i\g_5 S^{-1}(k_{\bar{\eta}})\right) -i\left[ m_{q_1}(\mu_R) +m_{q_2}(\mu_R) \right] \G^j_5(p,P), 
\end{equation}
the $m_q$ being the renormalised current quark masses, and $j$ indexing the isospin components. Any truncation procedure should be consistent with equation \refeq{eq:AVWTI}.

\begin{figure}[t]
  \centering
  \begin{equation*}
    \eqfig{0.2}{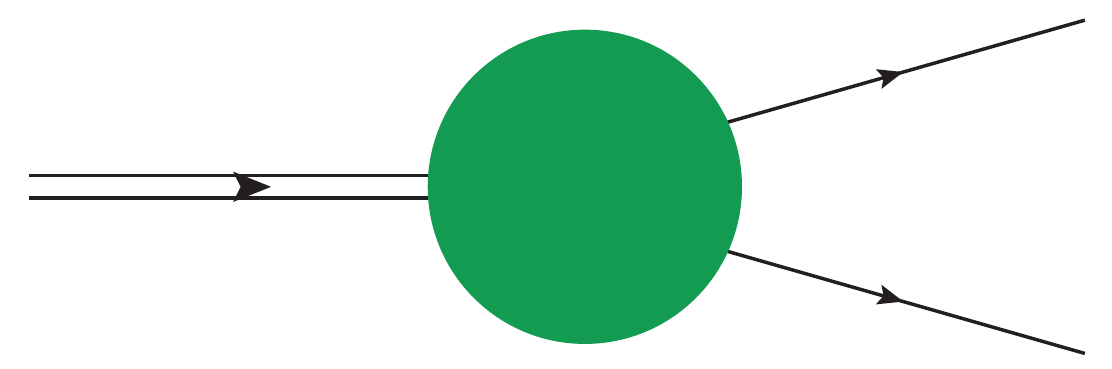} = \eqfig{0.05}{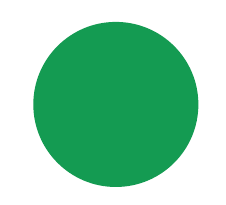} - \eqfig{0.25}{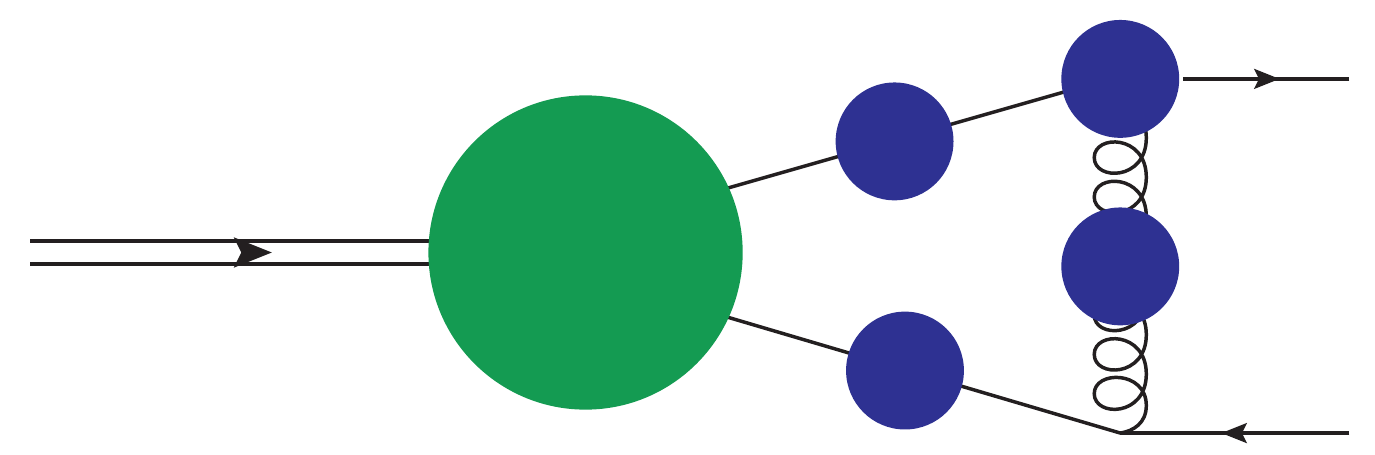} + \eqfig{0.25}{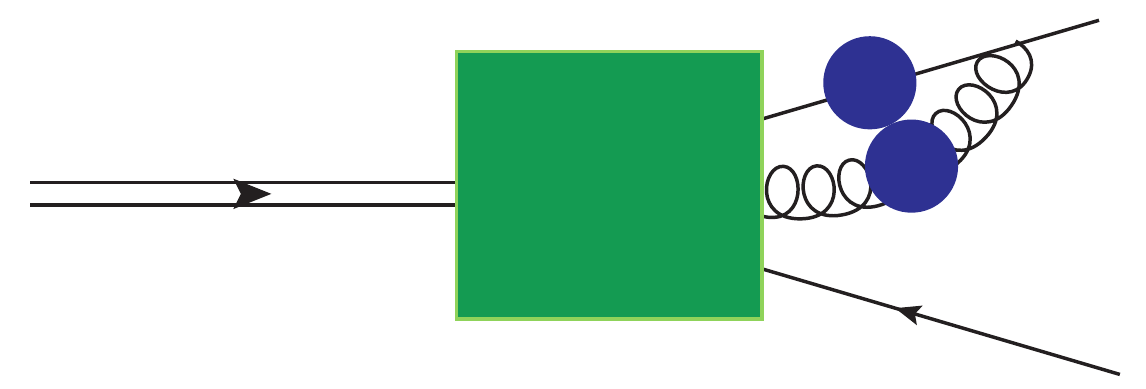}
  \end{equation*}
  \caption{Graphical representation of the inhomogeneous \bse for the axial-vector vertex. The blue circles indicate that the considered Green functions are dressed. The large green circles correspond to $\G_{5\mu}$, the green colour encoding the fact that this objetc obey to an inhomogeneous \bse. The small green cirle is the inhomogene term $\g^\mu \g_5$, and the green square $\Lambda_{5\mu\nu}$.  This graphical representation also work for the pseudoscalar vertex.}
  \label{fig:AVBSE}
\end{figure}

A possible strategy consists in solving the \bs equation for the pseudoscalar and axial-vector vertices, and then getting the pion \bs amplitude as a pole contribution of the two previous ones \cite{Maris:1997hd}. Indeed, both $\G_{5\mu}^i(p,P)$ and $\G_5^i(p,P)$ fulfil inhomogenous \bs equations. For instance in the case of the axial-vector vertex, one gets \cite{Bender:2002as}:
\begin{eqnarray}
  \label{eq:AVBSE}
  \G_{5\mu}(p,P) &=&  Z_2\g_\mu\g_5 - Z_1~g^2\int \frac{\textrm{d}^4k}{(2\pi)^4} D_{\alpha\beta}(p-k)\frac{\lambda^a}{2}\g^\alpha S(-k_{\bar{\eta}})\G_{5\mu} (k,P)S(k_\eta)\frac{\lambda_a}{2}\G^\beta(q_{\bar{\eta}}, k_{\bar{\eta}})\nonumber \\
  & &+ Z_1~ g^2\int \frac{\textrm{d}^4k}{(2\pi)^4} D_{\alpha\beta}(p-k)\frac{\lambda^a}{2}\g^\alpha S(k_{\eta})\frac{\lambda_a}{2} \Lambda_{5\mu\beta}(k,q,P).
\end{eqnarray}
This equation is illustrated on figure \ref{fig:AVBSE}.
A similar equation can be derived in the case of the pseudoscalar vertex. Truncation schemes has been developed \cite{Munczek:1994zz, Bender:1996bb,Chang:2009zb} in order to solve these equations consistently with the AVWTI \refeq{eq:AVWTI}.
Two main approximations are usually made to solve the \dses \cite{Chang:2012cc}. The first one, called Rainbow Ladder (RL), consists in approximating the kernel of the gap equation as:
\begin{equation}
  \label{eq:GapKernelRL}
  Z_1g^2 D_{\mu\nu}(p-k) \Gamma^\nu_a (k,p) = (p-k)^2 \mathcal{G}\left((p-k)^2\right) D_{\mu\nu}^{\textrm{free}}(p-k)\g^\nu.
\end{equation}
In this approach, the free gluon propagator in the Landau gauge $D_{\mu\nu}^{\textrm{free}}(p-k)$ is dressed by a specific function $\mathcal{G}\left((p-k)^2\right)$ reproducing the asymptotic behaviour (see \eg \refscite{Maris:1997hd,Maris:1998hc,Qin:2011dd}), and the quark-gluon vertex is taken bare. This choice for the Gap equation kernel is compatible with a choice of $\Lambda_{5\mu\beta}=0$ in equation \refeq{eq:AVBSE}. Another truncation scheme has been recently developed to improve RL by incorporating the capacity to account non-perturbatively for dynamical chiral symmetry breaking (DCSB) in the integral equations connected with bound-states and is usually described as DB truncation~\cite{Chang:2009zb}. Therein, the quark-gluon vertex is modelled by using the Ball-Chiu Ansatz (BC) \cite{Ball:1980ay} and an additional component generating an Anomalous Chromomagnetic Moment (ACM) \cite{Chang:2012cc}:
\begin{equation}
  \label{eq:GapKernelDCSB}
  Z_1g^2 D_{\mu\nu}(p-k) \Gamma^\nu_a (k,p) = \mathcal{G}\left((p-k)^2\right) D_{\mu\nu}^{\textrm{free}}(p-k)Z_2\tilde{\G}^\nu(k,p),
\end{equation}
with
\begin{equation}
  \label{eq:DCSBQQGAnsatz}
  \tilde{\G}^\nu(k,p) = \G^\nu_{\textrm{BC}}(k,p) + \G^\nu_{\textrm{ACM}}(k,p).
\end{equation}
In this case, the contribution of $\Lambda_{5\mu\beta}$ cannot be ignored anymore. Indeed, DB truncation has been proven to be phenomenologically successful not only describing ground-state vector- and isospin-nonzero pseudoscalar-mesons constituted by light quarks, as RL, but in all the channels considered thus far~\cite{Qin:2011xq,Chang:2011ei,Chen:2012qr,Chang:2013pq}. It is worth anyhow to remark that the DB kernel for the quark Gap equation has been very recently proved to coincide with the equivalent kernel directly inferred from the analysis of gauge-sector (gluon and ghost propagators) Gap equations~\cite{Binosi:2014aea}, namely from an {\it ab-initio} analysis.

One should keep in mind that Green functions are quantities which remains gauge-dependent, and thus, the presentation below would be incomplete without a word on the choice of the gauge. Usually, the Landau gauge is chosen for multiple reasons. Among them, the Landau gauge is a fixed point of the renormalisation of the covariant gauge parameter. Then it is also the gauge which is the less sensitive to the chosen Ansatz for the quark-gluon vertex as emphasised in \refcite{Bashir:2008fk,Raya:2013ina}. Finally, it is also the gauge used in lattice computations.

\subsection{The Nakanishi representation}

Due to the decomposition of equations \refeq{eq:Propagator} and \refeq{eq:PionDiracStructure}, solving the \dses consists in computing five scalar functions parameterising the quark propagator and the pion \bs amplitude ($H_\pi$ is usually neglected \cite{Chang:2013pq}). This is possible numerically, at the cost of discrete momentum variables. Thus one can expect to get numerical grids of functions with a given number of points in $p^2$. Even if performing such a computation is already a success, such grids may nevertheless be inadequate to compute other non-perturbative functions. Therefore, interpolating functions have been introduced and successfully fitted on the numerical results here presented.

\subsubsection{Interpolating functions}

In the case of the quark propagator, it has been chosen \cite{Chang:2013pq} to use a parameterisation involving complex conjugates poles:
\begin{equation}
  \label{eq:PropaConjPol}
  S(k) = \sum_{j}^N \left[ \frac{z_j}{i \g \cdot k + m_j}+\frac{z_j^*}{i \g \cdot k + m_j^*}\right],
\end{equation} 
where $z_j$ and $m_j$ are fitted on the grids of solutions. Fits are satisfactory for $N\ge 2$.

The functions associated with the Dirac structure of the pion \bs amplitude \refeq{eq:PionDiracStructure} can be 
\begin{figure}[t]
  \centering
  \includegraphics[width=0.5\textwidth]{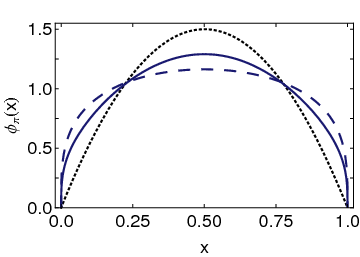}
  \caption{Pion DA computed at renormalisation scale $\mu_R = 2~\GeV$. The dashed curve corresponds to the Rainbow Ladder truncation scheme of equation \refeq{eq:GapKernelRL}, the solid curve to the improved kernel of equation \refeq{eq:GapKernelDCSB}, and the dotted curve is the asymptotic DA. This figure comes from \refcite{Chang:2013pq}.}
  \label{fig:PionDA}
\end{figure}
described using the so-called Nakanishi representation \cite{Nakanishi:1963zz}:
\begin{equation}
  \label{eq:NakanishiRepresentation}
  \chi (k,P) = \mathcal{N}\int_{-1}^{1} \textrm{d}z \int_0^\infty \textrm{d}\g \frac{\varphi _l^{[n]}(z,\g)}{\left(\gamma + \tilde{m}^2 - \frac{1}{4}m_H^2 - k^2-P \cdot k z - i\epsilon\right)^{n+2}},
\end{equation}
with $m_H$ being the mass of the considered hadron, $\mathcal{N}$ a normalisation constant, $\tilde{m}$ the mass of the constituents, and $\varphi _l^{[n]}(z,\g)$ is such that:
\begin{equation}
  \label{eq:DummyN}
   \lim_{\g \rightarrow \infty} \frac{\varphi_l^{[n]}(z,\g)}{\g^n} = 0.
\end{equation}
The Nakanishi representation has been used in modern \ds and \bs studies of mesons, see \eg \refscite{Karmanov:2005nv, Chang:2013pq}. For instance, the authors of \refcite{Chang:2013pq} split the functions $E_\pi$, $F_\pi$, and $G_\pi$ into infrared and ultraviolet dominant contributions, and parameterise them using the Nakanishi representations. This method has allowed the computation of the pion DA, yielding a significantly wider distribution in terms of the momentum fraction than the asymptotic one, as shown on figure \ref{fig:PionDA}. Very recently, the same approach has been also followed to compute distribution amplitudes for bound-states involving strange and charm quarks~\cite{Shi:2015esa,Ding:2015rkn}.

\subsubsection{An algebraic model}
\label{sec:AM}

The Nakanishi representation has been the starting point of a algebraic model suggested in \refcite{Chang:2013pq}, for both the quark propagator and the \bs amplitude. Focusing on the $\g_5$ term of the \bs amplitude, the algebraic model yields:
\begin{eqnarray}
S( p ) 
& = & \big[ - i \gamma \cdot p + M \big] \Delta_M( p ^2 ), \label{eq:TMQuarkPropagator} \\
\Delta_M( s )
& = & \frac{1}{s + M^2}, \label{eq:TMPropagatorMassTerm} \\
\Gamma_\pi( k, P )
& = & i \gamma_5 \frac{M}{f_\pi}M^{2\nu} \int_{-1}^{+1} \mathrm{d}z \, \rho_\nu( z ) \ 
\left[\Delta_M( k_{+z}^2 )\right]^\nu, \label{eq:TMBSA} \\
\rho_\nu( z ) & = & \frac{1}{\sqrt{\pi}}\frac{\Gamma(\nu+3/2)}{\G(\nu+1)}( 1 - z^2 )^\nu, \label{eq:TMSmearing} \\
k_{+z} & = & k-\left(\frac{1-z}{2} - \eta\right)P \label{eq:TMkDef}.
\end{eqnarray}
Two parameters can be identified here\footnote{We remind the reader that the physics cannot depends on $\eta$ because of translational invariance.}: $\nu$ which controls the shape of the pion \bs amplitude and $M$ which can be seen as an effective dressed quark mass. For $\nu=1$, the analytic computation of the pion DA with this algebraic model is possible and leads to the asymptotic results \cite{Chang:2013pq}. Indeed, the pion DA defined in equation \refeq{eq:PionDADef} can be seen as the projection of the \bs wave function:
\begin{equation}
  \label{eq:DefDAProj}
  f_\pi \varphi_\pi (u) = \textrm{Tr}\left[ Z_2 \int \frac{\textrm{d}^4k}{(2\pi)^4}~\delta(k_\eta\cdot n-u P\cdot n)~\g \cdot n~ \g_5~ \chi_\pi(k,P)\right],
\end{equation}
$k_\eta$ being defined in equation \refeq{eq:Defketa}.

The results obtained for the pion DA are very encouraging, both numerically and analytically. Indeed, the algebraic parameterisation based on a simple Nakanishi representation \refeq{eq:TMQuarkPropagator}-\refeq{eq:TMkDef} allows to compute analytically the asymptotic pion DA. Then fitting the parameters of equations \refeq{eq:PropaConjPol} and \refeq{eq:NakanishiRepresentation} on the numerical solution of the DSEs provides a more realistic model for the pion DA, deforming the asymptotic one. A similar approach has been developed in \refcite{Mezrag:2014tva} for GPDs. The idea is to use the algebraic parameterisation to highlight the main features of GPD modelling through the Nakanishi representation, paving the way for numerical computations based on equations \refeq{eq:PropaConjPol} and \refeq{eq:NakanishiRepresentation}.


\section{Covariant computation of the pion GPD}
\label{sec:CovariantComputations}
\subsection{Pion GPD Mellin moments}

In this section, we review the modelling approach, within the previously described DSEs-BSE framework, used in \refcite{Mezrag:2014tva} and references therein, based on a covariant computation of a non-perturbative matrix element which is generally performed through the so-called impulse approximation. 

\subsubsection{Local Impulse Approximation}

\begin{figure}[b]
  \centering
  \includegraphics[width=0.3\textwidth]{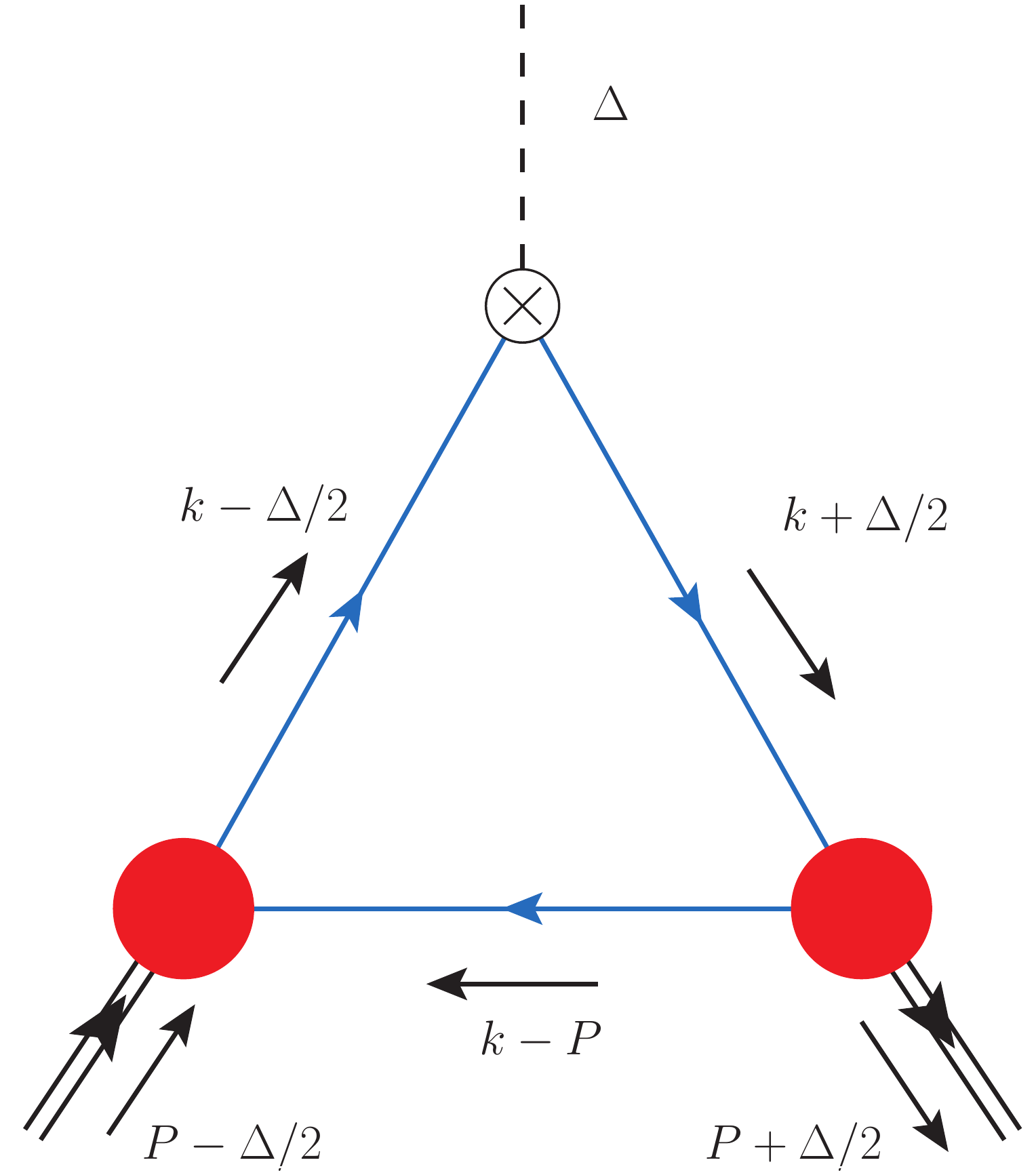}
  \quad
  \includegraphics[width=0.32\textwidth]{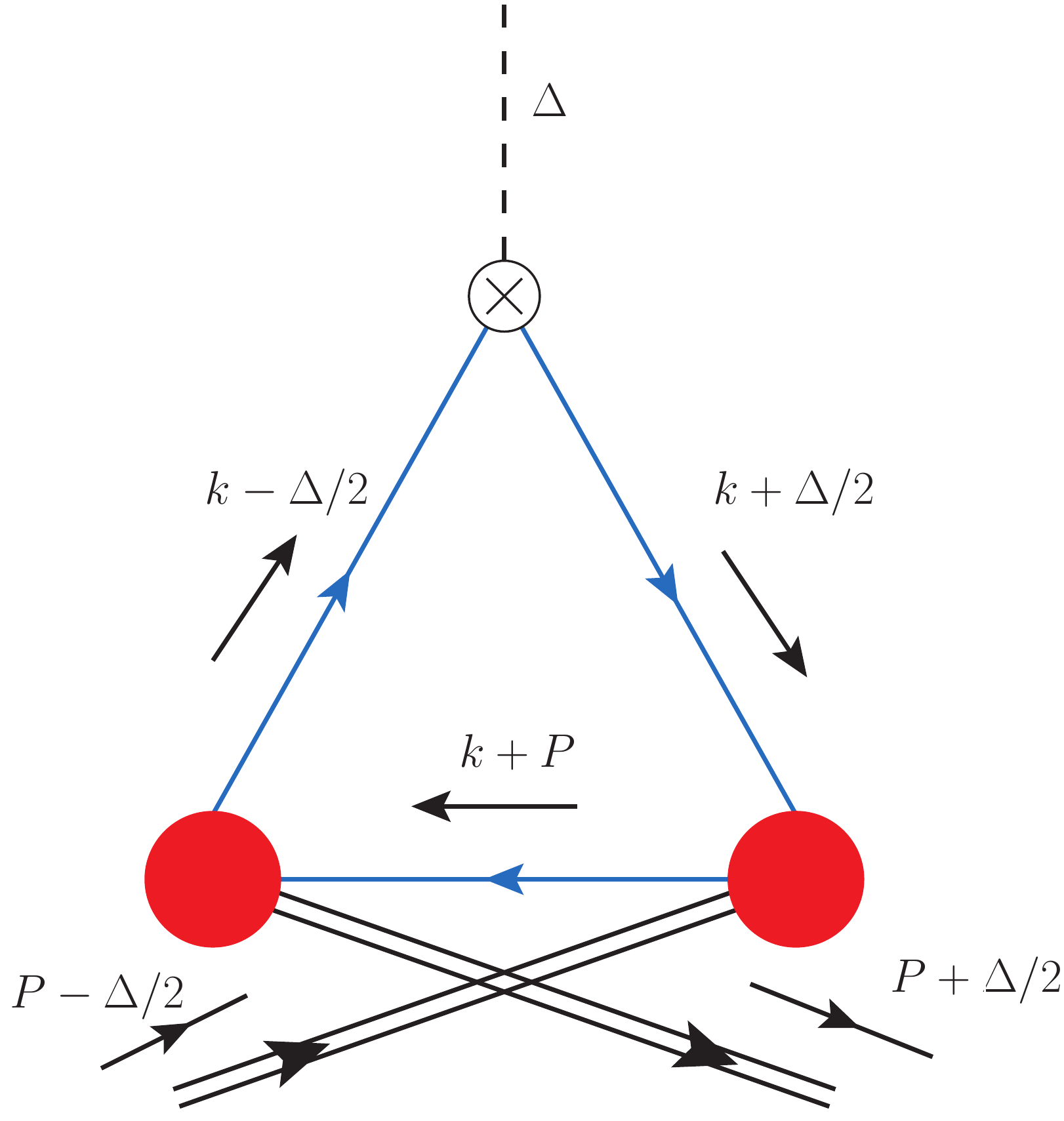}
  \caption{Triangle diagram approximation. Left-hand side: case of the quark GPD. Right-hand side: case of the anti-quark GPD.}
  \label{fig:TriangleDiagrams}
\end{figure}

Widely used to compute form factors, the impulse approximation corresponds in practice to model the considered matrix element using a triangle diagram as shown on figure \ref{fig:TriangleDiagrams} (e.g. see refs.~\cite{Broniowski:2003rp,Broniowski:2007si,Tiburzi:2002tq,Theussl:2002xp,Bissey:2003yr,VanDyck:2007jt,Frederico:2009fk}). However, this approximation cannot be applied directly to the GPD itself, due to the fact that, contrary to the form factor, the GPD is defined through a non-local matrix element. Consistently modelling this operator is not a trivial task, but the problem of non-locality can be circumvented. Indeed, as seen in equation \refeq{eq:MMandLocalMatrixElement}, the Mellin moments of the pion GPD are directly proportional to a \emph{local} matrix element in terms of the covariant local twist-two operators in equation \eqref{eq:LocalOperator}. 
Yet, as emphasised in section \ref{sec:DSEs}, working in a DSEs-BSE framework means working in the Landau gauge, and thus one should take the covariant derivative instead of the partial one in equation \refeq{eq:MMandLocalMatrixElement}. Expanding the covariant derivative leads to:
\begin{equation}
  \label{eq:CovariantDerivativeExpansion}
  (\overleftrightarrow{D} \cdot n)^m = \frac{1}{2^m}\sum_{j=0}^m\binom{m}{j}(\overrightarrow{\partial} \cdot n - n \cdot \overleftarrow{\partial})^j(2ig~n \cdot A)^{m-j},
\end{equation}
where the gluon contributions explicitely appears. However, the choice done here is to focus only on the term $j=m$, which corresponds to neglect the gauge link in the definition of the GPDs. The reader may find this a bold approximation, but it has been argued in \refcite{Kopeliovich:2011rv} that the gauge link in twist-two operators along the lightcone brings only numerically negligible contributions.

The overall normalisation of the triangle diagram is controlled by the local twist-two operator. A normalisation condition of the \bs amplitude based on charge conservation in the ladder approximation has been introduced by  Mandelstam \cite{Mandelstam:1955sd}. It relies on the computation of a pion form factor at vanishing momentum transfer within the impulse approximation. The value of the form factor is then fixed to $1$. Consequently, as the form factor is the Mellin Moment of the GPD $H$ for $m=0$, it consists in fixing $\mathcal{M} _0(0,0)$ to $1$.
Using the Ward-Takahashi identity (WTI):
\begin{equation}
  \label{eq:EMWTI}
i \Delta^\mu \Gamma_\mu(k+\frac{\Delta}2,k-\frac{\Delta}2) \ = \ S^{-1}(k+\frac{\Delta}2) - S^{-1}(k-\frac{\Delta}2), 
\end{equation}
Mandelstam's condition was shown \cite{Lurie:1965, Nishijima:1967} to be equivalent to the canonical normalisation of the \bs amplitude given in equation \refeq{eq:BSNormalisation}.

\subsubsection{Computations of Mellin Moments}

\begin{figure}[b]
  \centering
    \includegraphics[width = 0.48\textwidth]{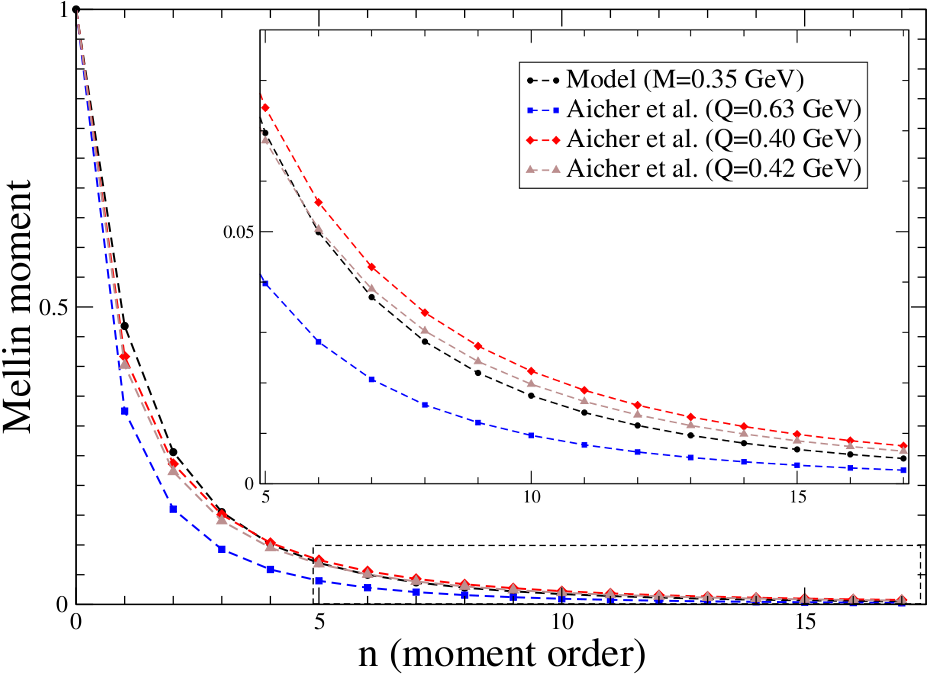}
      \caption{Mellin moments from equation \refeq{eq:MellinMomentDirectDiagram} and obtained with the parameterisation of \refcite{Aicher:2010cb} run with DGLAP equation down to $Q = 0.40~\GeV\xspace$ and $0.42~\GeV$.}
  \label{fig:PDFMoments}
\end{figure}

Following the Mandelstam approach which ensures charge conservation, the $m=0$ local twist-two operator is assimilated to the vector current vertex $\Gamma_\mu$. Higher-$m$ operators are based on the first one, taking into account the action of the $(i\overleftrightarrow{D} \cdot n)^m$ on the incoming and outgoing quarks. In the algebraic model presented in \refsec{sec:AM}, when applied to the quark fields, these operators yield: $(k \cdot n)^m$.
Consequently, the computation of the Mellin moments $\MM$ of the pion quark GPD leads to:
\begin{eqnarray}
\label{eq:TriangleDiagrams} 
2 ( \Pn )^{m+1} \MM & = &  \textrm{Tr}_{\textrm{CFD}}\left[ \int \frac{\mathrm{d}^4k}{(2\pi)^4}(k\cdot n)^m \, \tau_- i\bar{\Gamma}_\pi\left((1-\eta)\left(k+\Dd\right)+\eta(k-P),P+\Dd \right) \right. \nonumber \\
& &  S( k + \Dd ) \ i~n \cdot \Gamma(k+\Dd,k-\Dd) ~ S( k-\Dd ) \nonumber \\
& &  \left. \tau_+ i\Gamma_\pi\left(\eta(k-P)+(1-\eta)\left(k-\Dd\right),  P-\Dd \right) S( k - P ) \right],
\end{eqnarray}
for the leftmost diagram of figure \ref{fig:TriangleDiagrams} (a similar equation works for the rightmost one), where 
$S$ and $\Gamma_\pi$ have been defined in equations \refeq{eq:TMQuarkPropagator} and \refeq{eq:TMBSA}, respectively. $n \cdot \Gamma(k+\Dd,k-\Dd) $ stands for the dressed quark photon vertex projected on the lightcone, and the $\tau^\pm$ are linear combinations of the Pauli matrices $\tau^i$:
\begin{equation}
  \label{eq:DefTauPM}
  \tau^\pm = \tau^1 \pm i \tau ^2.
\end{equation}
To ensure the normalisation of the pion GPD computed in the impulse approximation, one need to use a vertex and a propagator consistent with the WTI \refeq{eq:EMWTI}. Relying on equation \refeq{eq:TMQuarkPropagator} for the quark propagator, the WTI yields:
\begin{equation}
  \label{eq:WTIQuarkPropaTM}
  \ S^{-1}(k+\frac{\Delta}2) - S^{-1}(k-\frac{\Delta}2) = i\Delta \cdot \gamma ,
\end{equation} 
and thus, the minimal Ansatz required to fulfil the normalisation condition is $\G^\mu=\g^\mu$. Then, the DSEs-BSE inspired algebraic parametrisation in equations~(\ref{eq:TMQuarkPropagator}-\ref{eq:TMkDef}) can be applied to \eqref{eq:TriangleDiagrams} (details of the computation can be found in \refcite{Mezrag:2014tva,Mezrag:2015mka}) 
and leads to:
\begin{eqnarray}
\MM & = & 
\frac{M^2}{2 \pi^2 f_\pi^2} \int_0^1 \mathrm{d}x \, \mathrm{d}y \, \mathrm{d}u \, \mathrm{d}v \, \mathrm{d}w \, \int_{-1}^{+1} \mathrm{d}z \, \mathrm{d}z' \, \delta( x + y + u + v + w - 1 ) x^{\nu - 1} y^{\nu - 1} \rho( z ) \rho( z' ) \nonumber \\
& &  \frac{M^{4\nu}}{2} \left[ \frac{\Gamma( 2 \nu + 1 )}{\Gamma( \nu )^2} \left( \big( f \Delta \cdot n + g P \cdot n \left( \left( \Dd \right)^2 - P^2 \right) - 2 P\cdot n \left( \Dd \right)^2 \right) \frac{1}{(M')^{2 \nu + 1 }} \right. \nonumber \\
& & + \frac{\Gamma( 2 \nu )}{\Gamma( \nu )^2} \frac{1}{2} \left( P\cdot n + \Dd\cdot n \right) \delta( v ) \frac{1}{(M')^{2\nu}} + \frac{\Gamma( 2 \nu )}{\Gamma( \nu )^2} \frac{1}{2} \left( P\cdot n - \Dd\cdot n \right) \delta( w ) \frac{1}{(M')^{2\nu}} \nonumber \\
& & \left. + \frac{\Gamma( 2 \nu )}{\Gamma( \nu )^2} \left( f \Delta\cdot n + g P\cdot n \right) \delta( u ) \frac{1}{(M')^{2\nu}} \right] \frac{( f \Delta\cdot n + g P\cdot n )^m}{2 (P\cdot n)^{m+1}} 
\label{eq:MellinMomentDirectDiagram}
\end{eqnarray}
where:
\begin{eqnarray}
  \label{eq:Shortcutf}
  f = f(x,y,v,w,z,z') & = & \frac{1}{2} \left( -\frac{1+z'}{2}y + \frac{1+z}{2}x + v-w \right), \\
  \label{eq:Shortcutg}
  g = g(x,y,u,z,z') & = & \left(\frac{1-z'}{2}\right ) y + x \frac{1-z}{2} + u,   \\
  \label{eq:ShortcutM}
  M'^2(t,P^2,x,y,u,v,w,z,z') & = & M^2 +\frac{t}{4}\left( -4f^2 + y\left( \frac{1+z'}{2}\right)^2 + x \left( \frac{1+z}{2} \right )^2 +v + w \right ) \nonumber \\  
& &  +P^2 \left( -g^2 + \left(\frac{1-z'}{2}\right)^2 y + \left(\frac{1-z}{2}\right)^2 x + u \right ).
\end{eqnarray}
A similar expression can be derived for the rightmost diagram of figure \ref{fig:TriangleDiagrams} and both can be numerically evaluated to produce, for instance, the results plotted in figure \ref{fig:PDFMoments}. 

\subsubsection{Comparison to available data}

Unfortunately, no experimental data are currently available for off-forward kinematics on a pion target, despite some studies on virtual targets \cite{Amrath:2008vx}. Therefore, the present model can be compared only to PDF and form factors experimental data. 

Owing to equation \refeq{eq:SumRulesFF}, the form factor can be computed by taking $m=0$ in equation \refeq{eq:MellinMomentDirectDiagram}.
\begin{figure}[t]
  \centering
  \begin{tabular}{cc}
  \includegraphics[width= 0.48\textwidth]{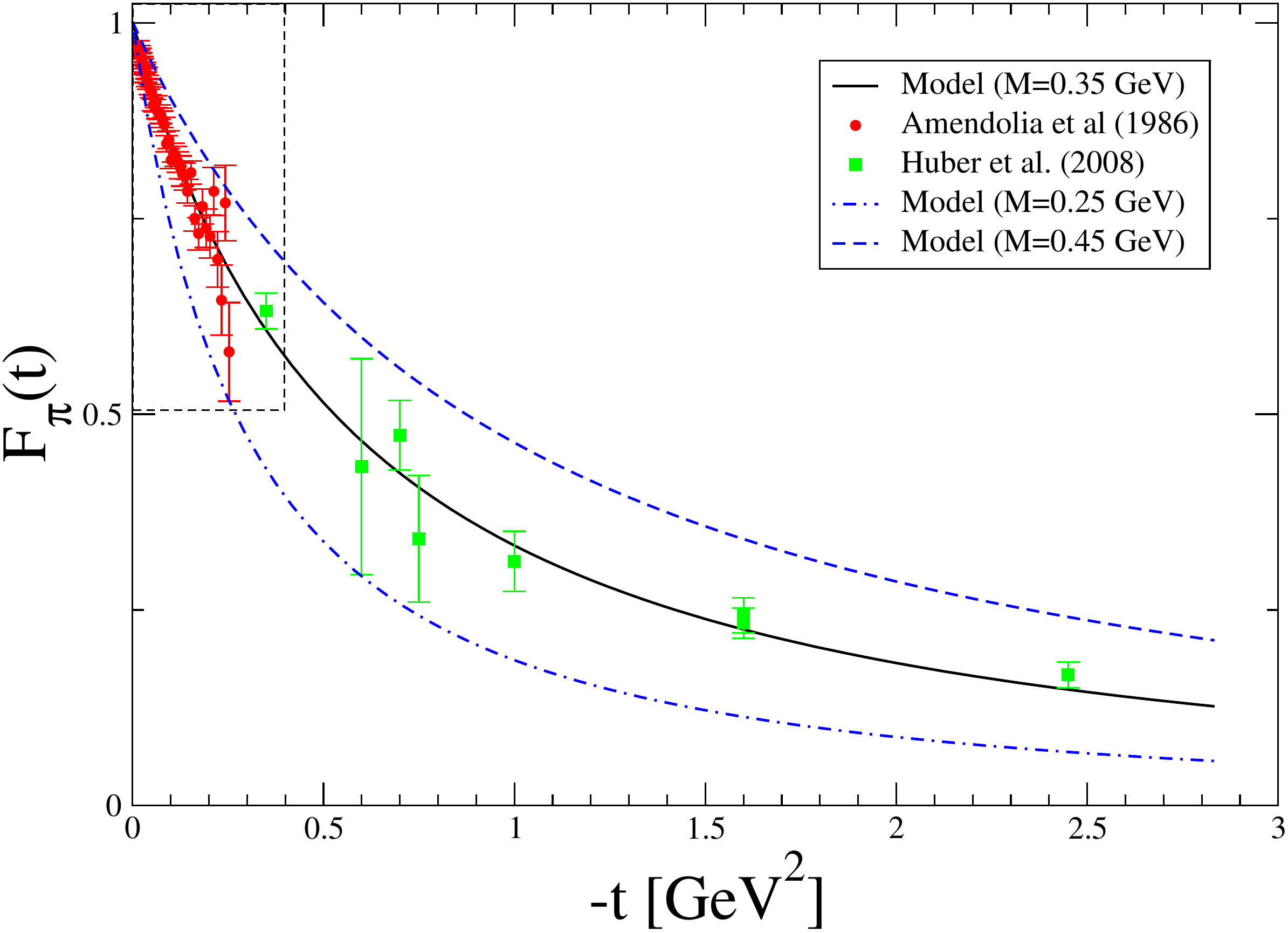} 
  &
  \includegraphics[width= 0.48\textwidth]{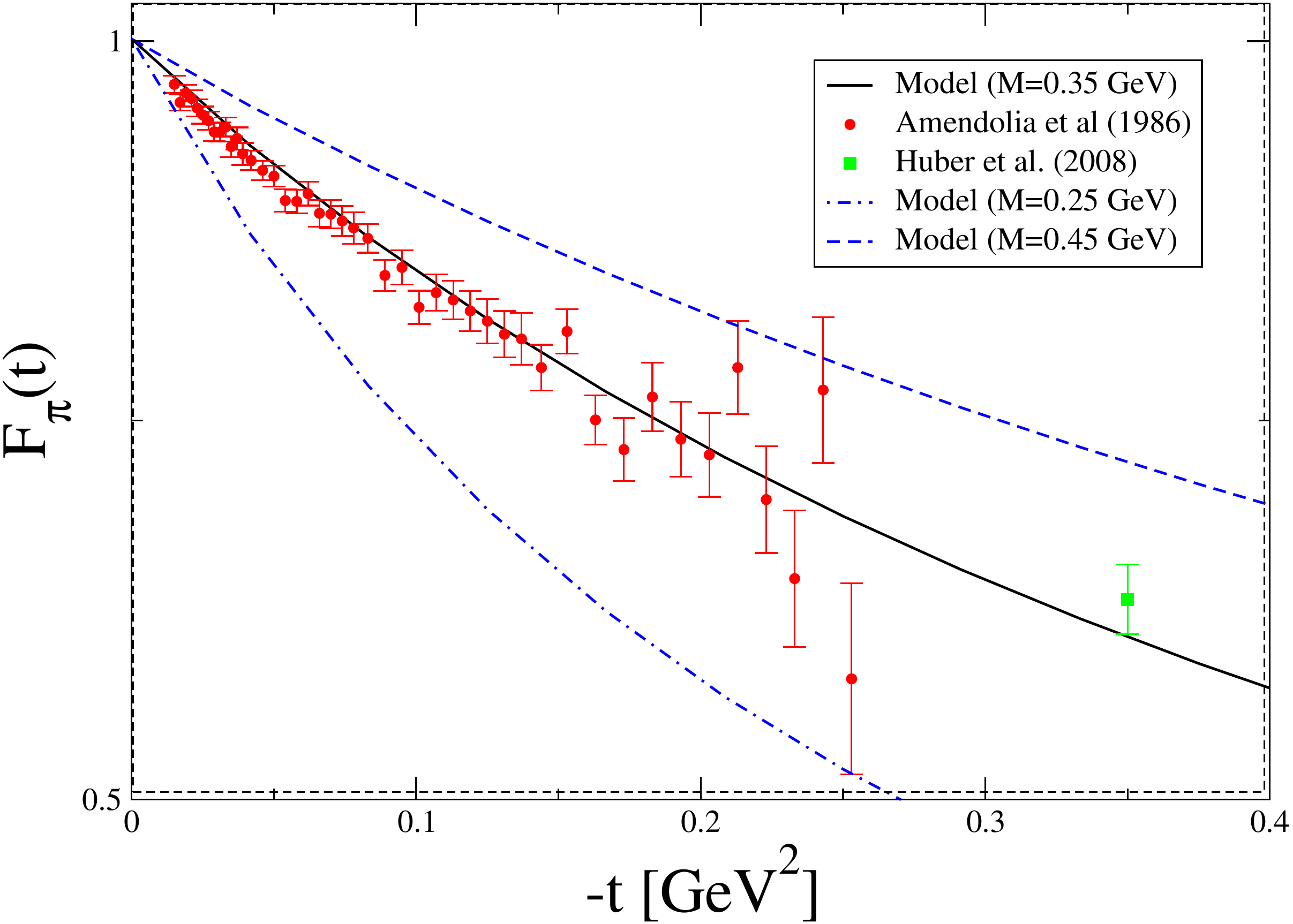} 
  \end{tabular}
  \caption{Left-hand side: the pion form factor $F_\pi$ computed at $M$ = 0.35~\GeV (solid black line), 0.25~\GeV (dot-dashed blue line) and 0.45~\GeV (dashed blue line), with $\nu = 1$ for the three cases. Experimental data are taken from \refcite{Amendolia:1986wj,Huber:2008id}. Right-hand side: zoom of the dashed square in the leftmost plot allowing to emphasise the constraint provided by the large number of data points in the low-momentum region.}
  \label{fig:FormFactor}
\end{figure}
In this approach, two parameters enter the game, as explained in section \ref{sec:AM}: $\nu$ is fixed to 1 and 
only $M$ remains as a free parameter, which drives the $t$-behaviour of the model. This can be easily highlighted since after proper normalisation, equation \refeq{eq:MellinMomentDirectDiagram} can be explicitly written in terms of $\theta = t/M^2$, instead of $t$ and $M$. 
As argued in \refcite{Mezrag:2014tva}, the fit of the $m=0$ moment to the available experimental data of \refscite{Amendolia:1986wj,Huber:2008id} allows for a precise determination of the parameter $M$, although the interest of such a fit is limited here, as our main purpose is to explore the possibilities offered by the DSE framework with regard to GPDs modelling. However, figure \ref{fig:FormFactor} shows that data are well described for $M\simeq 0.35~\GeV$, which is a typical constituent quark mass. The sensitivity of the model with respect to $M$ is highlighted by plotting two additional curves: one for $M = 0.25~ \GeV$ and one for $M = 0.45~ \GeV$. The reader should note at this point that the form factor is providing a mass scale to the algebraic model, whereas when solving DSEs, the mass scale is intrinsic to the result. 

The charge radius of the pion can also be determined here. The NA7 collaboration gives in \refcite{Amendolia:1986wj} its experimental result:
\begin{equation}
\label{eq:PionElectricChargeRadiusNa7}
\cond{r_\pi^2}^{\textrm{exp}} = \left. -6 \frac{dF_\pi}{dt}\right|_{t=0}= 0.439 \pm 0.008~\textrm{fm}^2.
\end{equation}
The model would reach agreement with the NA7 Collaboration value for $M = 339 \pm 3~\MeV$, which is close to the choice of $M = 350~\MeV$.

Besides the pion form factor, the pion PDF has also been measured in the large-$x$ region. Recent analyses taking into account gluon resummation \cite{Aicher:2010cb} have provided the community with phenomenological parameterisations of the pion PDF. Consequently, it is possible to compute the Mellin moments of these phenomenological parameterisations, and to compare them with the ones computed in equation \refeq{eq:MellinMomentDirectDiagram}. However, the comparison can be done only at the same scale, and the scale of the Mellin moments of equation \refeq{eq:MellinMomentDirectDiagram} is \emph{a priori} unknown. Moreover, as the PDF is taken at $\theta = 0$, the dependence in $M$, the only mass scale directly available, vanishes. The idea developed in \refcite{Mezrag:2014tva} is thus to consider the scale $\mu_F$ as a free parameter. The Mellin moments of the phenomenological parameterisations of \refcite{Aicher:2010cb} have thus been evolved to lower scales. The relevant scale is then selected when the evolved phenomenological Mellin moments are in agreement with the ones computed through equation \refeq{eq:MellinMomentDirectDiagram}. As shown on figure \ref{fig:PDFMoments} this corresponds to a low scale, close to the one chosen for $M$ and thus highlighting the consistency of this approach.

\begin{figure}[h]
  \begin{center}
  \includegraphics[width=0.40\textwidth]{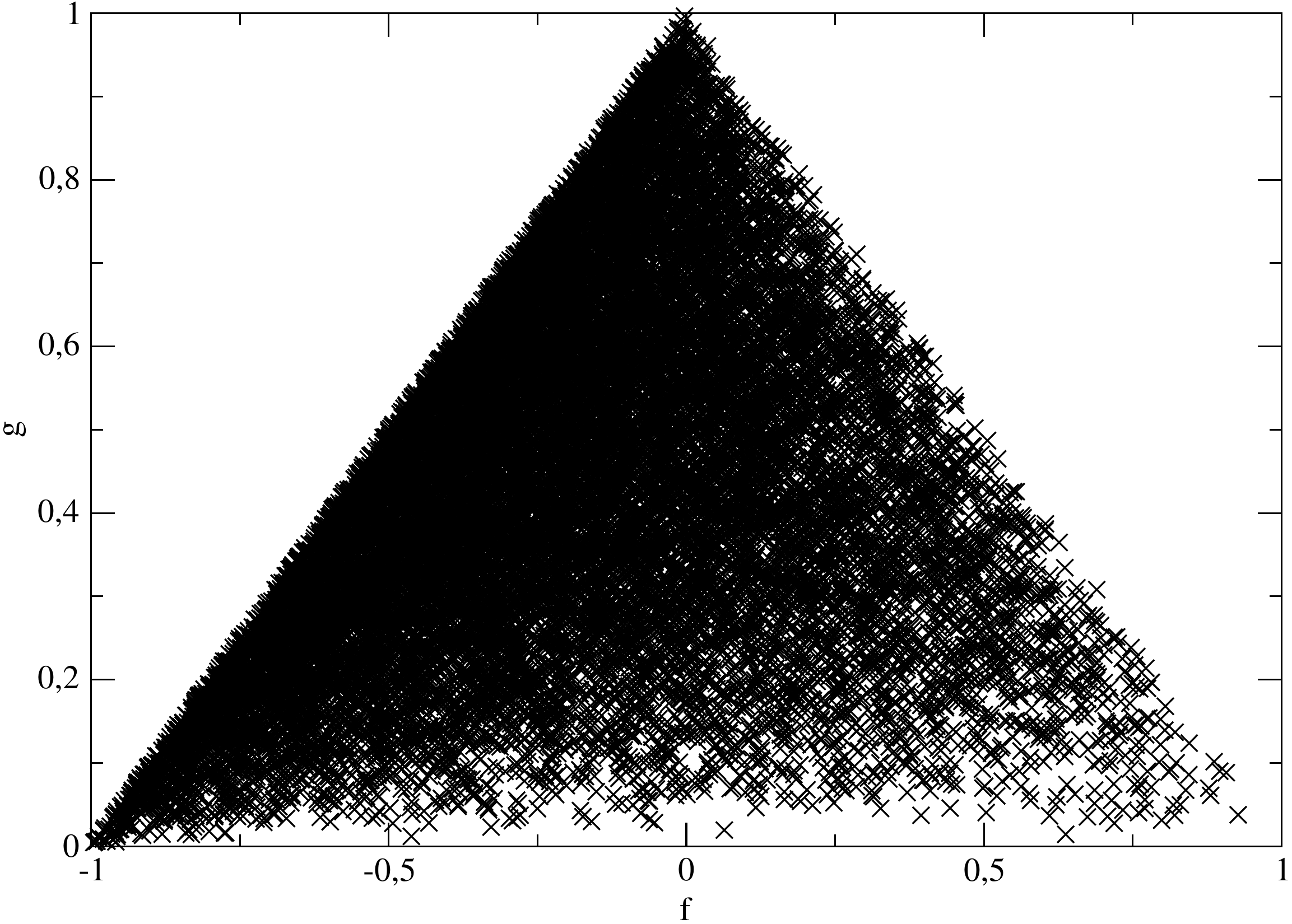}
  \end{center}
  \caption{Right-hand side: Values of $f$ and $g$ when generating the Feynman parameters in a Monte-Carlo approach using $2.10^5$ points. The shape of a half-rhombus clearly appears.}
  \label{fig:MonteCarloRhombus}
\end{figure}

\subsection{From Mellin moments to GPDs}

Going from the Mellin moments to the pion GPD itself is \textit{a priori} possible through the so-called Mellin-Barnes inverse transform \cite{Mueller:2005ed}. However, this technique has been performed for models much simpler than the integrals in equations \refeq{eq:MellinMomentDirectDiagram} only, and is hardly practicable here. Therefore, one should turn to other approaches.

\subsubsection{Tensorial structure and Double Distributions}

\begin{figure}[b]
  \centering
  \begin{tabular}[h]{l r}
    \includegraphics[width=0.4\textwidth]{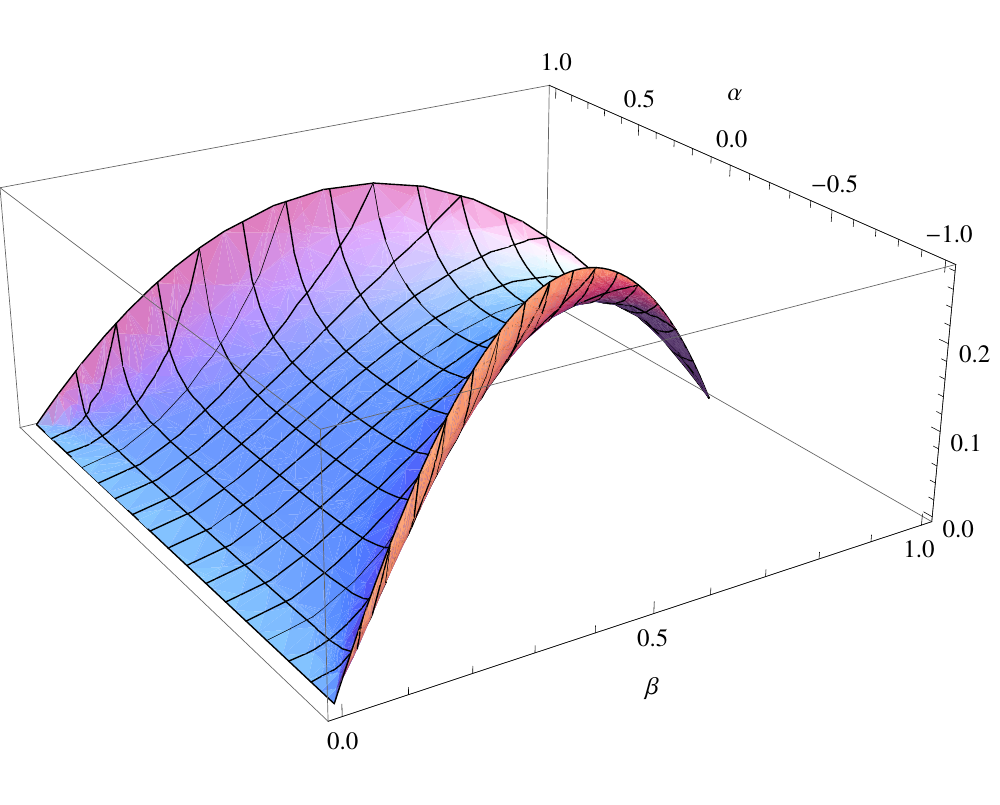} & \includegraphics[width=0.4\textwidth]{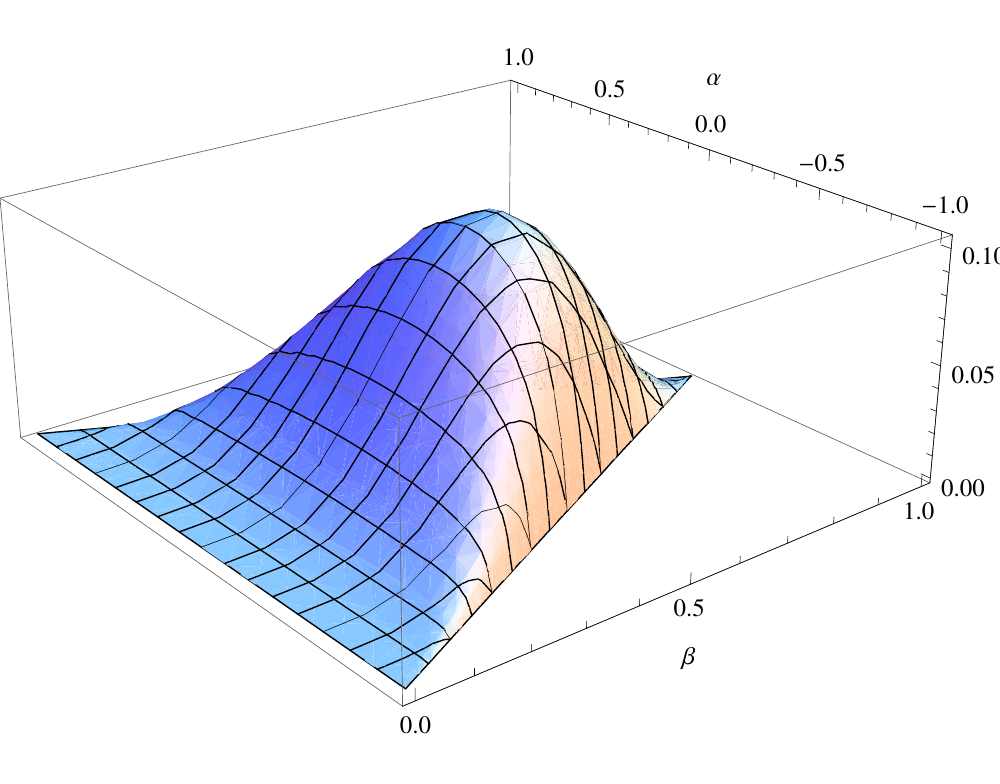}
  \end{tabular}
  \caption{DDs $F$ for $\nu = 1$ (left-hand side) and $\nu =2$ (right-hand side).}
  \label{fig:DDFs}
\end{figure}
\begin{figure}[t]
  \centering
  \begin{tabular}[h]{l r}
 \includegraphics[width=0.4\textwidth]{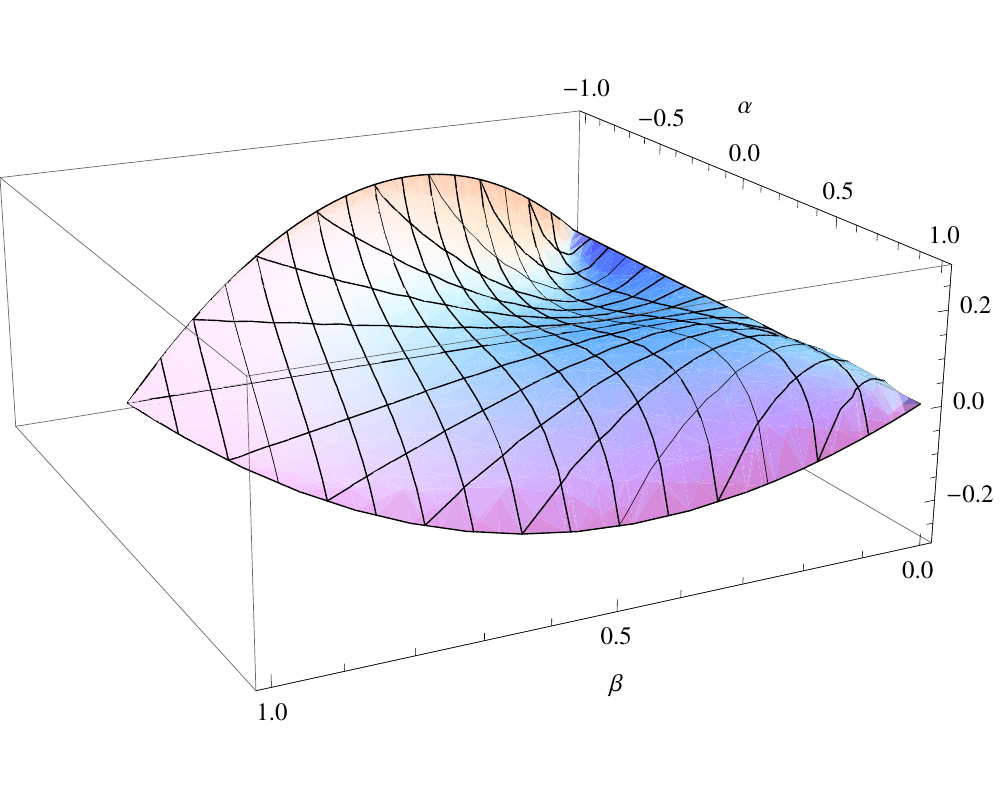} & \includegraphics[width=0.4\textwidth]{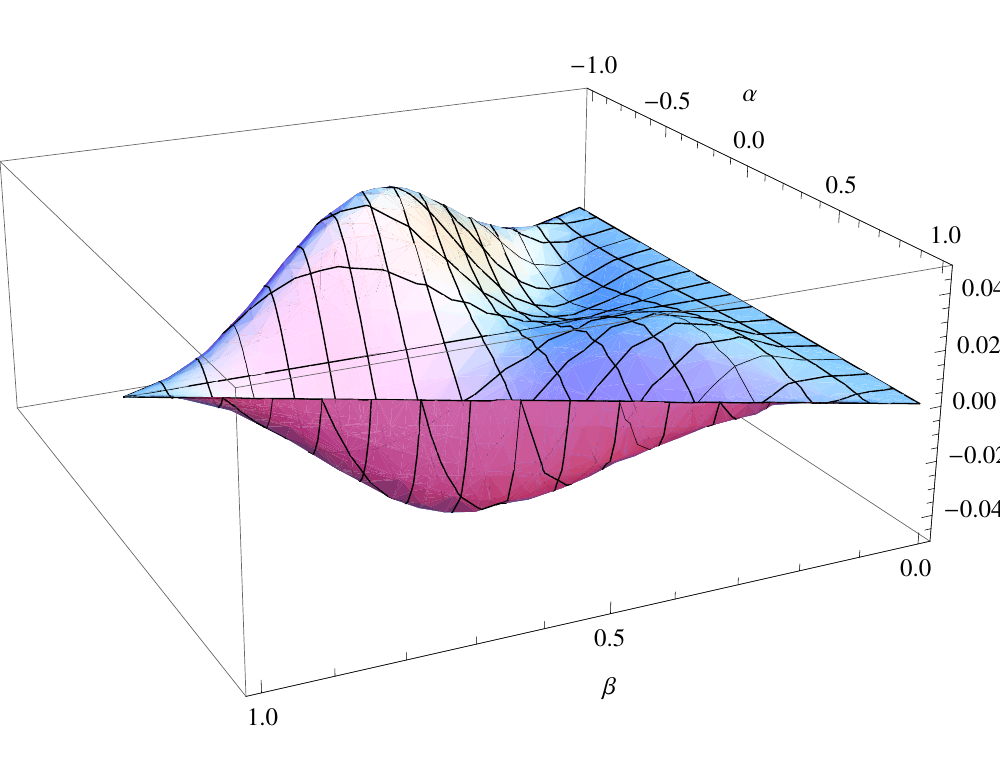} 
  \end{tabular}
  \caption{DDs $G$ for $\nu = 1$ (left-hand side) and $\nu =2$ (right-hand side).}
  \label{fig:DDGs}
\end{figure}

Instead, one can try to express the Mellin moments of equation \refeq{eq:MellinMomentDirectDiagram} in the same way as in equation \refeq{eq:GPDMellinMomentViaDD}, in order to identify the DDs $\DDF$ and $\DDG$. This can actually be done, providing that one can identify the relevant variable $\beta$ and $\alpha$ with the good support $\Omega$, described in equation \refeq{eq:DDSupport}. From equation \refeq{eq:MellinMomentDirectDiagram}, the natural candidates are:
\begin{equation}
  \label{eq:DDNaturalVariables}
  \left \{
    \begin{array}[h]{l c r}
      \beta & = & g\\
      \alpha & = & -f
    \end{array}
    \right .  ,
\end{equation}
where $f$ and $g$ have been defined in equations \refeq{eq:Shortcutf} and \refeq{eq:Shortcutg}. The proof that $f$ and $g$ describe actually a half-rhombus can be found in \cite{Mezrag:2014tva}. But it is plain when looking at figure \ref{fig:MonteCarloRhombus}. Identifying the DDs $\DDF$ and $\DDG$ requires then a change of variables from the Feynman parameters to the DD variables $\beta$ and $\alpha$, \ie following the pattern:
\begin{equation}
\int_0^1 \mathrm{d}x \, \mathrm{d}y \, \mathrm{d}u \, \mathrm{d}v \, \mathrm{d}w \, \int_{-1}^{+1} \mathrm{d}z \, \mathrm{d}z' \, \delta( x + y + u + v + w - 1 ) \phi( x, y, u, v, w, z, z' ) = \int_{\Omega} \mathrm{d}\beta \, \mathrm{d}\alpha \Phi( \beta, \alpha ).
\label{eq:ChangeOfVariableRhombus}
\end{equation}
Such a change of variable is non-trivial, but has been identified and performed in \refcite{Mezrag:2014tva} leading to the DDs illustrated in figures \ref{fig:DDFs} and \ref{fig:DDGs} for $t=0$.

\subsubsection{GPDs reconstruction: successes and weaknesses}

From the DDs, it is possible to compute the pion GPD through the relation \refeq{eq:RelationDDGPDH}. The results for $\nu =1$, $\xi \ge 0$ and $t=0$ yields in the DGLAP region:
\small
\begin{eqnarray}
  \label{eq:HDGLAP}
   H^u_{x \ge \xi}(x,\xi,0) & = &  \frac{48}{5}\left\{\frac{3 \left(-2 (x-1)^4 \left(2 x^2-5 \xi ^2+3\right) \log (1-x)\right)}{20 \left(\xi ^2-1\right)^3}\right. \nonumber \\
  & &  \frac{3 \left(+4 \xi  \left(15 x^2 (x+3)+(19 x+29) \xi ^4+5 (x (x (x+11)+21)+3) \xi ^2\right) \tanh ^{-1}\left(\frac{(x-1) \xi }{x-\xi ^2}\right)\right)}{20 \left(\xi ^2-1\right)^3} \nonumber \\
  & & +\frac{3 \left(x^3 (x (2 (x-4)x+15)-30)-15 (2 x (x+5)+5) \xi ^4\right) \log \left(x^2-\xi ^2\right)}{20 \left(\xi ^2-1\right)^3} \nonumber \\
  & & +\frac{3\left(-5 x (x (x (x+2)+36)+18) \xi ^2-15 \xi ^6\right) \log \left(x^2-\xi ^2\right)}{20 \left(\xi ^2-1\right)^3} \nonumber \\
 & & +\frac{3 \left(2 (x-1) \left((23 x+58) \xi ^4+(x (x (x+67)+112)+6) \xi ^2+x (x ((5-2 x) x+15)+3)\right)\right)}{20 \left(\xi ^2-1\right)^3}\nonumber \\
 & & +\frac{3\left(\left(15 (2 x(x+5)+5) \xi ^4+10 x (3 x (x+5)+11) \xi ^2\right) \log \left(1-\xi ^2\right)\right)}{20 \left(\xi ^2-1\right)^3} \nonumber \\
 & &+\left.\frac{3\left(2 x (5 x (x+2)-6) +15 \xi ^6-5 \xi ^2+3\right) \log \left(1-\xi ^2\right)}{20 \left(\xi ^2-1\right)^3}\right\},
\end{eqnarray}
\normalsize
and:
\small
\begin{eqnarray}
  \label{eq:HERBL}
  H^u_{|x| \le \xi}(x,\xi,0) & = &  \frac{48}{5}\left\{\frac{ 6\xi (x-1)^4 \left(-\left(2 x^2-5 \xi ^2+3\right)\right) \log (1-x)}{40 \xi  \left(\xi ^2-1\right)^3}\right. \nonumber \\
 & & +\frac{6\xi \left(-4 \xi  \left(15 x^2 (x+3)+(19 x+29) \xi ^4+5 (x (x (x+11)+21)+3) \xi ^2\right) \log (2 \xi )\right)}{40 \xi  \left(\xi ^2-1\right)^3} \nonumber \\
 & & +\frac{6\xi (\xi +1)^3 \left((38 x+13) \xi ^2+6 x (5 x+6) \xi +2 x (5 x (x+2)-6)+15 \xi ^3-9 \xi +3\right) \log (\xi +1)}{40 \xi  \left(\xi ^2-1\right)^3} \nonumber \\
 & & +\frac{6\xi(x-\xi )^3 \left((7 x-58) \xi ^2+6 (x-4) x \xi +x (2 (x-4) x+15)+15 \xi ^3+75 \xi -30\right) \log (\xi -x)}{40 \xi \left(\xi ^2-1\right)^3} \nonumber \\
 & & +\frac{3(\xi -1) (x+\xi ) \left(4 x^4 \xi-2 x^3 \xi  (\xi +7)+x^2 (\xi  ((119-25 \xi ) \xi -5)+15)\right)}{40 \xi  \left(\xi ^2-1\right)^3} \nonumber \\
 & & +\left.\frac{3(\xi -1) (x+\xi)\left(x \xi  (\xi  (\xi  (71 \xi +5)+219)+9)+2 \xi  (\xi  (2 \xi  (34 \xi +5)+9)+3)\right)}{40 \xi  \left(\xi ^2-1\right)^3}\right\},
\end{eqnarray}
\normalsize
in the ERBL one.
\begin{figure}[t]
  \centering
  \includegraphics[width=0.5\textwidth]{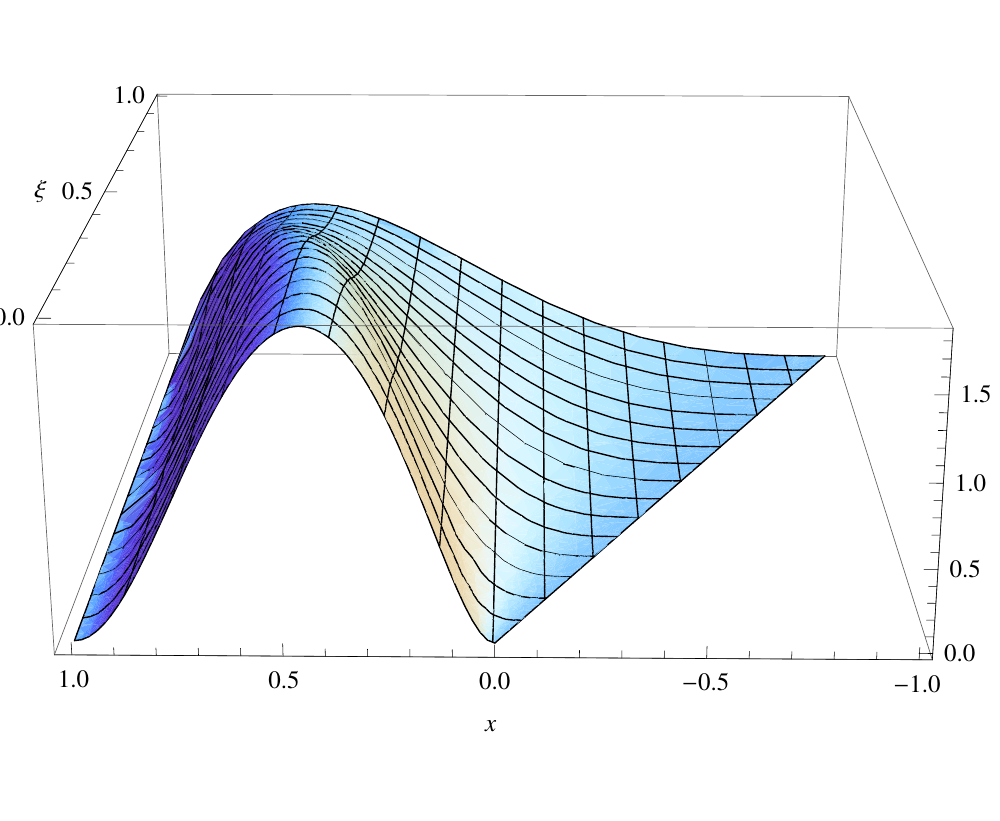}
  \caption{The pion GPD $H^q$ computed for $\nu =1$  at vanishing $t$.}
  \label{fig:GPDnu1}
\end{figure}
These results are illustrated on figure \ref{fig:GPDnu1}. Equations \refeq{eq:HDGLAP} and \refeq{eq:HERBL} deserve several comments. First, several properties listed in section \ref{sec:HadronStructure} are fulfilled as a direct consequence of the DD properties. It is the case of the support property, the parity in $\xi$, and the polynomiality property, despite the presence of logarithms in the contributions of the DGLAP and ERBL region to final results. The forward limit can be computed, leading to the following formula for the PDF computed in the triangle diagram approach:
\begin{equation}
  \label{eq:PDFTriangle}
  q_\pi^{\textrm{Tr}}(x) = \frac{72}{25} \Big( ( 30 - 15 x + 8 x^2 - 2 x^3 )  x^3 \log x + ( 3 + 2 x^2 ) ( 1 - x )^4 \log( 1 - x ) + ( 3 + 15 x + 5 x^2 - 2 x^3 ) x ( 1 - x ) \Big).
\end{equation}
The result on the pion PDF is analysed in section \ref{sec:BeyondImpulse}.

Still one property is not fulfilled here: the positivity. Indeed, as $q_\pi^{\textrm{Tr}}(x) \rightarrow 0$ when $x \rightarrow 0$, one should have:
\begin{equation}
  \label{eq:PositivityTrouble}
  |H(x,\xi ,t)|  \le  \sqrt{H\left(\frac{x-\xi}{1-\xi},0,0\right)H\left(\frac{x+\xi}{1+\xi} ,0,0\right)} \rightarrow 0 \textrm{~when~} x \rightarrow \xi .
\end{equation}
Thus, $H^q$ should vanish on the line $x = \xi$. From figure \ref{fig:GPDnu1}, this is obviously not the case. The reason why the positivity property is violated in this approach remains unknown, but one cannot exclude that it might be a facet of the algebraic model used for the computations and that, when using the full solution of the DSEs, the problem might not exist any more.


\section{Chiral properties and soft pion theorem}
\label{sec:SPT}

As stated above, the pion GPD model building cannot be constrained by experimental results for an off-forward kinematics. However, the verification of the soft pion theorem is a challenging test that a proper model must pass beyond the forward kinematics, at $\xi=1$ and $t=0$. We stress now the minimal requirements needed for a DSE-BSE inspired model to succeed in fulfilling the theorem. 

\subsection{Consequences of the Axial-Vector Ward Takahashi Identity}

Both $\Gamma_\mu^5(k,P)$ and $\G_5(k,P)$ can be written in terms of the pion \bs amplitude $\G^j_\pi(k,P)$ \cite{Maris:1997hd}:
\begin{eqnarray}
  \label{eq:ParameterisedAxialVectorVertex}
  \Gamma_\mu^{5j}(k,P) & = & \frac{\tau^j}{2}\g_5 \left[\g_\mu F_{AV}(k,P) + \g \cdot k k_\mu G_{AV}(k,P) -\sigma_{\mu\nu}k^{\nu}H_{AV}(k,P) \right] \nonumber \\
  & & + \tilde{\G}_{\mu}^{5j}(k,P) + \frac{f_\pi P_\mu}{P^2+m^2_\pi}\G_\pi^j(k,P),
\end{eqnarray}
and
\begin{eqnarray}
  \label{eq:ParameterisedAxialVertex}
  i\Gamma^{5j}(k,P) & = & \frac{\tau^j}{2}\g_5 \left[iE_{A}(k,P) + \g \cdot P F_{A}(k,P) + \g \cdot k k \cdot P G_{A}(k,P) \right. \nonumber \\
    & & \left. +\sigma_{\mu\nu}k^{\nu}P^{\mu}H_{A}(k,P) \right]  + \frac{\rho_\pi}{P^2+m^2_\pi}\G_\pi^j(k,P),
\end{eqnarray}
with $\rho_\pi$ being a constant and $j$ the isospin index.  $E,F,G$ and $H$ are scalar functions with no singularities at $P^2=-m_\pi^2$.  $\tilde{\G}_{\mu}^{5j}$ is also a non-singular contribution to the axial-vector vertex at the pion mass. Therefore, working in the chiral limit, one can relate the axial-vector and axial vertices to the \bs amplitude through:
\begin{eqnarray}
  \label{eq:BSAAV}
  \lim_{P\rightarrow 0} P^\mu \Gamma_\mu^{5j}(k,P) & = & f_\pi \G_\pi^j(k,0), \\
  \label{eq:BSAA}
  \lim_{P^2\rightarrow 0} iP^2 \Gamma_5^{j}(k,P) & = & \left. \rho_\pi \G_\pi^j(k,P)\right|_{P^2 = 0}.
\end{eqnarray}
Injecting equation \refeq{eq:BSAAV} within the AVWTI defined in equation \refeq{eq:AVWTI} one gets:
\begin{equation}
  \label{eq:AVWTIChiralLimit}
  f_\pi \G_\pi^j(k,0) = i\g_5\frac{\tau^j}{2}S^{-1}(k) + iS^{-1}(k)\g_5 \frac{\tau^j}{2},
\end{equation}
in the chiral limit. Using the expansion of $\G_\pi^j(k,0)$ in equation \refeq{eq:PionDiracStructure} and of $S^{-1}(k)$ in equation \refeq{eq:Propagator}, it is possible to simplify equation \refeq{eq:AVWTIChiralLimit} in a Goldberger-Treimann-like relation \cite{Goldberger:1958vp,Maris:1997hd}:
\begin{equation}
  \label{eq:GoldbergerTreiman}
  f_\pi E_\pi(k,0) = B_q(k^2) = M_q(k^2)A_q(k^2).
\end{equation}
Consequently, the internal structure of the pion is directly related to the running dressed quark mass $M_q$.

\subsection{Soft Pion Theorem}

In the chiral limit, the AVWTI plays a key role in recovering the so-called soft pion theorem described in section \ref{sec:HadronStructure}. As it has been highlighted in \refcite{Mezrag:2014jka}, when properly taken into account, the AVWTI allows the computation of the pion GPDs to fulfil the soft pion theorem within the impulse approximation. To show this, one first has to look at the kinematics in which the soft pion theorem applies:
\begin{eqnarray}
  \label{eq:kinematics}
  \xi = 1 & \Rightarrow & \Delta^+ = -2P^+, \nonumber \\
   m_\pi^2 = 0 & \Rightarrow & (P+\Dd)^2 = 0 \Rightarrow \Delta_\perp = 0 \Rightarrow P^- = 0, \nonumber \\
   t = 0 & \Rightarrow & \Delta^- = 0.
\end{eqnarray} 
Therefore, on figure \ref{fig:TriangleDiagrams}, the momentum of the outgoing pion $p_2 = P+\frac{\Delta}{2}$ vanishes, whereas the incoming pion is a lightlike particle as $p_1 = 2P^+$. At vanishing momentum, one can consider the \bs amplitude of the outgoing pion as the limit of the axial-vector vertex as shown in equation \refeq{eq:BSAAV}. On the other hand, the momentum of the incoming pion being such that $p_1^2 = (2P)^2 = \Delta^2$ allows one to use equation \refeq{eq:BSAA} to express the \bs amplitude in terms of the axial vertex. More formally, one can write:
\begin{align}
  \label{eq:TriangleSPT1}
   \mathcal{M}_m&(\xi=1,t=0)  =  \lim_{t\rightarrow 0} \lim_{\xi \rightarrow 1} \lim_{m_\pi^2 \rightarrow 0}\textrm{Tr}_{\textrm{CFD}}\left[ \int \frac{\mathrm{d}^4k}{(2\pi)^4}\frac{(k\cdot n)^m\tau_-}{2 ( \Pn )^{m+1}} \, i \frac{(P+\Dd)^\mu}{f_\pi}\bar{\Gamma}^5_{\mu}\left(\left(k-P\right),P+\Dd \right) \right. \nonumber \\
& \times  S( k - P ) \ i~n \cdot \Gamma(k-P,k+P) ~ S( k+P )\left. \tau_+ i \frac{i(P^2+(\Dd)^2)}{\rho_\pi} \Gamma_5\left(\left(k+P\right),P-\Dd\right) S( k - P ) \right].\nonumber \\
\end{align}
Injecting the AVWTI \refeq{eq:AVWTI} in equation \refeq{eq:TriangleSPT1} and taking the chiral limit leads to:
\begin{equation}
  \label{eq:AVWTISoftPion}
   (P+\Dd)^\mu\bar{\Gamma}^{5}_\mu\left(\left(k-P\right),P+\Dd \right) = i\g_5S^{-1}(k-P) + S^{-1}(k-P)i\g_5,
\end{equation}
for $P+\Dd \rightarrow 0$. Then injecting equation \refeq{eq:AVWTISoftPion} into equation \refeq{eq:TriangleSPT1}, one gets:
\begin{align}
  \label{eq:TriangleSPT2}
   \mathcal{M}_m&(\xi=1,t=0)  =   \lim_{t\rightarrow 0} \lim_{\xi \rightarrow 1}\textrm{Tr}_{\textrm{CFD}}\left[ \int \frac{\mathrm{d}^4k}{(2\pi)^4}\frac{(k\cdot n)^m\tau_-}{2 ( \Pn )^{m+1}} \, \frac{i}{f_\pi}  \right. \nonumber \\
&  \times  \left( i\g_5 \ i~n \cdot \Gamma(k-P,k+P) ~ S( k+P )~\tau_+ i \frac{i(P^2+(\Dd)^2)}{\rho_\pi} \Gamma_5\left(\left(k+P\right),P-\Dd\right) S( k - P ) \right.\nonumber \\
&   \left. \left. ~+ i\g_5 S( k - P ) \ i~n \cdot \Gamma(k-P,k+P) ~ S( k+P )\tau_+ i \frac{i(P^2+(\Dd)^2)}{\rho_\pi} \Gamma_5\left(\left(k+P\right),P-\Dd\right) \right) \right]. \nonumber \\
\end{align}
Consequently, one has to compute two contributions containing only two dressed vertices instead of three. To do so, it is necessary to ``unfold'' these vertices. Both of them fulfil an inhomogeneous \bs equation, and therefore in the RL approximation of equation \refeq{eq:GapKernelRL}, they can be seen as an infinite sum of gluons ladder:
\begin{equation}
  \label{eq:InhomogenousInterpretation}
  \vcenter{\hbox{\includegraphics[width=0.25\textwidth]{InBSA.pdf}}} = \sum_{n=0}^{\infty} \vcenter{\hbox{\includegraphics[width=0.25\textwidth]{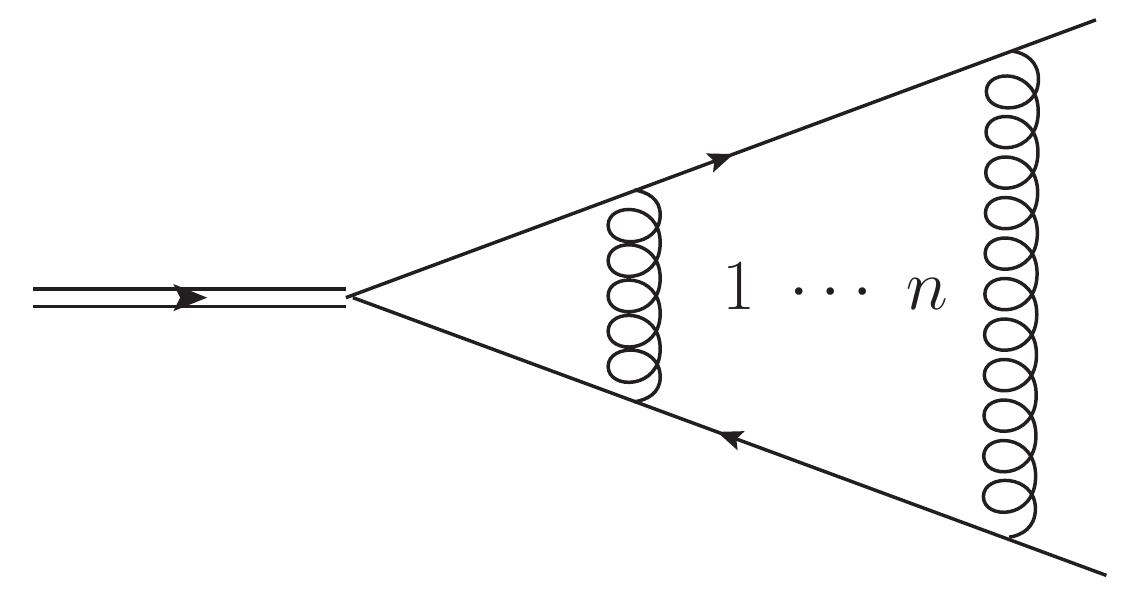}}} ,
\end{equation}
where $n$ stands for the number of considered gluon ladders, and $n=0$ correspond to the ``undressed'' case, \ie when no gluon is exchanged between the two quarks. The Dirac matrices $\g_5$ in equation \refeq{eq:TriangleSPT2} are thus trapped between two sets of gluon ladders. Nevertheless, relabelling the series leads to the following contributions:
\begin{equation}
  \label{eq:Reindexing}
  \sum_{n_2=0}^{\infty}\sum_{n_1=0}^{\infty} \vcenter{\hbox{\includegraphics[width=0.45\textwidth]{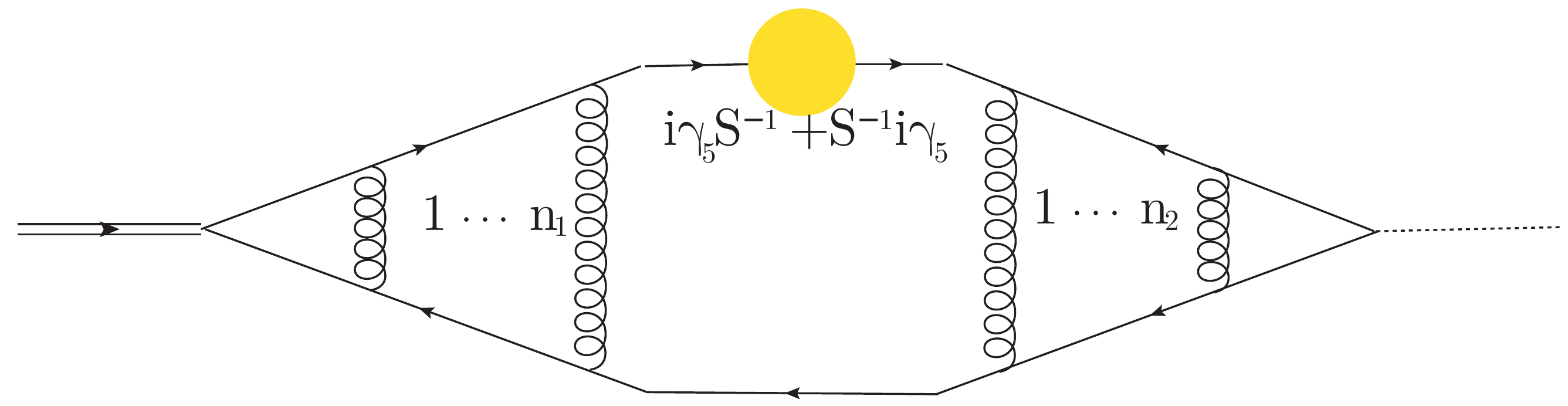}}} \propto ~ \sum_{n=0}^{\infty}\sum_{j=0}^{n}\quad\vcenter{\hbox{\includegraphics[width=0.1\textwidth]{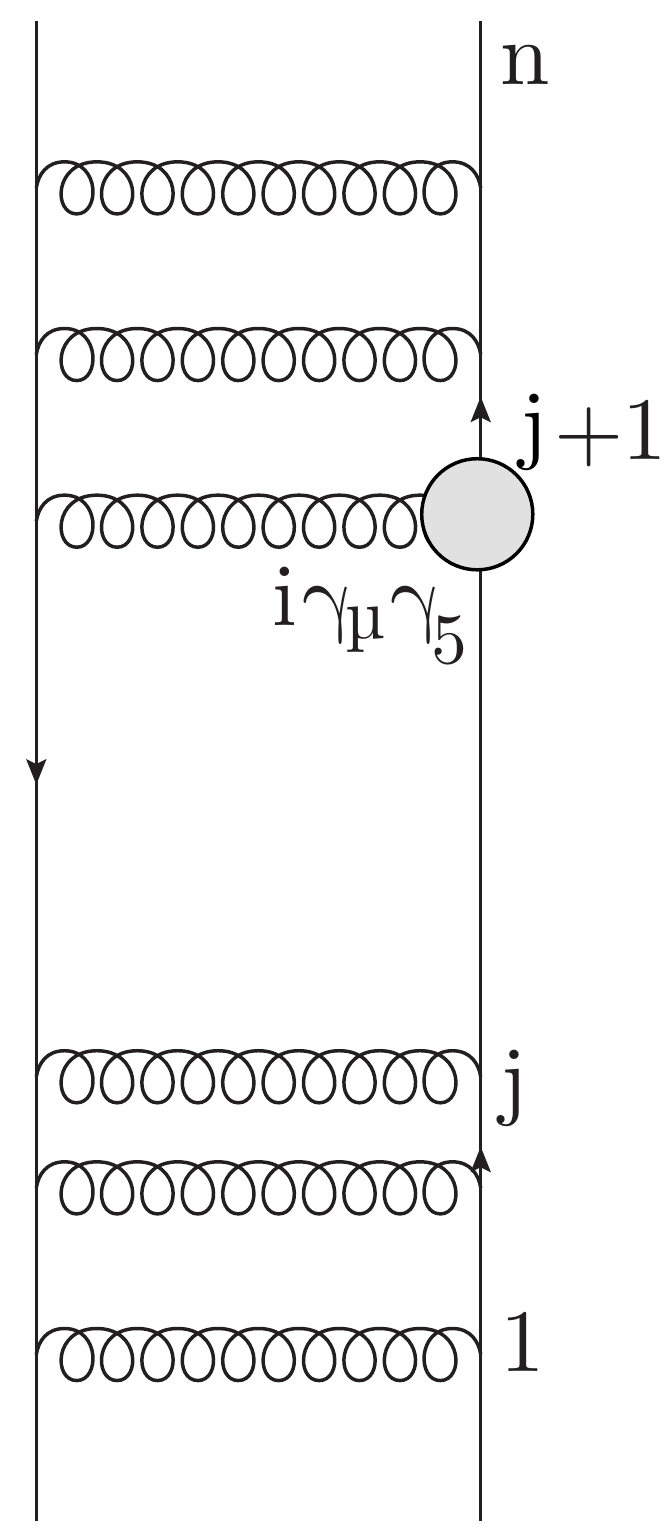}}} + \quad \vcenter{\hbox{\includegraphics[width=0.1\textwidth]{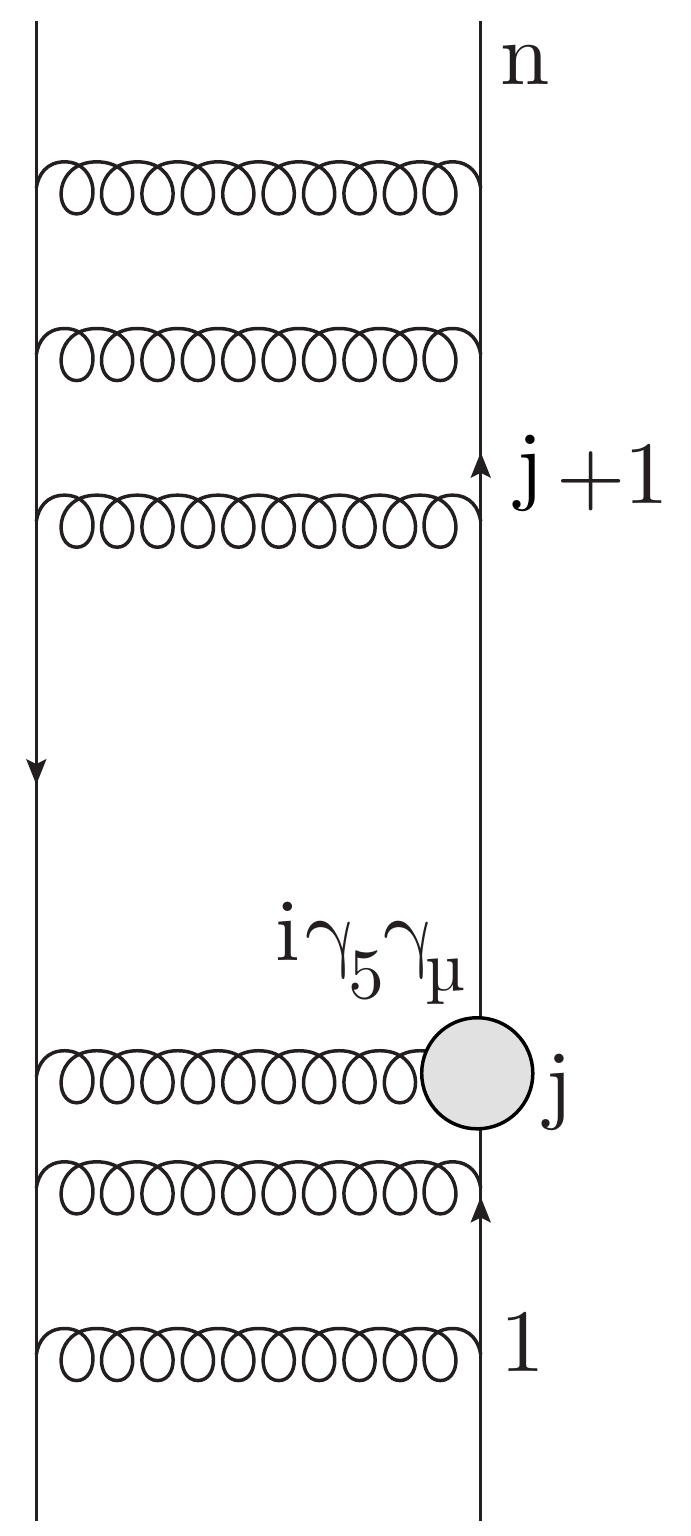}}},
\end{equation}
where $n$ stands now for the total number of gluon ladder in the diagram, and the $j$ index is such that the AVWTI is inserted between the $j^{th}$ and the $(j+1)^{th}$ gluon ladder. As $\g_5$ and $\g^\mu$ anticommute, for a given $n$, only two contributions on the right-hand side of equation \refeq{eq:Reindexing} do not cancel completely each other. In the case of the first ladder in equation \refeq{eq:Reindexing} it is the contribution for $j=n$, whereas in the case of the second ladder, it is the one for $j=0$. These two contributions have a very specific meaning. Indeed, in both case, one of the vertices is considered as ``undressed'', as there is no gluon ladder between the insertion and the vertex itself. On the other hand, this analysis being valid for any $n$, one can resum the contributions for $n$ from $0$ to infinity. This is exactly what is done in equation \refeq{eq:InhomogenousInterpretation}, allowing one to get back the dressed vertex in the RL approximation. As a result, equation \refeq{eq:TriangleSPT2} yields:
\begin{eqnarray}
  \label{eq:SoftPionTwoLastContributions}
   \mathcal{M}_m(\xi=1,t=0)  & = &  \lim_{t\rightarrow 0} \lim_{\xi \rightarrow 1} \textrm{Tr}_{\textrm{CFD}}\left[ \int \frac{\mathrm{d}^4k}{(2\pi)^4}\frac{(k\cdot n)^m\tau_-}{2 ( \Pn )^{m+1}} \, \frac{i}{f_\pi}  \right. \nonumber \\
& &  \times \left( i\g_5 \ i~n \cdot \g ~ S( k+P )~\tau_+ i \frac{i(P^2+(\Dd)^2)}{\rho_\pi} \Gamma_5\left(\left(k+P\right),P-\Dd\right) S( k - P ) \right .\nonumber \\
& &  \left. \left.~+ S( k - P ) \ i~n \cdot \Gamma(k-P,k+P) ~ S( k+P )\tau_+ i \frac{i(P^2+(\Dd)^2)}{\rho_\pi} i(\g_5)^2 \right) \right]. \nonumber \\
\end{eqnarray}
Taking the limit $\xi \rightarrow 1$ and $t \rightarrow 0$ leads to $(P^2+ (\Dd)^2) \rightarrow 0$. Thus in equation \refeq{eq:SoftPionTwoLastContributions}, the first contribution, depending on $\Gamma_5\left(k+P,P-\Dd\right)$ can be written in terms of the pion \bs amplitude through equation \refeq{eq:BSAA}. The second one actually vanishes, due to the lack of poles in $\Gamma(k-P,k+P)$ when  $(P^2+ (\Dd)^2) \rightarrow 0$, \ie there is no massless vector meson. One is therefore left with:
\begin{equation}
  \label{eq:SoftPionTheoremBadVar}
  \mathcal{M}_m(1,0) = \textrm{Tr}_{\textrm{CFD}}\left[\frac{1}{f_\pi} \int \frac{\mathrm{d}^4k}{(2\pi)^4}\frac{(k\cdot n)^m\tau_-}{2 ( \Pn )^{m+1}} \g_5  ~n \cdot \g ~ S( k+P )~\tau_+  \Gamma_\pi\left(\left(k+P\right),P-\Dd\right) S( k - P ) \right] .
\end{equation} 
Then, the following change of variables:
\begin{equation}
  \label{eq:SPTChangeofVariables}
  \left \{
  \begin{array}[h]{ccc}
    p_1 & = & 2P \\
    k' & = & k+P
  \end{array}
  \right. ,
\end{equation}
leads to: 
\begin{eqnarray}
  \label{eq:SoftPionChangeOfVar}
  \mathcal{M}_m(1,0) & = & \textrm{Tr}_{\textrm{CFD}}\left[ \frac{1}{f_\pi}\int \frac{\mathrm{d}^4k'}{(2\pi)^4}\frac{(k'\cdot n-P \cdot n)^m\tau_-}{2 ( \Pn )^{m+1}} \g_5  ~n \cdot \g ~\tau_+ ~ \chi_\pi(k',p_1)  \right] \nonumber \\
   & = & \textrm{Tr}_{\textrm{CFD}}\left[\frac{1}{f_\pi ~p_1 \cdot n} \int \frac{\mathrm{d}^4k'}{(2\pi)^4}\left( 2 \frac{k'\cdot n}{p_1 \cdot n} -1 \right)^m\g_5  ~n \cdot \g ~\tau_+ ~ \chi_\pi(k',p_1)  \right] \nonumber \\
   & = & \textrm{Tr}_{\textrm{CFD}} \left[ \int \textrm{d} u (2u-1)^m  \frac{1}{f_\pi}\int \frac{\mathrm{d}^4k'}{(2\pi)^4} ~\delta(u~p_1 \cdot n - k \cdot n)\g_5  ~n \cdot \g ~\tau_+ ~ \chi_\pi(k',p_1) \right] \nonumber \\
   & = & \int \textrm{d} u (2u-1)^m \varphi_\pi(u),
\end{eqnarray}
where $\varphi_\pi$ is the pion DA defined, as defined in equation \refeq{eq:DefDAProj} in terms of the \bs wave function. As a continuous compactly supported function is uniquely defined by its Mellin moments, one can conclude that:
\begin{equation}
  \label{eq:FinalSoftPion}
  H^q_\pi(x,1,0) = \frac{1}{2} \varphi_\pi \left( \frac{1+x}{2} \right ),
\end{equation}
in agreement\footnote{Depending on the conventions used, an additional factor 1/2 may appear in the literature, like for instance in \refcite{Polyakov:1998ze}. } with the literature \cite{Belitsky:2005qn}. In terms of isovector and isoscalar GPDs, one gets back equation \refeq{eq:SoftPionTheorem}.

Summarising the proof above, one can say that the impulse approximation is compatible with the soft pion theorem in the \emph{chiral limit}, under two main conditions. First, one has to work within a truncation scheme which fulfils the so-called AVWTI. Then, one can proove that the soft pion theorem is fulfilled when using the RL truncation scheme. One can expect that to go beyond the chiral limit, a more refined truncation scheme is needed, providing a way to better describe the Dynamical Chiral Symmetry Breaking than the RL approximation of equation \refeq{eq:GapKernelRL}. Indeed, from the proof presented above, we expect the fulfilment of the soft pion theorem by a DSE-BSE based GPD model, to be deeply related to the way the truncation scheme dynamically breaks chiral symmetry.


\section{Beyond the impulse approximation}
\label{sec:BeyondImpulse}
As seen in the previous sections, the impulse approximation allows the building of simple models of GPDs, which fulfil most of the theoretical constraints discussed in section 2. It is also consistent, in the relevant 
off-forward kinematic limit, with the soft pion theorem, within the RL truncation scheme and provided that the AVWTI is also verified. It remains nevertheless an approximation whose limitations can already be seen in the forward limit.

\subsection{Forward kinematics implications}

In the following, we will show how, according to \refcite{Chang:2014lva}, the valence-quark PDF appears distorted by the impulse approximation. Introducting the appropriate amendments, it is possible to go beyond this approximation highlight the basic features of the PDF.

\subsubsection{Asymmetry of the PDF}

Indeed, in the approach exposed in the previous section, one considers that the pion is composed of two effective dressed quarks. From discrete and isospin symmetries detailed in equations \refeq{eq:RelationIsoscalarChargedPions}-\refeq{eq:H_isoscalar_even}, it is possible to conclude that the two quark pion GPDs are such that: 
\begin{equation}
  \label{eq:QuarkSymmetry}
  H^u_{\pi^+}(x,\xi,t) = H_{\pi^+}^{\bar{d}}(x,\xi,t).
\end{equation}
In the forward case, for the valence-quark PDF where neither sea-quark nor gluon contributions do not play any role, momentum conservation adds a strong constraint. Valence distributions are normalised to $1$ and number densities become probability densities. Therefore, if one has a probability density to find a $u$ quark carrying a momentum fraction $x$, then, a $\bar{d}$ can be found with the same probability density but carrying a $1-x$ momentum fraction. Together with equation \refeq{eq:QuarkSymmetry}, momentum conservation imposes the pion two-body PDF to be symmetric under the transformation $x \leftrightarrow (1-x)$. On top of this, it has been also proved that, in a two-body framework, the forward limit for the GPD in the so-called overlap representation results in a symmetric PDF~\cite{Mezrag:2014jka}. However, the results obtained in equation \refeq{eq:PDFTriangle} slightly violate the $x \leftrightarrow (1-x)$ symmetry, as it will also be shown on figure \ref{fig:PDFTotal}. Then, when computing the pion GPD within the covariant approach and impulse approximation, one necessarily misses an important piece of information.  

\subsubsection{Additional symmetrising Contributions}

According to \refcite{Chang:2014lva}, the flaw in the valence-quark PDF comes from contributions related to gluons binding the quarks into the pion that, within the impulse approximation, are explicitly omitted. Therein, the drawback is  discussed and the appropriate correction to the usual PDF definition within the impulse approximation is described. The discussion is made there in terms of the virtual Compton scattering amplitude in the RL truncation scheme. \refcite{Chang:2014lva} presents a diagram analysis, which shows how the textbook {\it handbag} contribution can be approximated to the triangle diagram, typically the only retained. But it also shows how additional terms emerge and the role played by the latter in order to recover the basic features for the PDF.  

\begin{figure}[t]
  \centering
  \begin{equation*}
     \vcenter{\hbox{\includegraphics[width=0.2\textwidth]{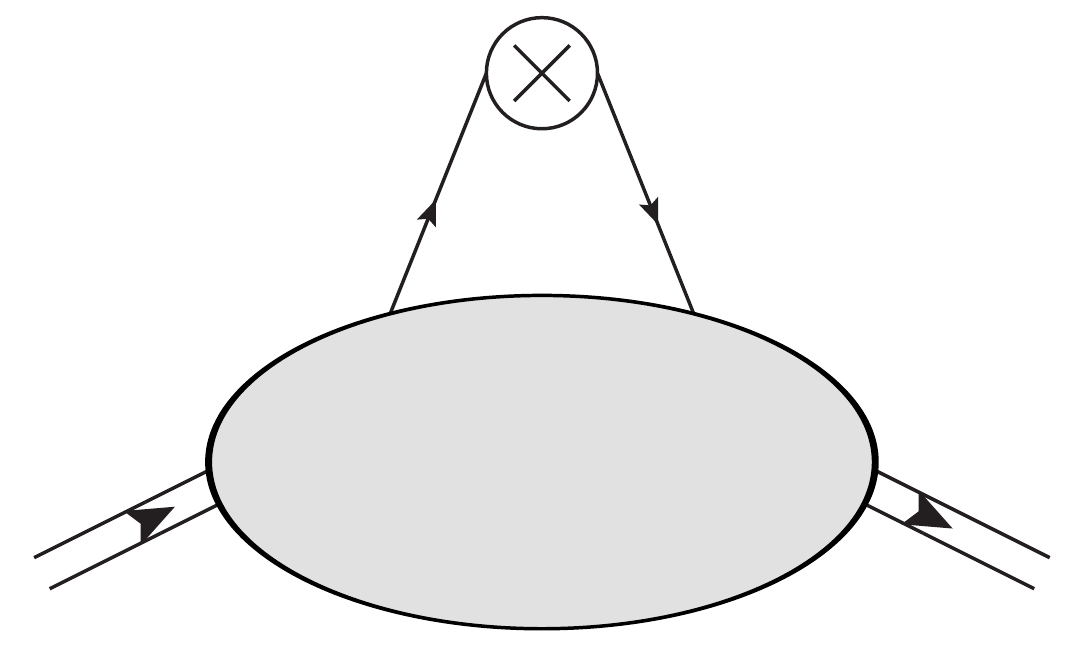}}} \approx  \vcenter{\hbox{\includegraphics[width=0.2\textwidth]{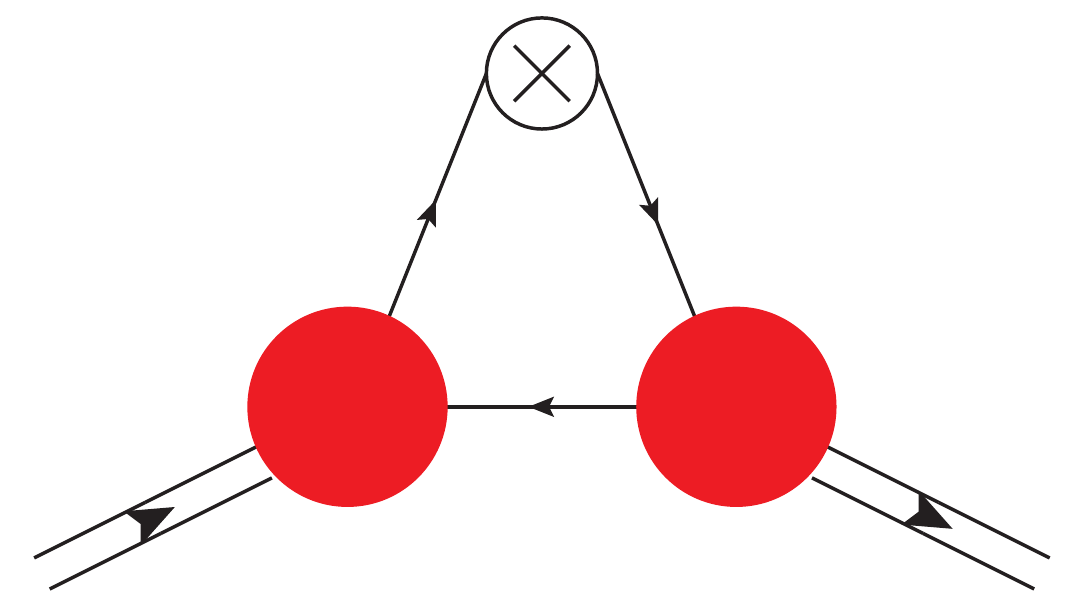}}} +  \vcenter{\hbox{\includegraphics[width=0.2\textwidth]{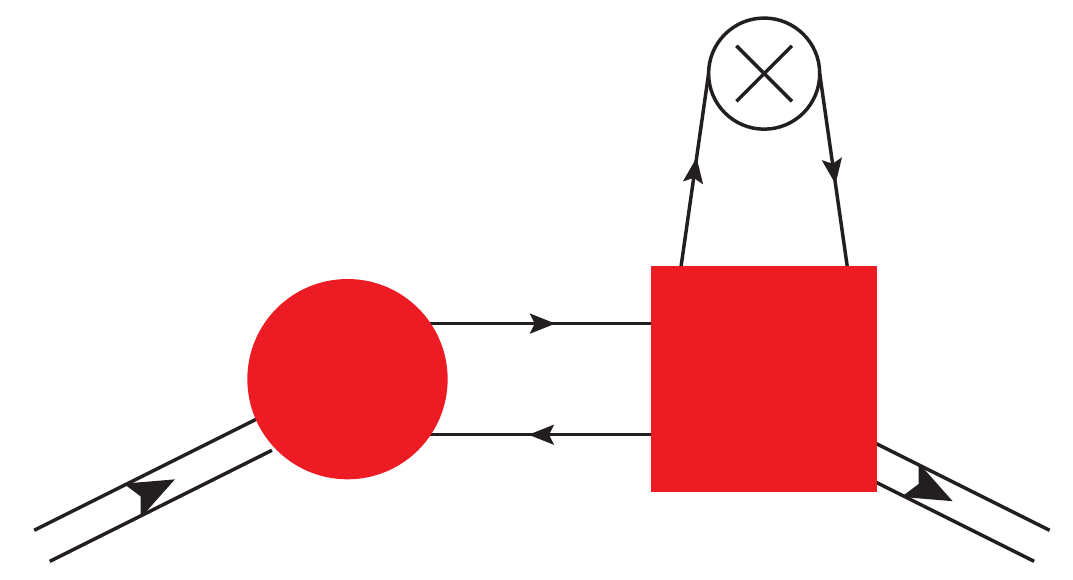}}} + \vcenter{\hbox{\includegraphics[width=0.2\textwidth]{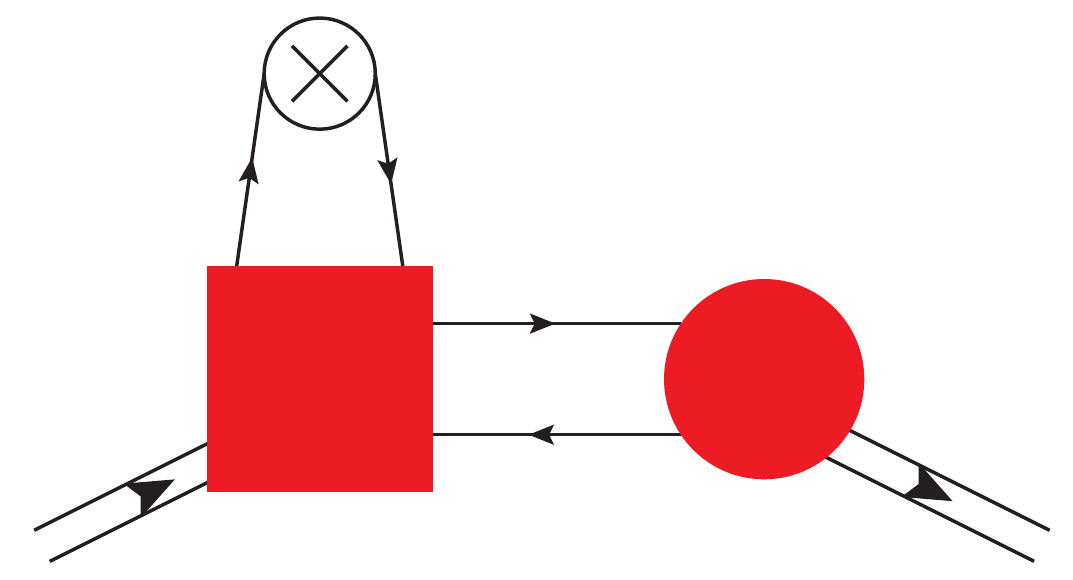}}}
  \end{equation*}
  \caption{Additional contributions to the triangle diagrams. The circle vertices correspond to the \bs amplitudes, whereas the squared vertices denote a new non-perturbative object related to the \bs amplitude but on which the twist-two operators also act.}
  \label{fig:AdditionalContributions}
\end{figure}

Here, we will follow the alternative approach shown in \refcite{Mezrag:2015mka} to both exhibit the failure of the impulse approximation and pinpoint the correct amendment in the forward limit, relying on the proper definition of the twist-two local operators. In the impulse approximation, the insertion of the twist-two local operators is solely done ``{\it between}'' two \bs amplitudes, leading to the triangle diagram. Nonetheless, as suggested on figure \ref{fig:AdditionalContributions}, one can also imagine that the operators can be directly inserted ``{\it inside}'' the \bs vertices, for either the incoming or outgoing hadrons. Precisely, such a direct insertion is needed to reproduce the omitted diagrams in the usual PDF computations within the impulse approximation~\cite{Mezrag:2014jka,Mezrag:2015mka}. 

In order to figure out how these new contributions are related to the \bs amplitude, one has to ``unfold'' the latter. Indeed, the RL kernel in equation \refeq{eq:GapKernelRL} suggests that the \bs amplitude is composed of a infinite number of gluon exchanges between the two quarks. In this view, it is then possible to insert the twist-two operator between any of these gluon exchanges. Therefore, the ``squared'' vertex can be considered as a sum over all the possible ways to include the local twist-two operators between two gluon ladders, \ie:
\begin{equation}
  \label{eq:SquaredVertex}
   \vcenter{\hbox{\includegraphics[width=0.20\textwidth]{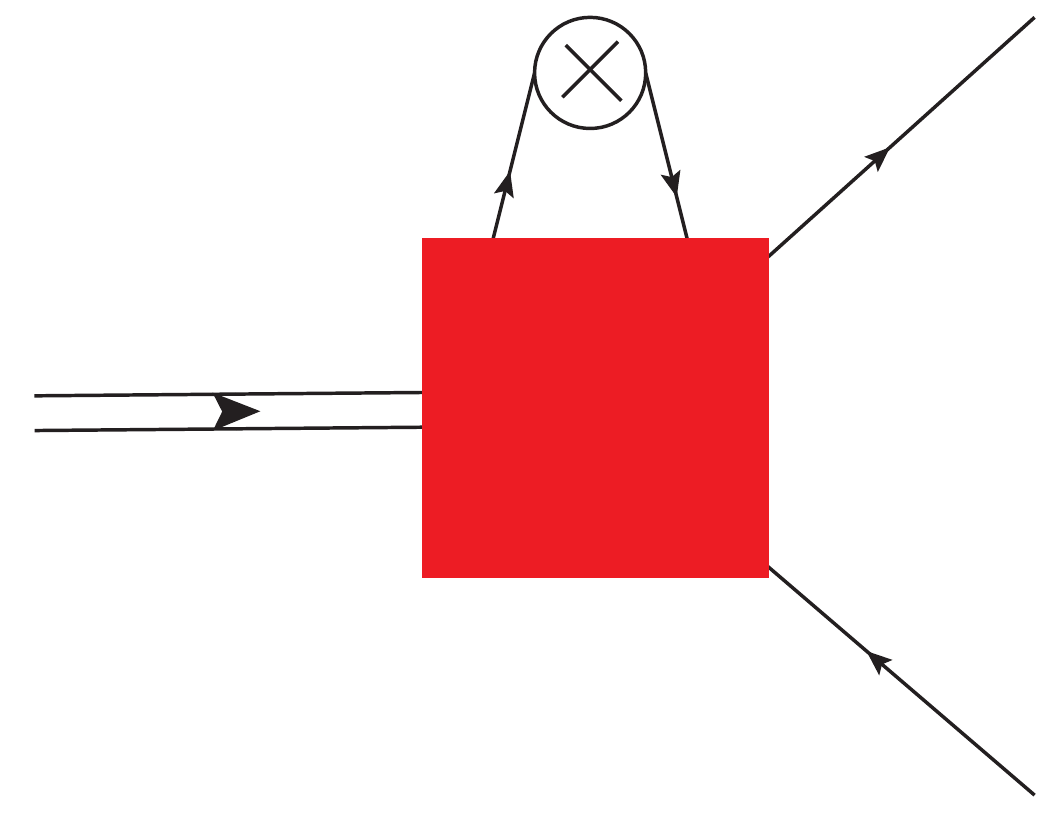}}} = \sum_{j=0}^{\infty} \vcenter{\hbox{\includegraphics[width=0.30\textwidth]{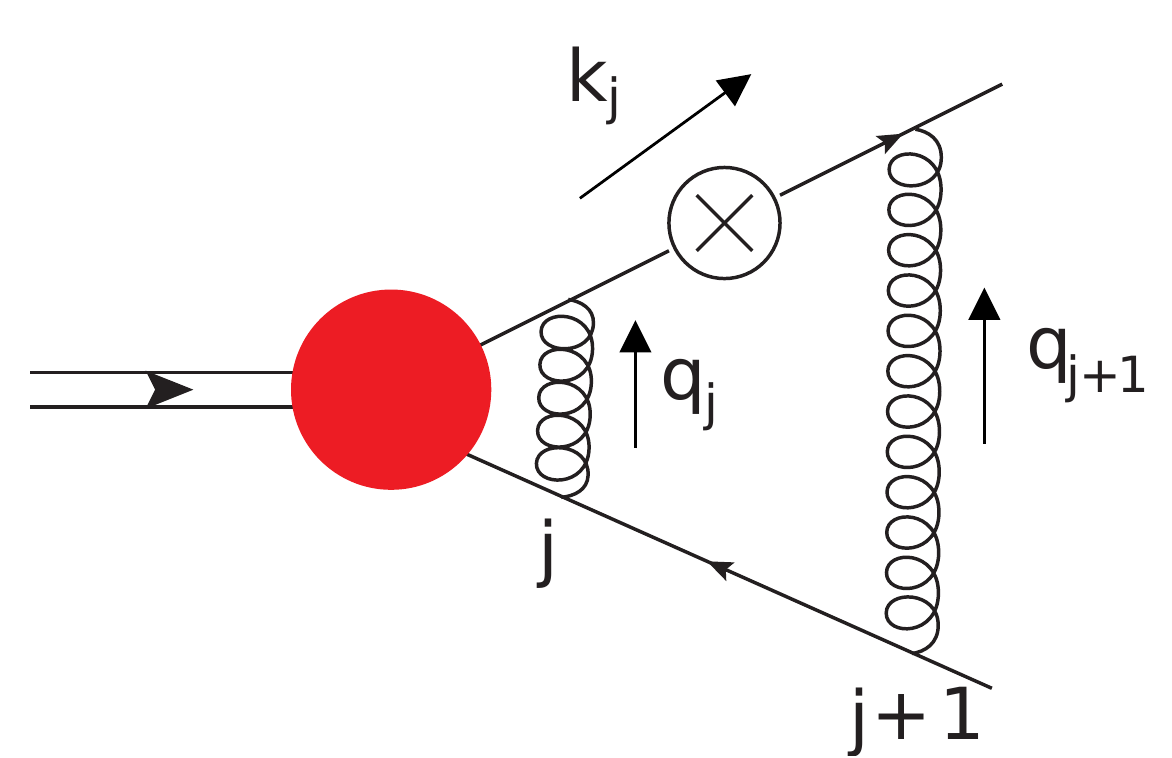}}} ,
\end{equation} 
where as in figure \ref{fig:TriangleDiagrams}, the cross indicate the insertion of the twist two operators between the $j^{th}$ and the $(j+1)^{th}$ gluons. In momentum space, this contribution can be written as:
\begin{equation}
  \label{eq:OperatorSquareV}
  n_\mu n_{\mu_1} \cdots n_{\mu_m} O^{\{\mu,\mu_1 \cdots \mu_m \}} \rightarrow (k \cdot n)^m \ i\ \G (k-\Dd, k+\Dd) \cdot n ,
\end{equation}
for non-vanishing values of $\Delta$. As before, $\G (k-\Dd, k+\Dd) \cdot n $ denotes the projection of the vector current on the lightcone. As previously, this vertex has to fulfil the WTI \refeq{eq:EMWTI}. In the forward case, this leads to:
\begin{equation}
  \label{eq:EMWTIForwardLimit}
  i \G^\mu(k_j,k_j) = \frac{\partial S^{-1}}{\partial k_j^\mu}(k_j),
\end{equation}
$k_j$ being the quark momentum as defined in equation \refeq{eq:SquaredVertex}. Denoting $q_j$ the momentum of the $j^{th}$ gluon exchanged, momentum conservation imposes $k_{j+1} = k_j+q_{j+1}$. Therefore, being given that $P$, the pion total momentum and $k$, the relative momentum between the quarks are given, the relevant ``internal variables'' of the \bs amplitude are the $\{q_j\}$s.
Therefore, one can formally rewrite the momenta $k_j$ as:
\begin{equation}
  \label{eq:Rewritekj}
  k_j = k - \sum_{\ell=j+1}^{\infty}q_\ell . 
\end{equation}
The derivative of the inverse propagator can be now related to the propagator itself through:
\begin{equation}
  \label{eq:NewTermRL}
  \frac{\partial S}{\partial k_j^\mu}(k_j) = - S(k_j) \frac{\partial S^{-1}}{\partial k_j^\mu}(k_j) S(k_j),
\end{equation}
we can next take advantage of the change of variable in equation \refeq{eq:Rewritekj} to, as follows,
\begin{equation}
  \label{eq:kjChangeOfVariable}
  \frac{\partial S}{\partial k_j^\mu}\left(k_j\right) = \frac{\partial S}{\partial k^\mu}\left(k_j \right),
\end{equation}
let the derivative act on the outgoing momentum $k$ instead of $k_j$. 
Then, as the derivative in equation \refeq{eq:kjChangeOfVariable} does not depend on $j$ anymore, the infinite sum over all the possible insertions can be reduced to the derivative of the \bs amplitude itself:
\begin{equation}
  \label{eq:FinalSquareVertex}
   \vcenter{\hbox{\includegraphics[width=0.20\textwidth]{SquareVertex.pdf}}} = -\frac{1}{2}(k \cdot n)^m n^\mu\frac{\partial \G_\pi^q}{\partial k^\mu}\left(k, P \right).
\end{equation}
The factor $1/2$ stands here to avoid double counting: isospin and crossing symmetries authorize the definition of the GPD by an insertion only sticking either on the quark (as here) or on the anti-quark line; however, the derivative of the 
\bs amplitude in \eqref{eq:FinalSquareVertex} acts on all the propagators of both the quark and the antiquark lines in \eqref{eq:SquaredVertex}, both yielding the same result; then, one needs to correct for an spurious factor $2$ in order to defining the additional proper contribution to the pion GPD. 

\subsubsection{Computations and results}

\begin{figure}[tb]
  \centering
  \includegraphics[width=0.60\textwidth]{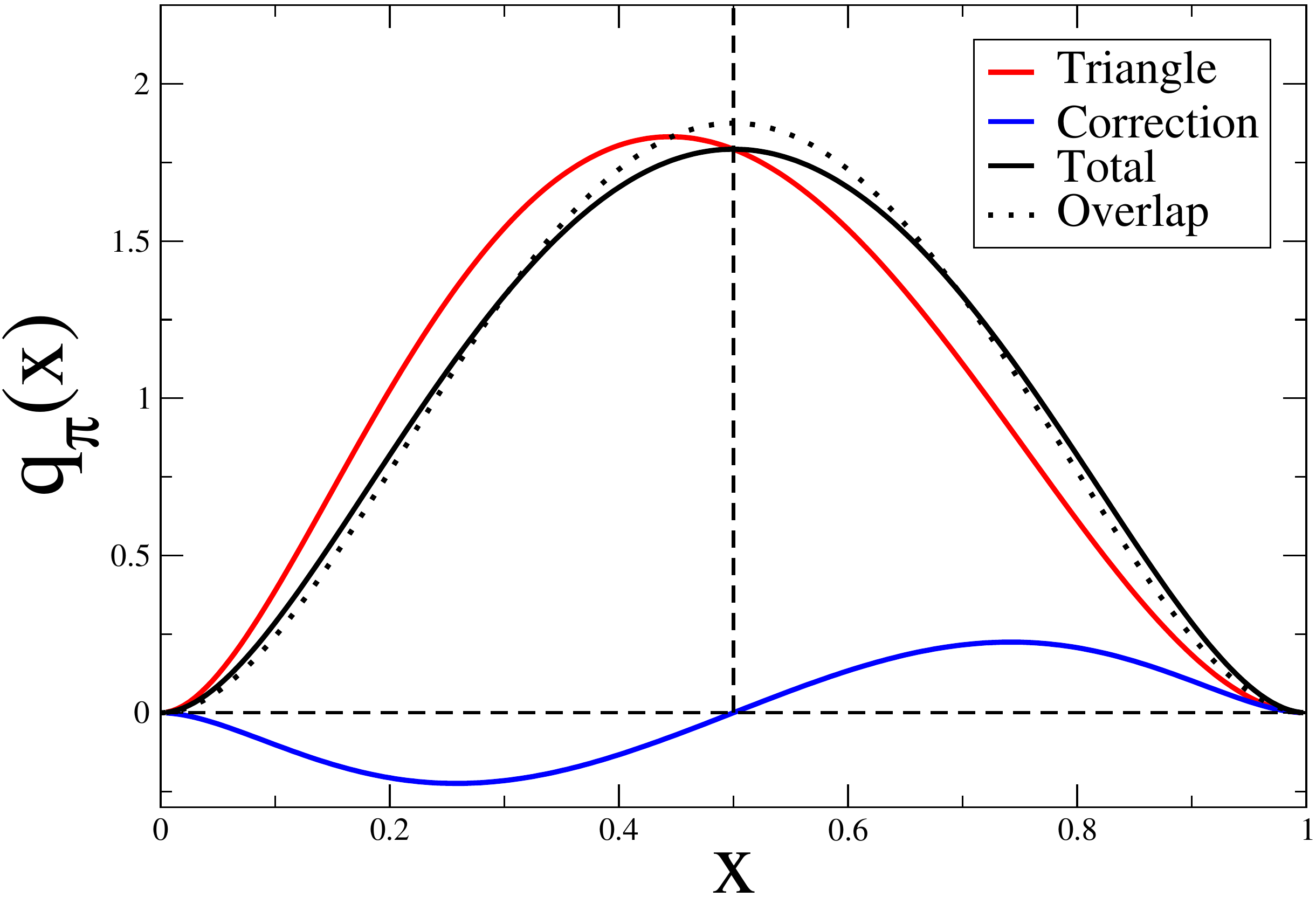}
  \caption{PDF contributions. In red the $q_\pi^{\textrm{Tr}}(x)$ coming from the triangle diagram, in blue the additional contribution taking into account the derivative of the \bs amplitude $q_\pi^{\textrm{ad}}(x)$, and in black solid line the total PDF $q_\pi^{\textrm{tot}}(x) $. The black dotted line corresponds to PDF obtained by applying the overlap representation that will be discussed in the next section.}
  \label{fig:PDFTotal}
\end{figure}

We can now plug equation \refeq{eq:FinalSquareVertex} for the {\it new} vertices in the additional contributions for pion PDF displayed in figure \ref{fig:AdditionalContributions} and apply next the computation techniques developed in section \ref{sec:CovariantComputations}. Owing to equation \refeq{eq:TMBSA}, the derivative of the \bs amplitude yields:
\begin{equation}
  \label{eq:DeriveBSA}
  \frac{\partial \G_\pi^q}{\partial k^\mu}\left(k, P \right) =  -2\nu \int \textrm{d}z \frac{M^{2\nu}\rho(z)\left(k^\mu - (\frac{1-z}{2}-\eta)P^\mu)\right)}{\left[\left(k-(\frac{1-z}{2}-\eta) P\right)^2 +M^2 \right]^{\nu+1}} .
\end{equation}
In the forward limit, both the two additional diagrams shown on figure \ref{fig:AdditionalContributions} give the same results and one can thus define the additional contribution to the PDF Mellin moments as:
\begin{equation}
  \label{eq:AdditionalTermMM}
  \mathcal{M}_m^{\textrm{ad}}(0,0) = \textrm{Tr}_{\textrm{CFD}}\left[ -\int \frac{\mathrm{d}^4k}{(2\pi)^4}\frac{(k\cdot n)^m}{2 (P\cdot n)^{m+1}} \, \tau_- i\bar{\Gamma}_\pi\left(k,P\right)  S( k ) \tau_+ i n^\mu\frac{\partial\Gamma_\pi}{\partial k^\mu}\left(k,  P \right) S( k - P ) \right].
\end{equation}
Then, following the procedure described in section \ref{sec:CovariantComputations}, one can obtain from this additional contribution a supplement to the DDs that had been previously computed from \eqref{eq:MellinMomentDirectDiagram}, denoted now $F^{\textrm{ad}}(\beta,\alpha,0)$ and $G^{\textrm{ad}}(\beta,\alpha,0)$. Using equations \eqref{eq:RelationDDGPDH} and \eqref{eq:ForwardLimit}, these new terms result in the additional contribution to the pion PDF:
\begin{eqnarray}
  \label{eq:AdditionalTermPDF}
  q_\pi^{\textrm{ad}}(x) & = & \int_{-1+x}^{1-x} \textrm{d}\alpha F^{\textrm{ad}}(x,\alpha,0) =  \frac{72}{25} \Bigg(-\left(2 x^3+4 x+9\right) (x-1)^3 \log (1-x) \nonumber \\
  & & + x^3 (2 x ((x-3) x+5)-15) \log (x) -x (x-1)(2x-1) ((x-1) x-9) \Bigg) .
\end{eqnarray}
Adding this component to the one coming from the impulse approximation \refeq{eq:PDFTriangle}, one gets the total PDF:
\begin{align}
  \label{eq:PDFTotal}
  q_\pi^{\textrm{tot}}(x) = &q_\pi^{\textrm{Tr}}(x) + q_\pi^{\textrm{ad}}(x) =  \frac{72}{25} \Bigg((-2x(1-x)-3x+15) x^3 \log (x)\nonumber \\
 &-\left(-2x(1-x)-3(1-x)+15\right) (x-1)^3 \log (1-x) -2 x  (x-1)((x-1) x+6)\Bigg) ,
\end{align}
which, as a remarkable feature, is symmetric under the $x\leftrightarrow 1-x$ exchange. Then, the contribution from the two additional diagrams, including the ``squared vertex'' defined by \eqref{eq:FinalSquareVertex}, on figure \ref{fig:AdditionalContributions}, exactly cancels the asymmetry noticed before. This is illustrated by the black solid curve displayed in figure \ref{fig:PDFTotal}.

A few further remarks are appropriate here. The contribution resulting from the derivative of the \bs amplitude does leave the pion form factor unaltered. This is plain on figure \ref{fig:PDFTotal}, as it is antisymmetric within the $x\leftrightarrow 1-x$ exchange. Therefore, the normalisation of the PDF can be determined only through the triangle diagram contribution, which is fully consistent with Mandelstam normalisation procedure, as explained in section \ref{sec:DSEs}. One can also check the momentum sum rule which here yields:
\begin{equation}
  \label{eq:MomentumSumRules}
  \int_0^1 \textrm{d}x~\left[ q_\pi^{\textrm{Tr}}(x) + q_\pi^{\textrm{ad}}(x)\right] = \frac{117}{250} + \frac{8}{250} = \frac{1}{2},
\end{equation}
fully in agreement with a valence-quark two-body picture, as both quark and antiquark carry on average $1/2$ of the total momentum. It is worthwhile to emphasise that the triangle diagram, by its own, is not able to fulfil this sum rule. Furthermore, in ref.~\cite{Chang:2015ela}, some corrections to the RL truncation originally used to define the PDF are included to make possible for some momentum its shifting from dressed quarks to sea-quarks and gluons and, thus, obtain a realistic PDF beyond the two-body picture which compares well with available experiment. A last comment on the large-$x$ behaviour for the present PDF model is also in order. Indeed, one is left with:
\begin{equation}
  \label{eq:LargeX}
  q_\pi^{\textrm{tot}}(x) = \frac{216}{5} (1-x)^2 + O\left((1-x)^3\right),
\end{equation}
when $x \rightarrow 1$. This is in agreement with the power law derived through the parton model \cite{Ezawa:1974wm,Farrar:1975yb}. Either in perturbative QCD \cite{Brodsky:1994kg,Ji:2004hz}, or in \dses computations \cite{Maris:2003vk,Bloch:1999rm,Hecht:2000xa}, a large-$x$ behaviour such that $q_\pi(x)\sim (1-x)^{2+\gamma}$ is predicted, with $\gamma >0$.

\subsection{Three-dimensional sketch of the pion}

The additional diagrams including the vertex in equation \refeq{eq:FinalSquareVertex} yield a new contribution to the pion GPD in the forward case which allows for the restoration of the $x \leftrightarrow 1-x$ symmetry. Its generalisation to the off-forward kinematics regime is a hard task and remains an open question. The relation of these additional terms with the positivity property, violated in our computations done within the impulse approximation, is also unclear.  However, at vanishing $\xi$, as described in \refcite{Mezrag:2014jka}, one can obtain an approximated result for the GPD and its associated impact-parameter dependent distribution, see equation~\eqref{eq:3DDensity}, which provides a qualitatively sound picture of the pion’s dressed-quark structure at an hadronic scale. Indeed, 
one can write the non-skewed pion GPD as:
\begin{equation}
  \label{eq:GPDCorrelationStructure}
  H^q_\pi(x,0,t) = H_\pi^q(x,0,0)\mathcal{N}(t)C^q_\pi(x,t)F^q_\pi(t),
\end{equation} 
where $F^q_\pi(t)$ is the quark contribution to the pion form factor, $C^q(x,t)$ encodes the correlations between $x$ and $t$, and $\mathcal{N}^q(t)$ is a normalisation factor such that:
\begin{equation}
  \label{eq:DefN}
  1 = \mathcal{N}^q(t) \int \textrm{d}x~H_\pi^q(x,0,0)C^q_\pi(x,t).
\end{equation}
It is then possible to build an insightful pion GPD Ansatz based on the previous calculations. As already shown on figure \ref{fig:FormFactor}, the impulse approximation provides a form factor in good agreement with experimental data. Yet, the vector current vertex can be appropriately dressed to account for its extended nature \cite{Roberts:2011wy}. The latter is proved not to generate new $(x,t)$ correlations and only affect slightly the $t$-behaviour of the form factor, such that the mass-scale $M$ for the algebraic model, see equations (\ref{eq:TMQuarkPropagator}-\ref{eq:TMkDef}), appears just shifted from 0.35 to 0.40 GeV in order to reproduce experimental data~\cite{Mezrag:2014jka}. The PDF $q_\pi^{\textrm{tot}}(x) = H_\pi^q(x,0,0)$ has been also computed in the previous section and the result is given in equation \refeq{eq:PDFTotal}. Within this framework, the only missing piece is $C^q_\pi(x,t)$, which should not be identically one in realistic approaches \cite{Burkardt:2002hr}.

In order to get sensible insights of the $(x,t)$ correlations, computations of an additional $F$-type DD have been performed using the following approximation for the Mellin moments~\cite{Mezrag:2014jka}, which is a generalisation of equation \refeq{eq:AdditionalTermMM}:
\begin{eqnarray}
  \label{eq:AdditionalTermGPD}
   \mathcal{M}^{\textrm{ad}}_m(\xi,t)  & = &\frac{1}{2}\int \frac{\textrm{d}^4k}{(2\pi)^4}\textrm{Tr}_{\textrm{CDF}}\bigg[ n \cdot \frac{\partial \bar{\Gamma}_\pi}{\partial \bar{k}}(k+\Dd,P+\Dd)S(k-P)\Gamma_\pi(k-\Dd,P-\Dd)S(k-\Dd) \nonumber \\
  & & + \bar{\Gamma}_\pi(k+\Dd,P+\Dd)S(k-P) n \cdot \frac{\partial \Gamma_\pi}{\partial k}(k-\Dd,P-\Dd)S(k+\Dd)\bigg]\frac{(k\cdot n)^m}{(P \cdot n)^{m+1}} .\nonumber \\
\end{eqnarray}
This expression provides the proper forward limit, incorporating the contributions from the additional diagrams in figure \ref{fig:AdditionalContributions}, as can be easily seen by an explicit evaluation at $\Delta \to 0$ and comparison with equation \eqref{eq:AdditionalTermMM}. Nevertheless, the resulting non-forward GPD is also flawed by the approximated extension to the off-forward kinematics defined by \eqref{eq:AdditionalTermGPD}. In particular, in the limit given by $\xi=1$ and $t=0$, equation \eqref{eq:AdditionalTermGPD} gives a non-vanishing contribution, whereas from the previous section, we expect this term to vanish. Indeed, the impulse approximation already provides the soft pion theorem in the RL truncation scheme, as soon as the AVWTI is well implemented. However, The results from 
\eqref{eq:AdditionalTermGPD} make possible a thoughtful study of the structure for the $(x,t)$ correlations and, finally, lead to an additional component for the $F$-type DD such that the total $F^{\textrm{tot}}(\beta, \alpha, t)$ can be seen as:
\begin{eqnarray}
  \label{eq:FTotalStructure}
  F^{\textrm{tot}}(\beta, \alpha, t) & = &  F^{\textrm{Tr}}(\beta,\alpha,t) +F^{\textrm{ad}}(\beta,\alpha,t) \nonumber \\
  & = &\phi^2(\beta,\alpha,t) \left[F^{\textrm{sym}}(\beta,\alpha)+\frac{t}{4M^2} V(\beta,\alpha)\phi(\beta,\alpha,t) \right] ,
\end{eqnarray}
where $F^{\textrm{sym}}$ is the contribution leading to the symmetric PDF $q_\pi^{\textrm{tot}}(x)$, and $\phi$ is given by:
\begin{equation}
  \label{eq:Phi}
  \phi(\beta,\alpha,t) = \frac{1}{1 + \frac{t}{4M^2}  (1 + \alpha - \beta) (1 - \alpha - \beta)}.
\end{equation}
Lightcone considerations (see \refcite{Mezrag:2014jka}), suggest that $V(\beta,\alpha)$ shows a pathological behaviour, presumably due to the simple modelling done in equation \refeq{eq:AdditionalTermGPD}.

These considerations allow an Ansatz for $C^q_\pi(x,t)$ based for simplicity on $\phi(\beta, \alpha = 0,t)$:
\begin{equation}
  \label{eq:ModelCpi}
  C^q_\pi(x,t) = \phi(x,0,t)^2 = \frac{1}{\left(1+\frac{t}{4M}(1-x)^2\right)^2} \ ,
\end{equation} 
which is shown~\cite{Mezrag:2014jka} not to differ very much from the same obtained with a heuristic model for light-cone wave function applied to obtain the GPD in the overlap representation~\cite{Burkardt:2002hr}. 
One should note that at large-$x$, $C^q_\pi(x,t) \rightarrow 1$, and thus, the perturbative behaviour previously highlighted remains. 
\begin{figure}[t]
  \centering
  \includegraphics[width=0.45\textwidth]{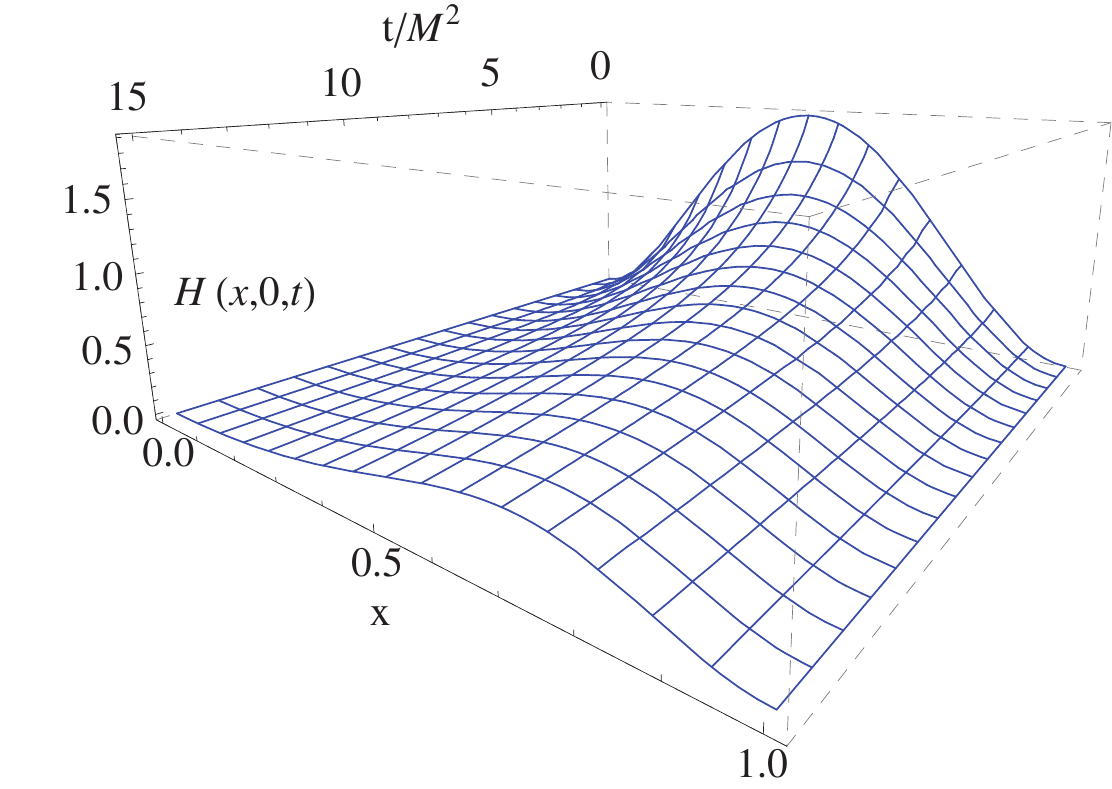}
  \quad
  \includegraphics[width=0.45\textwidth]{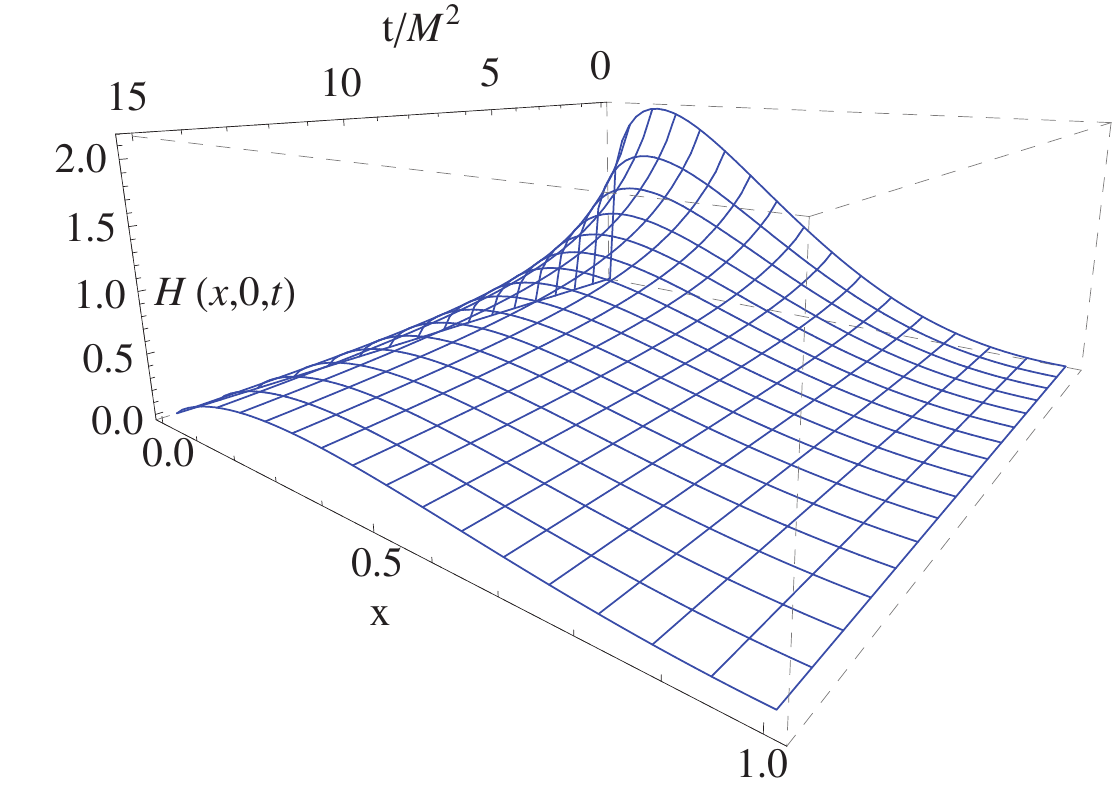}
  \caption{$H(x,0,t)$ using the correlation model of equation \refeq{eq:ModelCpi}. Left-hand side: Original scale $\mu_F = 0.51~\GeV$. Right-hand side: Evolved GPD at scale $\mu_F = 2~\GeV$. Plots are from \refcite{Mezrag:2014jka}. }
  \label{fig:GPDCorrelationModel}
\end{figure}
Using equation \refeq{eq:GPDCorrelationStructure} it is then possible to compute the GPD itself. The result is shown on the left-hand side of figure \ref{fig:GPDCorrelationModel}. Starting from a symmetric distribution, as $t$ grows, the GPD becomes more and more asymmetric, the maximum being shifted toward the  large-$x$ region. In addition, the distribution is also flattened as $t$ grows. Evolution does not change this latter point significantly, but it shifts the entire distribution to the small-$x$ region, as shown on the right-hand side of figure \ref{fig:GPDCorrelationModel}. The PDF is therefore no more symmetric. This can be easily understood, as evolution unravels the substructure of dressed quarks in terms of sea-quarks and gluons, which carry a smaller amount of the total pion momentum. 

\begin{figure}[h]
  \centering
  \includegraphics[width=0.45\textwidth]{F3Uqxbz0.pdf}
  \quad
  \includegraphics[width=0.45\textwidth]{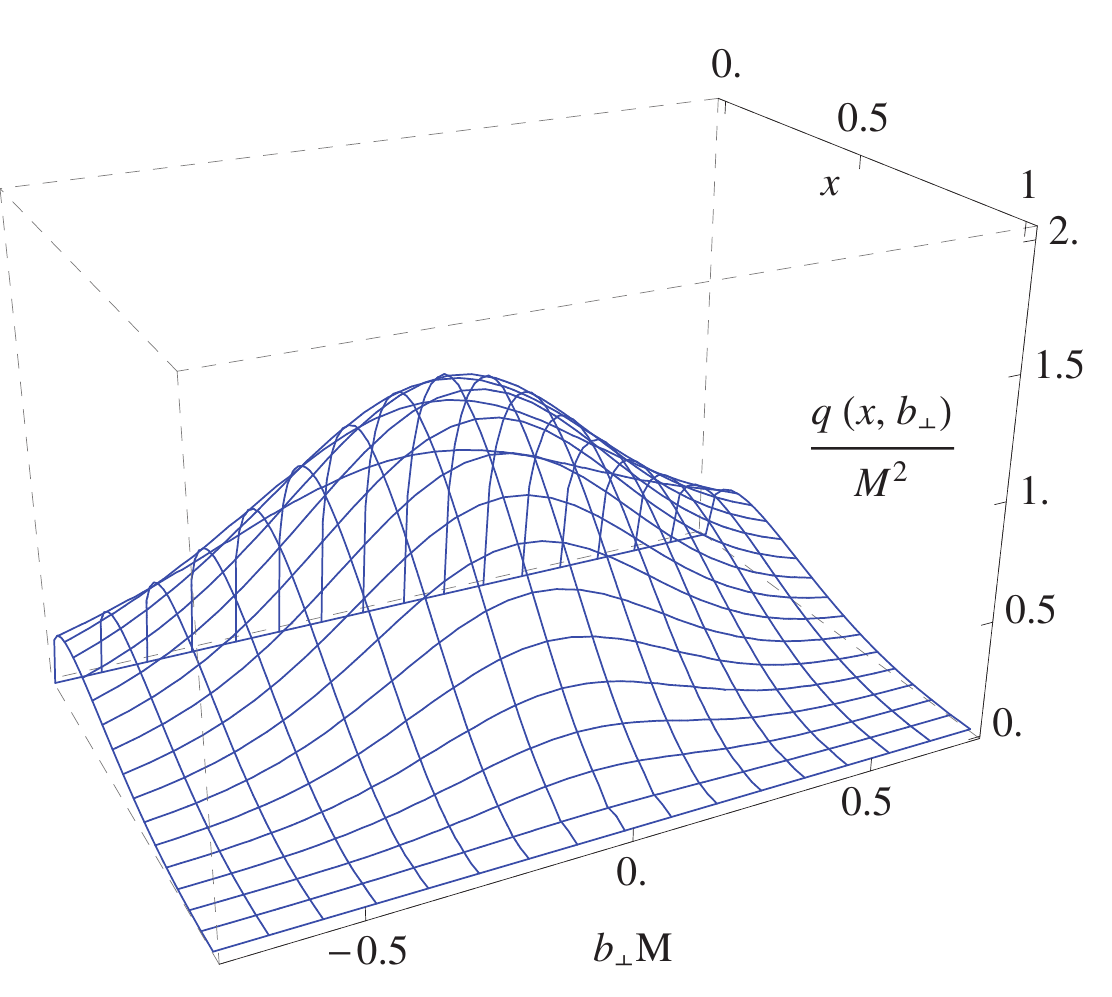}
  \caption{$q(x,\vperp{b})$ using the correlation model of equation \refeq{eq:ModelCpi}. Left-hand side: Original scale $\mu_F = 0.51~\GeV$. Right-hand side: Final scale at $\mu_F = 2~\GeV$. }
  \label{fig:3DPlots}
\end{figure}

Equation \refeq{eq:3DDensity} allows to compute the 3D dressed quark density in the transverse plane. As shown on figure \ref{fig:3DPlots}, when $x$ goes to 1, the distribution becomes narrower. Indeed, in this case, the considered parton carries most of the available momentum. Therefore, it defines the barycentre of the pion in the transverse plane, and the probability to find it far from $|\vperp{b}|=0$ collapses. On the other hand, when $x$ leaves the neighbourhood of $1$, this constraint is released, and as expected, the distribution becomes broader. This is fully consistent with the widening observed when evolving the impact space parameter distribution, since, as highlighted previously, sea quarks and gluons coming from the structure of the dressed quarks carry smaller fractions $x$ of the pion momentum.

\begin{figure}[h]
  \centering
  \includegraphics[width=0.50\textwidth]{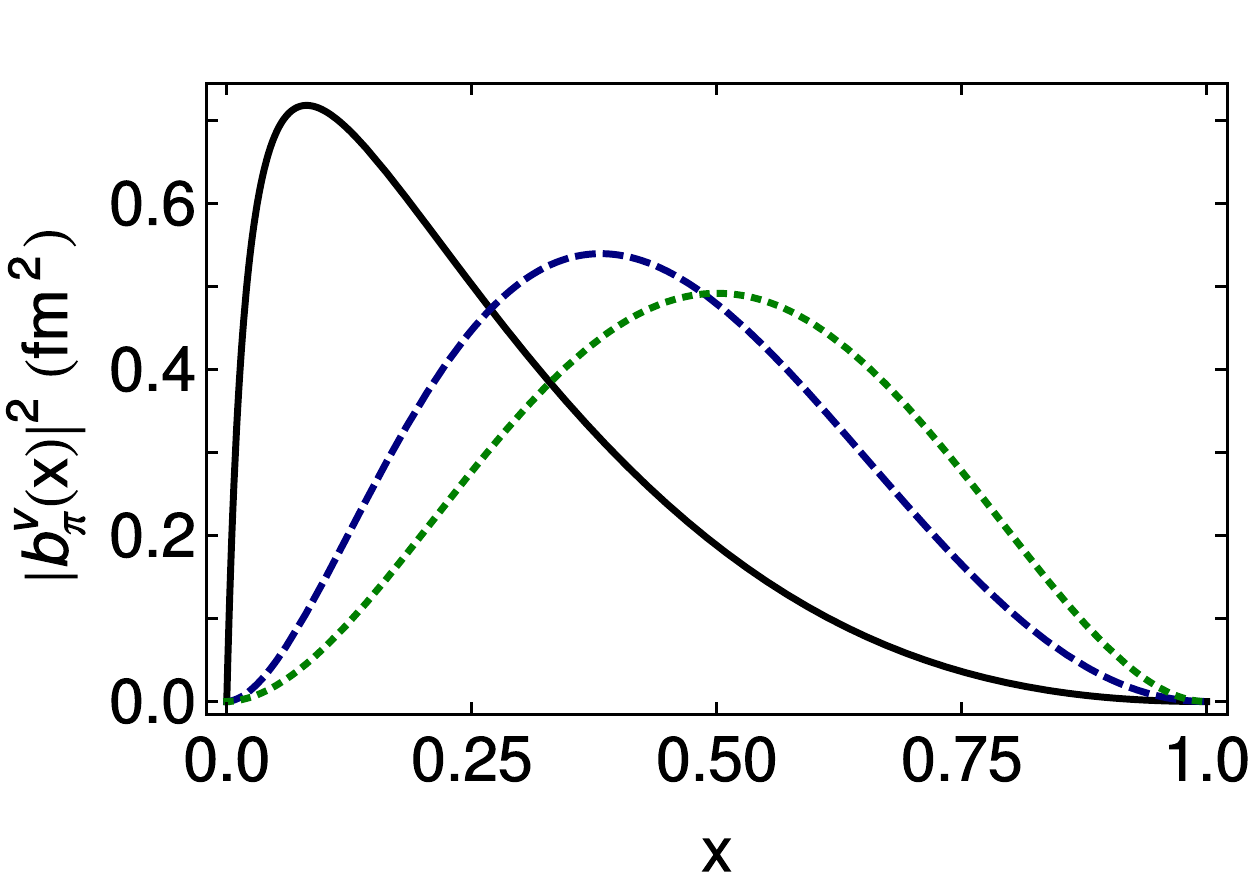}
  \caption{Distribution of the pion's mean-square transverse extend $\left \langle |\vperp{b}(x)|^2 \right \rangle $ from equation \refeq{eq:SecondMomentBperp} as a function of the longitudinal momentum fraction $x$. The short-dashed green curve is obtained with $C^q_\pi(x,t)=1$, the blue long-dashed one with $C^q_\pi(x,t)$ given in equation \refeq{eq:ModelCpi}, and the solid line is the evolution of the dashed one from $0.51~\GeV$ to $2~\GeV$. This figure comes from \refcite{Mezrag:2014jka}.  }
  \label{fig:SecondMomentBperp}
\end{figure}

These points are more visible on figure \ref{fig:SecondMomentBperp}, which shows the second moment of the distribution in $\vperp{b}$ computed as:
\begin{equation}
  \label{eq:SecondMomentBperp}
  \left \langle \vperp{b}(x)^2 \right \rangle = \int \textrm{d}^2\vperp{b}~ \rho^q(x,\vperp{b}) \vperp{b}^2 .
\end{equation}
Interestingly enough, correlations lead to a significant deviation from a symmetric $\left \langle \vperp{b}(x)^2 \right \rangle $ distribution with respect to $x\leftrightarrow 1-x$, emphasising the importance of such correlations in our understanding of hadrons.

\subsection{Strength and weakness of the covariant approach}

The covariant modelling of the pion GPD reveals itself to be very encouraging. First, it allows computations in the DSE-BSE framework, due to the use of the Nakanishi representation. Such a representation has been used here to define an algrebraic and insightful model of the \bs amplitude and of the quark propagator. The pion GPDs computed within the algebraic model fulfils the required GPDs theoretical constraints, with the highlighted exception of positivity. However, the comparison with available experimental data is satisfying. It also sheds light on the complications related to the use of a momentum dependent \bs amplitude. Indeed, the study of the soft pion theorem made here has stressed the importance of the so-called AVWTI, which relates the \bs amplitude with the quark propagators, constraining the $k$-dependence of the running quark mass. In the forward case, the absence of the $x \leftrightarrow 1-x$ symmetry highlights the intrinsic limitations of the impulse approximation in the case of momentum-dependent vertices. Once again, we would like to emphasise that the last two points are somehow invisible for the class of theories using momentum independent vertices, like for instance Nambu-Jona-Lasinio type models \cite{Nambu:1961tp}.

On the other hand, at the time of writing, the restoration of the $x\leftrightarrow 1-x$ has led to the consideration of additional terms, which expressions are known only in the forward case. This situation make difficult any correct prediction of the $\xi$-behaviour of the GPDs. These might also be responsible for the restoration of the positivity property. However they have to vanish in the soft pion limit, due to the consistency of the impulse approximation in this case. With no other clues, different strategies must be considered. Among the possibilities, the lightcone formalism provide an alternative way to model GPDs, as explained in the next section.


\section{Lightcone formalism}
\label{sec:LightconeFormalism}
In previous sections, a covariant approach have been followed to model the valence quark GPD, based first on the impulse approximation and including then some additional contributions needed to amend flaws already present in the forward limit. Pinpointing the proper corrections beyond the impulse approximation is a hard task for off-forward kinematics. Therefore, an alternative approach is desirable. In particular, the so-called overlap representation~\cite{Burkardt:2000za,Diehl:2000xz,Burkardt:2002hr,Diehl:2003ny} has been invoked in ref.~\cite{Mezrag:2014jka} to illustrate the symmetries about the momentum fraction carried by quarks in the two-body problem and provide then with an insightful guide to model its correlations with the momentum transfer. As we will see, if we are provided with a sensible lightcone wave function, the overlap representation opens an avenue to model GPDs, in such a way that the $x \leftrightarrow 1-x$ symmetry of the forward limit is, by construction, fulfilled. In the following, we hightlight how the DSEs allows to compute GPDs using the overlap representation.

\subsection{Lightcone wave functions}

Before describing GPDs in the lightcone formalism, we start here by reminding the reader some basics about lightcone physics.

\subsubsection{Basic facts on lightcone formalism}

At a given lightcone time, for instance $z^+ = 0$, let us split the fields $\psi^q(z)$ into a dynamically independent component, usually called the good field and denoted as $\phi^q(z)$, and a so-called ``bad'' component, which can be deduced from the good one. The good component is a projection of the Dirac field using $\Pi^+ = \frac{1}{2}\g^-\g^+$. It can be written in terms of a Fourier transform of creation and annihilation operators through:
\begin{align}
  \label{eq:GoodFieldDef}
  \phi^q(z^-,\vperp{z}) = \int \frac{\textrm{d}k^+}{k^+} \frac{\textrm{d}^2\vperp{k}}{16\pi^3} \Theta(k^+)\sum_\mu& \left[b_q(w)u_+(w)e^{-i(k^+z^- - \vperp{k}\cdot \vperp{z})} \right. \nonumber \\
  & \left. + d^\dagger_q(w)v_+(w)e^{i(k^+z^- - \vperp{k}\cdot \vperp{z})}\right],
\end{align}
where $\mu$ denotes here the polarisation and $w = (k^+,\vperp{k}, \mu)$. The colour indices are omitted for brevity. $\Theta$ denotes here the Heaviside function, and $u_+$ and $v_+$ are the ``good'' component of the spinors obtained after projecting using $\Pi^+$. The same procedure can be applied to define the transverse component of the gluon field on the lightcone. The creation and annihilation operators follows the usual anticommutation relations:
\begin{eqnarray}
  \label{eq:Anticommutation}
  \left\{b_{q'}(w'),b_q^\dagger(w) \right\} & = & \left\{d_{q'}(w'),d_q^\dagger(w) \right\} \nonumber \\
  & = & 16 \pi^3 k^+ \delta(k'^+-k^+)\delta^{(2)}(\vperp{k}'-\vperp{k})\delta_{\mu'\mu}\delta_{q'q},
\end{eqnarray}
and it is assumed that they act on a trivial perturbative vacuum $\ket{0}$ to generate partonic states:
\begin{eqnarray}
  \label{eq:Quanta}
  \ket{q;w} & = & b^\dagger_q(w)\ket{0} , \nonumber \\
  \ket{\bar{q};w} & = & d^\dagger_q(w)\ket{0}.
\end{eqnarray}

\subsubsection{Hadrons in the lightcone formalism}

Then, according to the definitions of the partonic Fock states \refeq{eq:Quanta}, it is possible to expand a hadron state $\ket{H;P,\lambda}$ of momentum $P$ and polarisation $\lambda$ on a Fock space of a given number $N$ of partons of quantum numbers generically denoted $\beta$, $\ket{N,\beta,k_1 \cdots k_N}$:
\begin{equation}
  \label{eq:HadronFockDecomposition}
  \ket{H;P,\lambda} = \sum_{N,\beta} \int [\textrm{d}x]_N [\textrm{d}^2\vperp{k}]_N \Psi_{N,\beta}^\lambda (\Omega)\ket{N,\beta,k_1 \cdots k_N},
\end{equation}
where $\Psi_{N,\beta}^\lambda (\Omega)$ is the $N$ partons lightcone wave function (LCWF), and $\Omega$ denotes the set of kinematic variables associated with the considered partons:
\begin{equation}
  \label{eq:DefOmega}
  \Omega = (x_1,\textbf{k}_{\perp 1}, \cdots ,x_N,\textbf{k}_{\perp N}).
\end{equation}
The measure terms in equation \refeq{eq:HadronFockDecomposition} are given through: 
\begin{eqnarray}
  \label{eq:DefMeasuresLCx}
  [\textrm{d}x]_N & = & \prod_{i=1}^N \textrm{d}x_i \ \delta \left(1 - \sum_{i=1}^Nx_i\right), \\
  \label{eq:DefMeasuresLCk}
  [\textrm{d}^2\vperp{k}]_N & = & \frac{1}{(16\pi^3)^{N-1}}\prod_{i=1}^N\textrm{d}^2\textbf{k}_{\perp i}\ \delta^2 \left(\sum_{i=1}^N \textbf{k}_{\perp i} -\vperp{P}\right),
\end{eqnarray}
and the canonical normalisation of the LCWFs yields:
\begin{equation}
  \label{eq:NormalisationLCWF}
  \sum_{N,\beta}\int [\textrm{d}x]_N [\textrm{d}^2\vperp{k}]_N |\Psi_{N,\beta}^\lambda (\Omega)|^2 =1.
\end{equation}
The lightcone wave functions can be defined in terms of non-local matrix elements. Following \refcite{Burkardt:2002uc}, one can introduce the pion wave function with anti parallel quark helicity as:
\begin{equation}
  \label{eq:LCWFDef}
  \Psi_{\uparrow\downarrow}(x,\vperp{k}) = \frac{1}{2P^+}\int \textrm{d}z^- \textrm{d}\vperp{z} e^{ixP^+z^-}e^{-i\vperp{k}\vperp{z}}\left.\bra{0}\bar{d}(0)\g^+ \g^5 u(z)\ket{\pi, P}\right|_{z^+=0},
\end{equation}
and the parallel quark helicity as:
\begin{equation}
  \label{eq:LCWFDefH1}
  ik^i \Psi_{\uparrow\uparrow}(k^+,\vperp{k}) = \frac{1}{2P^+} \int \textrm{d}z^-\textrm{d}^2\vperp{z} e^{ik^+z^--i\vperp{k}\vperp{z}}\left. \bra{0}\bar{d}(0)\sigma^{+i}\g_5\, u(z)\ket{\pi^+(P)}\right|_{z^+=0} .
\end{equation}
These wave functions allow one to write the pion two-body Fock states as: 
\begin{align}
  \label{eq:ContributionHadronicState}
  \left.\ket{\pi^+,P}\right|_{\uparrow\downarrow}^{\textrm{2-body}} = \int \frac{\textrm{d}^2\vperp{k}}{(2\pi)^3}\frac{\textrm{d}x}{\sqrt{x(1-x)}}&\Psi_{\uparrow\downarrow}(k^+,\vperp{k})\left[b^{\dagger}_{u\uparrow}(x,\vperp{k})d^{\dagger}_{d\downarrow}(1-x,-\vperp{k}) \right. \nonumber \\
& \left. +b^{\dagger}_{u\downarrow}(x,\vperp{k})d^{\dagger}_{d\uparrow}(1-x,-\vperp{k})\right]\ket{0},
\end{align}
and:
\begin{align}
  \label{eq:ContributionHadronicState2}
  \left.\ket{\pi^+,P}\right|_{\uparrow\uparrow}^{\textrm{2-body}} = \int \frac{\textrm{d}^2\vperp{k}}{(2\pi)^3}\frac{\textrm{d}x}{\sqrt{x(1-x)}}&\Psi_{\uparrow\uparrow}(k^+,\vperp{k})\left[(k_1-ik_2)b^{\dagger}_{u\uparrow}(x,\vperp{k})d^{\dagger}_{d\uparrow}(1-x,-\vperp{k}) \right. \nonumber \\
& \left. +(k_1+ik_2)b^{\dagger}_{u\downarrow}(x,\vperp{k})d^{\dagger}_{d\downarrow}(1-x,-\vperp{k})\right]\ket{0}.
\end{align}

\subsubsection{Pion lightcone and \bs wave functions}

The goal we pursue here is the derivation of the pion GPD from the pion two-body Fock states requiring the knowledge of the pion two-body lightcone wave functions which can be itself related to the \bs wave function introduced in equation \ref{eq:BSWF} by projecting it on the relevant Dirac structure and integrating over $k^-$:
\begin{align}
  \label{eq:RelationBSALCWF1}
  2P^+ \Psi_{\uparrow\downarrow}(k^+,\vperp{k}) = &  \int \frac{\textrm{d}k^-}{2\pi} \textrm{Tr}\left[\g^+\g_5 \chi(k,P) \right], \\
  \label{eq:RelationBSALCWF2}
  ik^i2P^+ \Psi_{\uparrow\uparrow}(k^+,\vperp{k}) = &  \int \frac{\textrm{d}k^-}{2\pi} \textrm{Tr}\left[\sigma^{+i}\g_5 \chi(k,P) \right].
\end{align}
Therefore, it is possible to compute the LCWFs, using the algebraic model from equations (\ref{eq:TMQuarkPropagator}-\ref{eq:TMkDef}), within the same model framework developed with the covariant approach in  previous sections. Thus, up to some ambiguities presumably related to the particular choice used for the Nakanishi representation of the \bs amplitude and to the definition of the vector insertion in the covariant approach, the result we might obtain for the GPD in the forward limit from these LCWFs should be close to the total PDF shown in figure \ref{fig:PDFTotal}. 

As the algebraic parameterisation is defined in euclidean space, one of the efficient ways to compute the LCWF is to proceed through the evaluation of the Mellin moments. As the LCWF are $\vperp{k}$-dependent, the $4$-vector $k$ is split as $k = q+\vperp{k}$, with $q = (q^+,q^-,0)$. Consequently, the Mellin moments yield:
\begin{eqnarray}
  \label{eq:MellinMomentLCWF}
  \int_0^1 \textrm{d}x~x^m \Psi_{\uparrow\downarrow}(x, \vperp{k}) & = & \int_0^1 \textrm{d}x~x^m \int \frac{\textrm{d}q^-\textrm{d}q^+}{(2\pi)^2} \textrm{Tr}\left[\gamma^+\gamma_5\chi_\pi(q+\vperp{k},P)\right] \delta (x P \cdot n - q \cdot n ) \nonumber \\
  & = & \int \frac{\textrm{d}^2q (q \cdot n)^m}{(2\pi)^2(P \cdot n)^{(m+1)}}\textrm{Tr}\left[\gamma^+\gamma_5\chi_\pi(q+\vperp{k},P)\right].
\end{eqnarray}
Computing the trace and integrating over $q$ using Feynman parameters, one can identify the scalar LCWF as:
\begin{equation}
  \label{eq:FinalLCWF}
  \Psi_{\uparrow\downarrow}(x, \vperp{k})= - \frac{\Gamma(\nu+1)}{\Gamma(\nu+2)}\frac{M^{2\nu +1}4^\nu R_\nu}{\left[\vperp{k}^2+M^2 \right]^{\nu+1}}x^\nu(1-x)^\nu .
\end{equation}
Using the same approach, the helicity-1 LCWF $\Psi_{\uparrow\uparrow}(k^+,\vperp{k})$ can be computed through its Mellin moments as:
\begin{equation}
  \label{eq:FinalLCWF2}
   \Psi_{\uparrow\uparrow}(x,\vperp{k}) = 2iM^{2\nu}\frac{\G(\nu+1)}{\G(\nu+2)}\frac{4^\nu R_\nu x^\nu(1-x)^\nu}{\left[\vperp{k}^2 +M^2\right]^{\nu+1}}
\end{equation}

\subsection{Overlap of pion wave functions}

The equations (\ref{eq:FinalLCWF},\ref{eq:FinalLCWF2}) appear to be the main ingredients needed to obtain the valence-quark pion GPD which, as follows, can be derived from the pion Fock states with a two-body truncation. 

\subsubsection{GPDs and LCWFs}

The lightcone formalism is a convenient approach to compute GPDs\footnote{We adapt here the derivation of the spin one-half case done in \refcite{Diehl:2000xz} to the spinless case.} \cite{Diehl:2000xz}, since the latter are defined through operators at a given lightcone time. These operators can then be written in terms of good components $\phi$ of the quark fields $\psi$:
\begin{equation}
  \label{eq:KeyIngredientGoodCompo}
  \left. \bar{\psi}^q\left(-\frac{z}{2}\right) \g^+ \psi^q \left(\frac{z}{2}  \right)\right|_{z^+=0} = \sqrt{2} \phi_q^\dagger \left(-\frac{z}{2}\right) \phi_q\left(\frac{z}{2}\right).
\end{equation}
The decomposition of equation \refeq{eq:HadronFockDecomposition} leads to the following expression for the pion GPD:
\begin{eqnarray}
  \label{eq:PionGPDLCDGLAP}
  H(x,\xi,t) & = & \sqrt{2} \sum_{N,N'}\sum_{\beta,\beta'} \int  [\textrm{d}\hat{x}']_{N'} [\textrm{d}^2\hat{\textbf{k}}_\perp']_{N'}  [\textrm{d}\tilde{x}]_N [\textrm{d}^2\tilde{\textbf{k}}_\perp]_N \Psi^{*}_{N',\beta'}(\hat{\Omega}')\Psi_{N,\beta}(\tilde{\Omega}) \nonumber \\
  & & \times \int \frac{\textrm{d}z^-}{2\pi}e^{iP^+z^-}\bra{N',\beta,k'_1 \cdots k'_N}\phi^{q\dagger} \left(-\frac{z}{2}\right)\phi^q\left( \frac{z}{2} \right) \ket{N,\beta, k_1 \cdots k_N}, \quad \quad
\end{eqnarray}
where the hat variables are given in the outgoing pion wave function frame, and the tilde ones in the incoming pion wave function frame. Other variables are considered in the GPD symmetric frame, \ie the frame such that $p_1=P-\Dd$ and $p_2 = P+\Dd$. In this latter frame, it is possible to define the average variables $\bar{k}_i$ as:
\begin{equation}
  \label{eq:AverageVariables}
  \bar{k}_i = \frac{1}{2} (k_i+k'_i ) \quad , \quad \bar{x}_i = \frac{\bar{k}^+_i}{P^+} ,
\end{equation}
which fulfil the momentum conservation sum rules:
\begin{equation}
  \label{eq:AverageVariableMomentumConservation}
  \sum_{i=1}^{N} \bar{x}_i = 1 \quad , \quad \sum_{i=1}^N \bar{\textbf{k}}_{\perp i} = P_\perp = 0.
\end{equation}
Therefore, active parton momentum variables, labelled as $j$, follow:
\begin{equation}
  \label{eq:ActivexConservation}
  x_j = \bar{x}_j + \xi \quad , \quad x'_j = \bar{x} - \xi \quad ,
\end{equation} 
and
\begin{equation}
  \label{eq:ActivePartonMomentumConservation}
  \textbf{k}_{\perp j} = \bar{\textbf{k}}_{\perp j} - \frac{\vperp{$\Delta$}}{2} \quad , \quad \textbf{k}'_{\perp j} = \bar{\textbf{k}}'_{\perp j} + \frac{\vperp{$\Delta$}}{2} \quad .
\end{equation}
On the other hands, the other partons variables, labelled $i$, are simply related through:
\begin{equation}
  \label{eq:SpectatorPartonMomentumConservation}
  k'_i = \bar{k}_i =k_i .
\end{equation}

In the incoming and outgoing momentum frames, one can also define the fraction of momentum $\tilde{x}_i$ and $\hat{x}'_i$ as:
\begin{equation}
  \label{eq:xDef}
  \tilde{x}_i = \frac{\tilde{k}_i^+}{p_1^+} \quad , \quad \hat{x}'_i = \frac{\hat{k}'_i}{p_2^+} .
\end{equation}
It is then possible, to boost these variables from their original frame, to the symmetric frame used here to compute GPDs. Doing so, one gets for the incoming variables:
\begin{equation}
  \label{eq:IncomingPionBoostedVariables}
  \begin{aligned}
    \tilde{x}_i = \frac{\bar{x}_i}{1+\xi}, & \quad \quad & \tilde{\textbf{k}}_{\perp i} & = &\bar{\textbf{k}}_{\perp i}+\frac{\bar{x}_i}{1+\xi}\frac{\vperp{$\Delta$}}{2},& \quad \quad \textrm{for}~i\neq j \\
    \tilde{x}_j = \frac{\bar{x}_j+\xi}{1+\xi},& \quad \quad &\tilde{\textbf{k}}_{\perp j} & = &\bar{\textbf{k}}_{\perp j}-\frac{1-\bar{x}_i}{1+\xi}\frac{\vperp{$\Delta$}}{2},&
  \end{aligned}
\end{equation}
and for the outgoing one:
\begin{equation}
  \label{eq:OutgoingPionBoostedVariables}
  \begin{aligned}
    \hat{x}'_i = \frac{\bar{x}_i}{1-\xi}, & \quad \quad & \hat{\textbf{k}}'_{\perp i} & = &\bar{\textbf{k}}_{\perp i}-\frac{\bar{x}_i}{1-\xi}\frac{\vperp{$\Delta$}}{2},& \quad \quad \textrm{for}~i\neq j \\
    \hat{x}'_j = \frac{\bar{x}_j-\xi}{1-\xi},& \quad \quad &\hat{\textbf{k}}'_{\perp j} & = &\bar{\textbf{k}}_{\perp j}+\frac{1-\bar{x}_i}{1-\xi}\frac{\vperp{$\Delta$}}{2}.&
  \end{aligned}
\end{equation}
Denoting now by $k$ the Fourier conjugate of $z$ in equation \refeq{eq:PionGPDLCDGLAP}, such as $k^+ = xP^+$, one immediately deduce that $x = \bar{x}_j$. In the case of the pion, equation \refeq{eq:PionGPDLCDGLAP} can be simplify in the DGLAP case to:
\begin{eqnarray}
  \label{eq:FinalPionGPDLightCone}
  H^q_\pi(x,\xi,t)|_{\xi \le x \le 1} & = & \sum_N \sqrt{1-\xi}^{2-N} \sqrt{1+\xi}^{2-N} \sum_{\beta = \beta'} \sum_{j} \delta_{s_jq} \nonumber \\
  & & \times \int [\textrm{d}\bar{x}]_N [\textrm{d}^2\bar{\textbf{k}}_\perp]_N \delta(x-\bar{x}_j) \Psi^*_{N,\beta'}(\hat{\Omega}') \Psi_{N,\beta}(\tilde{\Omega}).
\end{eqnarray}
The ERBL part can be derived similarly to the DGLAP one, but will be non-diagonal in $N$, as interactions between $\Psi_{N,\beta}(\hat{\Omega})$ and $\Psi_{N+2,\beta}(\tilde{\Omega})$ are expected. Therefore, in the case of a two-body truncation, no ERBL contribution can be computed directly using the lightcone formalism.

However, in the DGLAP case, the two-body truncation yields:
\begin{align}
  \label{eq:GPDPionValenceQuark}
   H^q_\pi(x,\xi,t)|_{\xi \le x \le 1}^{2-\textrm{body}} = &\int [\textrm{d}\bar{x}]_2 [\textrm{d}^2\bar{\textbf{k}}_\perp]_2 \delta(x-\bar{x}_j) \Psi^*_{\uparrow\downarrow}(\hat{\Omega}') \Psi_{\uparrow\downarrow}(\tilde{\Omega}) \nonumber \\
   & + \int [\textrm{d}\bar{x}]_2 [\textrm{d}^2\bar{\textbf{k}}_\perp]_2 \delta(x-\bar{x}_j)\left(\hat{k}_1+i\hat{k}_2 \right) \left(\tilde{k}_1-i\tilde{k}_2 \right) \Psi^*_{\uparrow\uparrow}(\hat{\Omega}') \Psi_{\uparrow\uparrow}(\tilde{\Omega}).
\end{align} 

\subsubsection{An example with the algebraic model}
It is then possible to inject equations \refeq{eq:FinalLCWF} and \refeq{eq:FinalLCWF2} in equation \refeq{eq:GPDPionValenceQuark}. Introducing the two Feynman parameters $u$ and $v$ and then integrating over $\vperp{k}$, one is left with two contributions coming from the different combinations of the spin of the dressed quarks. Using the fact that, in our euclidean and chiral approach, $t = \frac{\Delta_\perp^2}{1-\xi^2}$, one gets: 
\begin{equation}
  \label{eq:IntegrationKPerp}
  H^{\uparrow \downarrow}(x,\xi,t) = \frac{\G(2\nu+2)}{\G(\nu+2)^2}\int \textrm{d}u\textrm{d}v\ u^\nu v^\nu \delta\left(1-u-v \right)\frac{\left(2M^{2\nu}4^\nu R_\nu \right)^2\hat{x}^\nu(1-\hat{x})^\nu\tilde{x}^\nu(1-\tilde{x})^\nu}{\left(t\, uv\frac{(1-x)^2}{1-\xi^2} +M^2\right)^{2\nu+1}} ,
\end{equation}
and
\begin{align}
  \label{eq:IntegrationKPerpUpUp}
  H^{\uparrow \uparrow}(x,\xi,t)  = & \frac{\G(2\nu+2)}{\G(\nu+2)^2}\int \textrm{d}u\textrm{d}v\ u^\nu v^\nu \delta\left(1-u-v \right)\left(2M^{2\nu}4^\nu R_\nu \right)^2\hat{x}^\nu(1-\hat{x})^\nu\tilde{x}^\nu(1-\tilde{x})^\nu \nonumber \\
  & \times \left[\frac{\pi \frac{\G(2\nu)}{\G(2\nu+2)}}{\left(t\,uv\frac{(1-x)^2}{1-\xi^2} +M^2\right)^{2\nu}}
  -t\frac{uv\frac{(1-x)^2}{1-\xi^2}\pi \frac{\G(2\nu+1)}{\G(2\nu+2)}}{\left(t\, uv\frac{(1-x)^2}{1-\xi^2} +M^2\right)^{2\nu+1}}  \right] .
\end{align}

\begin{figure}[t]
  \centering
  \includegraphics[width=0.6\textwidth]{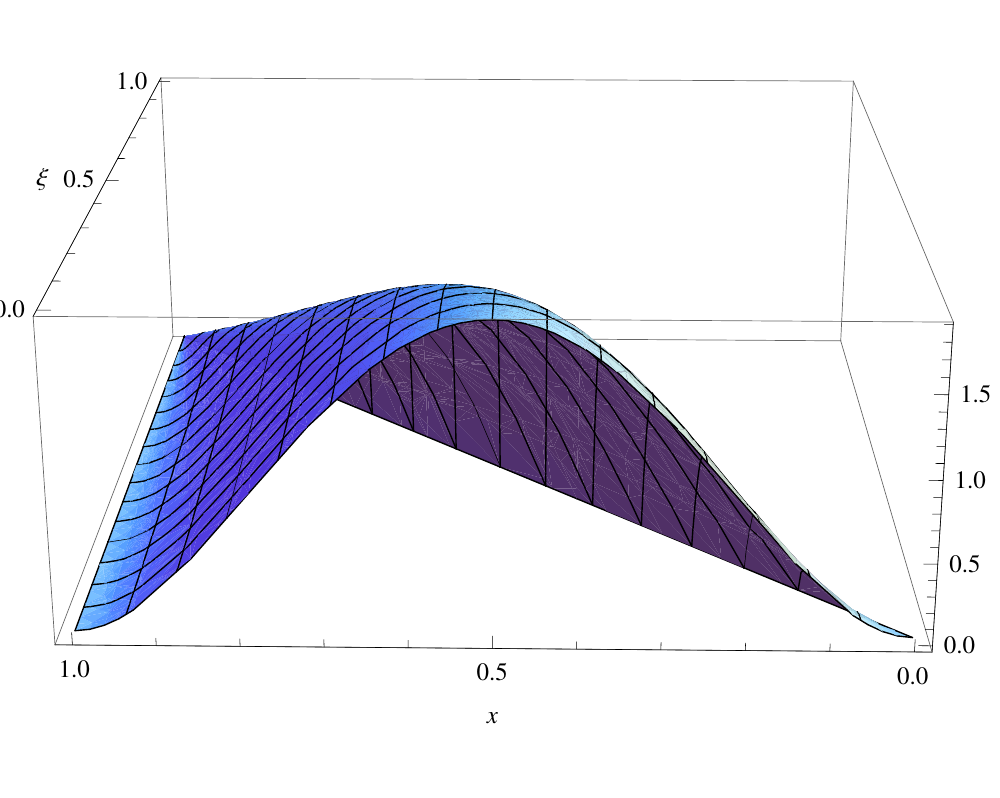}
  \caption{Pion GPD $H_\pi^q$ as a function of $x$ and $\xi$ obtained through the overlap of LCWFs in the DGLAP region.}
  \label{fig:DGLAPLCWF}
\end{figure}

One can now focus on the case $t=0$ where the pion quark GPD can be simplified as:
\begin{eqnarray}
  \label{eq:DGLAPLCFinal}
  \left. H_\pi^q (x,\xi,0)\right|_{\textrm{DGLAP}} & = & \kappa'_\nu \ \hat{x}^\nu(1-\hat{x})^\nu \tilde{x}^\nu (1-\tilde{x})^\nu \\
  \label{eq:DGLAPLCFinalExpanded}
  & = & \kappa'_\nu \frac{\left(1-x\right)^{2\nu}\left(x^2-\xi^2 \right)^\nu}{\left(1-\xi^2 \right)^{2\nu}},
\end{eqnarray}
$\kappa'_\nu$ is therein an overall normalisation constant. In the case $\nu = 1$, $\kappa'_\nu = 30$, and the result of the computation of the GPD in the DGLAP region is illustrated on figure \ref{fig:DGLAPLCWF}.

One can finally deduce, from equation \refeq{eq:DGLAPLCFinal}, the expression for the PDF computed with the overlap of LCWFs. This is done by taking $\xi = 0$ in addition to $t = 0$ leading to:
\begin{equation}
  \label{eq:PDFLCFinal}
  q_\pi(x) = \kappa'_\nu x^{2\nu}(1-x)^{2\nu} \ , 
\end{equation}
which should compare well with the PDF obtained in equation \refeq{eq:PDFTotal} within an equivalent framework and a covariant approach based on the impulse approximation and the amendment described in section~\ref{sec:BeyondImpulse}. 
The comparison is shown by the figure \ref{fig:PDFTotal}, where the agreement between black solid and dotted lines, which correspond to both the total covariant and overlap computations, is striking. 


Focusing on figure \ref{fig:DGLAPLCWF} and the related equation \refeq{eq:DGLAPLCFinal}, some comments must be made. The first tempting point is of course to compare the overlap results presented in figure \ref{fig:DGLAPLCWF} with the GPD obtained in the covariant approach through the impulse approximation. The latter result was shown on figure \ref{fig:GPDnu1}. One immediately notices that, the more off-forward, the more different the two DGLAP regions are, although they have been computed from the \emph{very same} \bs amplitude. The main difference lies in the fact that the overlap computations lead to a GPD which vanishes on the line $x = \xi$, whereas the triangle diagram computation predicts a non-vanishing value of the GPD on this line. As a consequence, the GPD computed in the overlap approach fulfils the positivity property defined in equation \refeq{eq:Positivity}. When $t$ vanishes, the formulae obtained for the GPD \refeq{eq:DGLAPLCFinal} and the PDF \refeq{eq:PDFLCFinal} lead to a relation stronger than the positivity condition, since:
\begin{equation}
  \label{eq:FullPositivityLC}
  \left. H_\pi^q (x,\xi,0)\right|_{\textrm{DGLAP}} = \sqrt{q_\pi(x_1)q_\pi(x_2)} .
\end{equation}
This relation between the GPD and the PDF is presumably a feature of the algebraic model defined in equations \refeq{eq:TMQuarkPropagator}-\refeq{eq:TMkDef}.

Nevertheless, the fact that the GPD vanishes on the line $x = \xi$ in a two-body approximation is not a surprise, and have been already highlighted in \refcite{Choi:2002ic} and \refcite{Ji:2006ea} for instance. Moreover, computations using a higher Fock state, including a gluon in addition to the dressed quarks, have been performed in \refcite{Ji:2006ea}. Using the same computation techniques within this three-body truncation than in two-body truncation, the authors showed that the three-body contribution leads to a non-vanishing GPD on the line $x=\xi$. Therefore, in this kinematic area, the GPD seems to be very sensitive to the truncation of the Fock space. This is of deep phenomenological interest (see section \ref{sec:Exclusive}), as the imaginary part of the CFF $\mathcal{H}$ is directly proportional to $H(\xi,\xi,t)$ at leading order.

\subsection{Overlap and polynomiality}

If the overlap representation offer a simple way to build a GPD model fulfilling the positivity property, the major drawback is that within a two-body truncation, it is not possible to get a proper model for the ERBL region. Therefore, it is a priori impossible to conclude on the polynomiality property of the Mellin moments. However, the fact that the LCWFs are derived from the \bs wave function suggests that the underlying Lorentz symmetry is preserved. And therefore, one should expect polynomiality to be fulfilled. 

The authors of \refcite{Hwang:2007tb} and \refcite{Muller:2014tqa} have shown that for a certain class of LCWFs, a simple change of variable allows the identification of the Double Distributions. From the DDs, it is then easy to compute the ERBL part of the GPD through the Radon transform \refeq{eq:RelationDDGPDH}. The authors of \refcite{Hwang:2007tb} use lightcone wave functions such that:
\begin{equation}
  \label{eq:MullerLCWF}
  \Psi (x,\vperp{k}) \propto \frac{M^{2p}}{\sqrt{1-x}}x^{-p} \left( M^2 - \frac{\vperp{k}^2+m^2}{x}-\frac{\vperp{k}+\lambda^2}{1-x}\right)^{-p-1},
\end{equation}
where $M$, $\lambda$ and $m$ are respectively the hadron, the spectator and the quark masses. The associated DD $e(\beta,\alpha,t)$, given in the so-called Pobylitsa scheme \cite{Pobylitsa:2002vi}, for the GPD $E(x,\xi,t)$ is then given by:
\begin{equation}
  \label{eq:MullerDDLC}
  e(\beta,\alpha,t) = N \frac{\left( \frac{m}{M} + \beta\right)\left( \left(1-\beta\right)^2-\alpha^2\right)^p}{\left[(1-\beta)\frac{m^2}{M^2} +\beta\frac{\lambda^2}{M^2}-\beta(1-\beta)-\left( \left(1-\beta \right)^2-\alpha^2 \right)\frac{t}{4M^2}\right]^{2p+1}},
\end{equation}
with $N$ an overall normalisation constant.
These results show that, even if it not possible to compute directly in the overlap approach a contribution for the ERBL region, it is possible to deduce one from the DGLAP area. And even more interesting, it is possible to get one which fulfils the polynomiality condition.  

Unfortunately, the wave functions of equations \refeq{eq:FinalLCWF} and \refeq{eq:FinalLCWF2} do not belong to the class of functions studied in \refcite{Hwang:2007tb} and \refcite{Muller:2014tqa}. Nevertheless, the Radon transform, relating DDs to GPDs, can be inverted. The general inversion formula is given by:
\begin{equation}
  \label{eq:FinalResultsInverse}
      f(\beta,\alpha)  =  \frac{2}{(2\pi)^2}\int_{-\frac{\pi}{2}}^{\frac{\pi}{2}}\textrm{d} \phi~\int_{-\infty}^{\infty}\textrm{d}\tau~\textrm{PV}\left[ \frac{1}{\tau}\right]\left.\left(\frac{\partial}{\partial s'}\mathcal{R}f(s'+\beta \cos(\phi)+\alpha \sin(\phi)-\tau,\phi)\right)\right|_{s'=0},
\end{equation}
where $f$ is the double distributions, and $\mathcal{R}f$ is its canonical Radon transform. PV stands for the principal value distribution defined as:
\begin{equation}
  \label{eq:PVDefinition}
  \int_{-\infty}^{\infty}\textrm{d}x~\textrm{PV}\left[\frac{1}{x}\right] u(x) = \int_0^{\infty} \textrm{d}x \frac{u(x)-u(-x)}{x} .
\end{equation}
However, as highlighted in \refscite{Teryaev:2001qm,Mezrag:2015mka}, this formula is numerically very sensitive to numerical noise. Moreover, it requires to know the GPDs on a extended support, including the DGLAP and ERBL region, but also the GPDs \emph{alter ego} in terms of crossing symmetry: the Generalised Distribution Amplitudes (GDA) \cite{Mueller:1998fv,Diehl:1998dk,Diehl:2000uv}. Therefore, a direct implementation of equation \refeq{eq:FinalResultsInverse} is not suitable for deriving the ERBL contribution from the DGLAP one. However, using the 1CDD scheme defined in equations \refeq{eq:F-1CDD} and \refeq{eq:G-1CDD}, one can relate the GPD to the canonical Radon transform $\mathcal{R}f$ of the function $f$ through:
\begin{equation}
  \label{eq:CanonicalRelationRadonTransform}
  \frac{\sqrt{1+\xi^2}}{x} H(x,\xi,t) = \mathcal{R}f = \int_\Omega \textrm{d}\beta \textrm{d}\alpha \delta\big(s-\beta \cos (\phi) - \alpha \sin (\phi)\big) f(\beta, \alpha,t),
\end{equation}
with $s = \frac{x}{\cos (\phi)}$ and $\xi = \tan (\phi)$. Then a careful mathematical study of the properties of the Radon transform shows that the double distribution $f$ can be uniquely defined on the entire rhombus, except for the line $\beta = 0$. The proof is left for a forthcoming work \cite{Mezrag:2016}. The consequences of these results are important, since one would be able to compute a proper contribution to the GPD in the ERBL region, within a two-body approximation. The deep cause for this comes from the constraints for the GPD imposed by Lorentz and discrete symmetries. The GPDs has to be the Radon transform of a certain DD. In particular, the Radon transform of a 1CDD $f(\beta,\alpha,t)$ can be recast as:
\begin{equation}\label{eq:fin}
\mathcal{R}f(s,\varphi,t) \ = \ \sum_{m=0}^\infty \sum_{l=0}^m g_{ml}(t) \ e^{i\; (-m+2l) \; \varphi} \, C^\alpha_m(s) \ ,
\end{equation}
where $g_{ml}(t)$ are the coefficients for the expansion in terms of Geigenbauer polynomials, $C^\alpha_m(s)$. Therefore, identifying properly the coefficients $g_{ml}(t)$, if possible, leads to the unique extension to the ERBL kinematical region for a GPD only known in the DGLAP domain. However, possible non-analytic contributions would not be visible and might spoil the previous results. The existence of such contributions is still debated~\cite{Muller:2015vha}. 

The Radon inversion remains an ill-posed mathematical problem in the sense of Hadamard. In practical, it means that uncertainties like for instance numerical noise, are amplified. Therefore, even if the ERBL region is a priori accessible, the numerical noise might make very hard the path to access it. Nevertheless, this problem may be overcome using proper algorithms. This study is left for a future work \cite{Mezrag:2016}.


\section{Conclusion}
\label{sec:Conclusion}

Experimental perspectives in hadron physics, and more specifically, in hadron structure will probably challenge our understanding of the non-perturbative sector of the strong interaction. However, without a proper connection of those data with the fundamental theory describing the strong interaction, one will miss a part of the reachable knowledge. In this context, we highlighted the role that GPDs can play. More precisely, we shed light on the possible achievements in terms of GPD modeling, using non-perturbative techniques such as the \dses. The recent progress in the development of DSE kernels and the use of the Nakanishi representation allow for optimism. Indeed, a two-body effective quark model of a pion GPD, generated dynamically through the DSEs and BSEs, will certainly be a valuable outcome in a very near future. However, such computations must take into account the important features analysed and discussed in the current paper. The algebraic model presented here has highlighted the specific role of the Axial-Vector Ward-Takahashi Identity in the fulfilment of the soft pion theorem. Any dynamical computations should therefore pay a specific attention to this point. In addition, the so-called impulse approximation has been shown beyond any reasonable doubt to be insufficient to ensure the preservation the symmetry related to the fraction of momentum carried by the quarks in the two-body system. If a proper amendment for this flaw resulting from the impulse approximation has been found in the forward case (\ie for the PDF), a piece of the pion GPD is still missing.

However, this problem may be overcome soon through the lightcone formalism. Indeed, recent results suggest that the pion GPDs can be computed after a two-body truncation, both in the DGLAP and ERBL regions. Using a suitable algorithm to compute the inverse Radon transform, it should be possible to obtain the pion GPD in its entire domain of definition. Such a breakthrough would then pave the way for a systematic study of the pion properties, including its 3D shape, in terms of choices of the DSEs kernels. 

On a longer time scale, this procedure should of course be extended to the proton, since it will be the main targeted hadron in the forthcoming JLab 12 experiments. Indubitably, the proton case will be significantly more difficult, as one has to deal with a three-body system at the lowest order. However, truncation schemes have also been developed for the Faddeev equations, and a first approximation of the proton in terms of a quark and a diquark may also be relevant. Therefore, it is realistic to consider the possibility of the emergence of DSE-based computations for the proton GPDs in forthcoming years. 

Notwithstanding the highlighted challenges, the computation of GPDs within a \ds framework promises to be an important tool for the interpretation of JLab 12 obervables, opening the path to a detailed knowledge on the nucleon structure and a deeper understanding of the large distance regime of the strong interaction.


\section*{Acknowledgements}
The authors would like to thank L. Chang, C.D. Roberts, F. Sabatié, S.M. Schmidt and P. Tandy with whom parts of the results described here have been derived. They are also grateful to L. Chang, I. Cloet, M. Defurne, J-F. Mathiot, B. Pasquini, B. Pire, C.D. Roberts, F. Sabatié, L. Szymanowski, J. Wagner and S. Wallon for their valuable discussions and comments. The authors are also thankful for the chance to participate in the workshops ``Non-Perturbative QCD 2012'',Matalasca$\tilde{\textrm{n}}$as, Spain, where this project originated, ``Many Manifestations of Nonperturbative QCD under the Southern Cross'', Ubatuba, Sao Paulo, where significant parts of this work were first presented and improvements discussed, as well as ``Non-Pertubative QCD 2014'', Punta Umbria, Spain, where the overlap representation project started being discussed.  

This work is partly supported by U.S. Department of Energy, Office of Science, Office of Nuclear Physics, under contract no.~DE-AC02-06CH11357, by the Commissariat à l’Energie Atomique, the Joint Research Activity ``Study of Strongly Interacting Matter'' (acronym HadronPhysics3, Grant Agreement n.283286) under the Seventh Framework Programme of the European Community, by the GDR 3034 PH-QCD ``Chromodynamique Quantique et Physique des Hadrons'', the ANR-12-MONU-0008-01 ``PARTONS'' and the Spanish ministry Research Project FPA2014-53631-C2-2-P.

\appendix

\section{Conventions and Notations}
\label{sec:AppendixA}

\subsection{Space-time and lightcone conventions}

Four-vectors are denoted with normal character \eg $p = (p^0,p^1,p^2,p^3)$, whereas three- and two-vectors are written in bold font: $\textbf{p}=(p^1,p^2,p^3)$.
The Minkowskian metric used through this text is:
\begin{equation}
  \label{eq:MetricDef}
  \eta_{\mu\nu} = \eta^{\mu\nu} = 
  \begin{pmatrix}
    1 & 0 & 0 & 0 \\
    0 & -1 & 0 & 0 \\
    0 & 0 & -1 & 0 \\
    0 & 0 & 0 & -1 
  \end{pmatrix}
  ,
\end{equation}
so that:
\begin{equation}
  \label{eq:ScalarProduct}
  p^2 = (p^0)^2 - \textbf{p}^2 = M^2.
\end{equation}

The lightcone variables for a four-vector $v$ are defined as:
\begin{equation}
  \label{eq:LCVariables}
  v^+ =\frac{1}{\sqrt{2}}(v^0+v^3) \quad , \quad v^- = \frac{1}{\sqrt{2}}(v^0-v^3)\quad \textrm{and} \quad \vperp{v} = (v^1,v^2).
\end{equation}

\subsection{Dirac Algebra}

The Dirac matrices obey the four-dimensional Clifford Algebra:
\begin{equation}
  \label{eq:CliffordAlgebra}
  \left\{\g^\mu,\g^\nu \right\} = 2\eta^{\mu\nu} .
\end{equation}
In all this work, the Weyl representation of the $\g^\mu$ is used:
\begin{equation}
  \label{eq:WeylRepresentation}
  \g^\mu = 
  \begin{pmatrix}
    0 & \bar{\sigma}^\mu \\
    \sigma^\mu & 0 
  \end{pmatrix}
  ,
\end{equation}
with:
\begin{equation}
  \label{eq:SigmaMuDef}
  \sigma^\mu = (1, \bm{\sigma}) \quad , \quad \bar{\sigma}^\mu = (1, -\bm{\sigma}),
\end{equation}
and the $\sigma^i$ are the Pauli matrices. The $\sigma^{\mu\nu}$ tensor is defined as:
\begin{equation}
  \label{eq:SigmaMuNu}
  \sigma^{\mu\nu} = \frac{i}{2} \left[\g^\mu, \g^\nu \right].
\end{equation}

\subsection{Euclidean space}

It is usually possible to define a Euclidean quantum field theory which is the counterpart of the Minkowskian one providing that:
\begin{eqnarray}
  \label{eq:EuclideanX}
  \int \textrm{d}^4x^M & = & -i \int  \textrm{d}^4x^E , \\
  \label{eq:EuclideanPartial}
  \g^M \cdot \partial^M  & = & i \g^E \cdot \partial^E \\
  \label{eq:EuclideanSlashed}
  p^M \cdot \g^M & = & -i p^E \cdot \g^E \\
  \label{eq:EuclideanScalarProduct}
  p^M \cdot q^M & = & - p^E \cdot q^E,
\end{eqnarray}
with $x^4 = ix^0$ and $\eta_{\mu\nu}^E = \delta_{\mu\nu}$, $\delta_{\mu\nu}$ being the four dimensional Kronecker symbol. The Euclidean Clifford algebra is given by:
\begin{equation}
  \label{eq:EuclideanClifford}
  \left\{ \g_\mu^E, \g_\nu^E \right \} = 2 \delta_{\mu\nu},
\end{equation}
leading to:
\begin{equation}
  \label{eq:EucldeanGamma}
  \g_4^E = \g^M_0, \quad \g_j^E = -i\g^M_j \quad \textrm{for } j = 1,2,3, \quad \g_5^E = -\g_1^E\g_2^E\g_3^E\g_4^E = \g^M_5 . 
\end{equation}
We follow these conventions for our computations in Euclidean space.

\subsection{Gordon Identity}
The Gordon identity relates different Dirac structures through:
\begin{equation}
  \label{eq:GordonIdentity}
  \bar{u}(p_2)\g^\mu u(p_1) = \frac{1}{2M}\bar{u}(p_2)(2P^{\mu} + i\sigma^{\mu\nu}\Delta_\nu)u(p_1),
\end{equation}
where $M$ is the mass of the considered hadron and $P$ and $\Delta$ are defined as in section \ref{sec:HadronStructure}.


\bibliographystyle{unsrt}
\bibliography{../Bibliography}

\end{document}